\documentclass[a4paper, notitlepage,12pt]{article}

\usepackage[polutonikogreek,english]{babel}

\usepackage[T1]{fontenc}
\usepackage[latin9]{inputenc}
\usepackage[a4paper]{geometry}
\usepackage{amsmath}
\usepackage{amssymb}
\usepackage{setspace}
\usepackage{esint}
\usepackage{subcaption}
\usepackage[mathscr]{eucal}
\usepackage{arydshln}
\usepackage[pdftex]{graphicx}
\usepackage{verbatim}
\usepackage{booktabs}
\usepackage{units}
\usepackage{lscape}
\usepackage{pdflscape}
\usepackage{rotating}
\usepackage{nicefrac}
\usepackage{amsopn}
\usepackage{placeins}
\usepackage{lipsum}
\usepackage{tabularx}
\usepackage{xcolor}
\usepackage{enumitem}
\usepackage{ctable}
\usepackage[authoryear]{natbib}
\usepackage{sectsty}

\allowdisplaybreaks
\newtheorem{theorem}{Theorem}
\newtheorem{assumption}{Assumption}

\newcounter{rmk}
\newenvironment{rmk}[1][]{\refstepcounter{rmk}\par\medskip\noindent
	\textbf{Remark~\thermk.\ #1} \rmfamily} {\smallskip}
\newenvironment{remark}{\smallskip \begin{rmk}}{\hfill $\diamondsuit$ \end{rmk}} 

\definecolor{astral}{RGB}{46,116,181}
\subsectionfont{\color{astral}}
\subsubsectionfont{\color{astral}}
\sectionfont{\color{astral}}

\usepackage[colorlinks=true,linkcolor=red,urlcolor=astral,citecolor=astral,anchorcolor=astral,
pdftex,breaklinks,pdfencoding=auto,psdextra,
bookmarksopenlevel=1,bookmarksopen=true]{hyperref}

\numberwithin{equation}{section} 

\def\sym#1{\ifmmode^{#1}\else\(^{#1}\)\fi}

\def\spacingset#1{\renewcommand{\baselinestretch}%
	{#1}\small\normalsize} \spacingset{2}

\usepackage{footmisc}

\begin{document}
	\doublespacing
	\newcolumntype{L}[1]{>{\raggedright\arraybackslash}p{#1}}
	\newcolumntype{C}[1]{>{\centering\arraybackslash}p{#1}}
	\newcolumntype{R}[1]{>{\raggedleft\arraybackslash}p{#1}}
	
\title{\color{astral} A simple but powerful tail index regression  \thanks{ \protect\doublespacing\small We thank Yannick Hoga, seminar participants at the Universidad Carlos III de Madrid, and participants of the XVth Workshop in Time Series Econometrics in Zaragoza  for useful comments and suggestions. Paulo M. M. Rodrigues gratefully acknowledges funding by Fundação para a Ciência e a Tecnologia  (UIDB/00124/2020, UIDP/00124/2020 and Social Sciences DataLab - PINFRA/22209/2016), POR Lisboa and POR Norte (Social Sciences DataLab,PINFRA/22209/2016).} }

\author{Jo{\~a}o Nicolau$^{a}$ and Paulo M. M. Rodrigues$^{b}$ \\
	$^{a}${\small {}ISEG, Universidade de Lisboa and CEMAPRE}\\
	$^{b}${\small {}Banco de Portugal and Nova School of Business and Economics}}

\date{\today }

\maketitle
\thispagestyle{empty}

\vspace{-1.25cm}

\begin{abstract}
\noindent This paper introduces a flexible framework for the estimation of the conditional tail index of heavy tailed distributions. In this framework, the tail index is computed from an auxiliary linear regression model that facilitates estimation and inference based on established econometric methods, such as ordinary least squares (OLS), least absolute deviations, or M-estimation. We show theoretically and via simulations that OLS provides interesting results. Our Monte Carlo results highlight the adequate finite sample properties of the OLS tail index estimator computed from the proposed new framework and contrast its behavior to that of tail index estimates obtained by maximum likelihood estimation of exponential regression models, which is one of the approaches currently in use in the literature. An empirical analysis of the impact of determinants of the conditional left- and right-tail indexes of commodities' return distributions highlights the empirical relevance of our proposed approach. The novel framework's flexibility allows for extensions and generalizations in various directions, empowering researchers and practitioners to straightforwardly explore a wide range of research questions.

\noindent \textbf{Keywords:} Conditional tail index, OLS estimation, MLE, extreme value
\smallskip{}

\noindent \textbf{JEL classifications:} C12, C22, G17. \bigskip
 
\end{abstract}
\thispagestyle{empty}

\doublespacing

\newpage

\section{Introduction}
The tail index is a crucial parameter in economic analysis (\citealp{Gabaix2009}). It provides a measure of the heaviness of the tails of a distribution, indicating the likelihood of extreme outcomes, which is essential for the understanding and managing of various extreme phenomena in fields, such as economics, finance, and environmental sciences.  This information is useful, for instance, for risk managers, investors, financial institutions, policy makers and central banks to understand the potential for the occurrence of tail events that may originate large losses and for the implementation of appropriate risk mitigation strategies (see e.g. \citealp{OordtZhou2016},  and \citealp{Gabaix2016}). The tail index also plays an important role in environmental analysis for the determination of the likelihood of extreme phenomena, such as hurricanes, floods, and earthquakes; see e.g. \citet{ConteKelly2021}. Hence, the development and evaluation of suitable tail index estimation approaches is of considerable relevance for many fields of knowledge.

Tail index estimation has a long history in the statistical literature (see, \textit{inter alia}, \citealp{BeirlantGoegebeurSegersTeugelsWaalFerro04}, and \citealp{Resnick07}, for detailed discussions).  One of the first tail index estimators was proposed by \citet{FisherTippett1928}, and almost half a century later  \citet{Hill75} introduced a conditional maximum likelihood approach, which has become one of the most widely used procedures in this literature. Since then, several other approaches have been proposed, each with its own advantages and limitations. Examples include,  kernel estimators (\citealp{CsorgoDeheuvelsMason85}), method of moments estimators (\citealp{DekkersEinmahlDeHaan1989}), least-squares estimators (\citealp{BeirlantVynckierTeugels96}), smooth functional approaches (\citealp{Drees1998}), regression based estimators (\citealp{GabaixIbragimov12}, and \citealp{NicolauRodrigues18}), and small sample estimators (\citealp{Huismanetal2001}).  

However, all of the approaches above focus on the estimation of the unconditional tail index, without considering the potential impacts of covariates that may influence extreme events. This is an important shortcoming since the use of covariates may allow for a more comprehensive understanding of the tail dynamics. Neglecting the impact of regressors, such as macroeconomic variables, market risk proxies, and firm-specific factors, among others, can lead to incomplete risk assessments and inadequate decision-making processes. 

The recognition of this limitation has led over the last two decades to the introduction of several conditional tail index estimation methods. For instance, \citet{BeirlantGoegebeur03}  develop an exponential regression model for Pareto-type distributed generalized residuals, which is estimated using a profile likelihood approach. \citet{BeirlantGoegebeur04} consider local polynomial maximum likelihood estimation, providing nonparametric estimates of the parameter functions and their derivatives up to the degree of the chosen polynomial. \citet{Chavez_DemoulinDavison2005} introduce smooth  non-stationary  generalized  additive  modeling  for  sample extremes, in which spline smoothers are incorporated into models for exceedances over high thresholds, and maximum penalized likelihood used for estimation. \citet{GardesGirard2010} use a nearest neighbor approach to construct new tail index estimators, and \citet{GardesGuillouSchorgen2012} propose a class of kernel-type estimators of the tail index of a heavy-tailed distribution in the presence of covariates, and establish its main asymptotic properties under very general conditions. Recently, \citet{MaWeiHuang2020} use a logarithmic function to link the tail index to a nonlinear predictor induced by covariates, which forms the nonparametric tail index regression model, and develop an approach based on local likelihood methods for estimation; and \citet{LLY22} develop  a novel semiparametric tail index regression model, and provide consistent estimators for both parametric and nonparametric components of the model.

Our contribution in this paper to this literature is threefold: First, we introduce a flexible framework, inspired in the exponential regression model of \citet{BeirlantGoegebeur03} and \citet{WangTsai09}, which can easily be used to estimate the conditional tail index using established econometric methods. Second, we show the appropriateness of using least-squares to estimate the conditional tail index in this new framework. We provide the necessary asymptotic properties of the least-squares estimators, and a Monte Carlo study that illustrates their suitable finite-sample properties and compares the results with those obtained based on the exponential regression based method. Third, we contribute with an empirical analysis of the impact of relevant covariates on commodities returns'  left- and right-tail dynamics.  

The remainder of the paper is organized as follows.  Section \ref{Pareto} introduces the novel conditional tail index estimation framework. It is shown that OLS can be easily implemented, and the resulting least-squares estimators' properties are contrasted with those of the maximum likelihood tail index estimator of \citet{WangTsai09} which has been extended by \citet{NicolauRodriguesStoykov23}.	Section \ref{MC} presents a Monte Carlo evaluation of the procedures  and a discussion of the simulation results. In Section \ref{EmpiricalApp} an empirical analysis of the impact of several covariates on 23 commodities' returns left- and right-tail indexes is provided; and Section \ref{conclusion} concludes the paper. All proofs of the results presented in the main text, as well as plots of the covariates used in the empirical section are collected in the Supplementary Material part of the paper.

\section{Conditional Tail Index Estimation \label{Pareto}}

Consider the time series $(y_t,\boldsymbol{x}_t)$, $t=1,\ldots,n$, where $y_t\in\mathbb{R}$ is the response variable and $\boldsymbol{x}_t=(x_{1t},\ldots,x_{Kt})^{\prime}\in\mathbb{R}^K$ is a $K$-dimensional vector of explanatory variables. Let $F_{y_t|\boldsymbol{x}_t,y_t>\mathit{w}_n}(y_t|\boldsymbol{x}_t \text{ and }y_t>\mathit{w}_n)\equiv F(y_t|\boldsymbol{x}_t,\mathit{w}_n)=P(y_t\leq y|\boldsymbol{x}_t \text{ and }y_t>\mathit{w}_n)$ be the cumulative distribution function (CDF) of $y_t$ conditional on $\boldsymbol{x}_t$ and $y_t>\mathit{w}_n$, where $\mathit{w}_n\in\mathbb{R}$ is the tail cut-off point, which possibly depends on the sample size $n$.

\begin{remark}\label{Remark1}
	\textit{When the CDF is Pareto the choice of the tail cut-off point $\mathit{w}_n$ is of no consequence, as results are always exact due to the nature of the underlying density, e.g., $\hat{\mathit{w}}_n=\min\{y_1,...,y_n\}$ in the case of the right tail.   However, when  Pareto-type distributions are considered $\mathit{w}_n$ is crucial for correct tail index estimation (see e.g. \citealp{Hoga2020}). In our empirical analysis, in Section \ref{EmpiricalApp}, we use the discrepancy measure proposed by \citet{WangTsai09} to determine $\mathit{w}_n$, which has been shown to also work well in the case of weakly dependent data by \citet{NicolauRodriguesStoykov23}. In what follows, for notational convenience and without loss of generality, we will refer to the tail threshold parameter as $\mathit{w}_n$, regardless of whether the distribution under analysis is Pareto or Pareto-type.}
\end{remark}

\newpage

\subsection{A new tail index regression}
\subsubsection{Model framework \label{2.1.1}} 

To introduce the novel tail index estimation framework, consider that $y_t,$ conditional on $\boldsymbol{x}_t$, is governed by a Pareto distribution, \textit{viz.},
\begin{equation}
	F(y_t|\boldsymbol{x}_t,\mathit{w}_n)=\begin{cases}
		1-(y_t/\mathit{w}_n)^{-\alpha(\boldsymbol{x}_t, \boldsymbol{\beta})} &\text{ if }  y_t \ge \mathit{w}_n \\
		0 &\text{ if }  y_t < \mathit{w}_n
	\end{cases}, \qquad t=1,\cdots,n, \label{Pareto_CDF}
\end{equation}
where $\alpha(\boldsymbol{ x}_t,\boldsymbol{\beta})=\exp(\boldsymbol{ x}_t^{\prime}\boldsymbol{\beta})$ (as in e.g.  \citealp[p.601]{BeirlantGoegebeur03}, \citealp{WangTsai09} and \citealp{NicolauRodriguesStoykov23}) and $\mathit{w}_n$ is the tail cut-off point.  

Our econometric approach requires evaluating quantities such as $\boldsymbol{x}%
_{t}\boldsymbol{x}_{t}^{\prime }$ and $\boldsymbol{x}_{t}y_{t}$ within the subsample
where $y_{t}>w_{n}$, hence conditioned on $y_{t}>w_{n}$. For instance, to
assess the expected value of $\boldsymbol{x}_{t}\boldsymbol{x}_{t}^{\prime }$, we will
focus on the conditional value $E\left( \left. \boldsymbol{x}_{t}%
\boldsymbol{x}_{t}^{\prime }\right\vert y_{t}>w_{n}\right) $ rather than $%
E\left( \boldsymbol{x}_{t}\boldsymbol{x}_{t}^{\prime }I_{\left(
	y_{t}>w_{n}\right) }\right) ,$ where $I_{(.)}$ is the indicator function. The
latter computes the expected value $E\left( \left. \boldsymbol{x}_{t}%
\boldsymbol{x}_{t}^{\prime }\right\vert y_{t}>w_{n}\right) $ weighted by the
probability of observing $\left( y_{t}>w_{n}\right) $, i.e., $E\left( 
\boldsymbol{x}_{t}\boldsymbol{x}_{t}^{\prime }I_{\left( y_{t}>w_{n}\right) }\right) =%
E\left( \left. \boldsymbol{x}_{t}\boldsymbol{x}_{t}^{\prime }\right\vert
y_{t}>w_{n}\right) P\left( y_{t}>w_{n}\right) $. This approach requires
considering the entire sample space. Conversely, the former, $E%
\left( \left. \boldsymbol{x}_{t}\boldsymbol{x}_{t}^{\prime }\right\vert
y_{t}>w_{n}\right) $, focuses strictly on the subset where $y_{t}>w_{n}$
(the new sample space), which is the setting adopted in this paper, as it
exclusively considers the impact of $\boldsymbol{x}_{t}$ on $y_{t}$ when $%
y_{t}>w_{n}$.

Specifically, define the subset of indices, $S:=\left\{ t\in T:y_{t}>w_{n}\right\}$, where $T=\left\{ 1,2,..,n\right\} .$ In the analysis that follows, we consider the data subsample $\left\{
\left( y_{\tau },\boldsymbol{x}_{\tau }\right) ,\tau \in S\right\} $, which consists of the pairs $\left\{ \left( y_{t},\boldsymbol{x}_{t}\right)
\right\} _{t=1}^{n}$ conditioned on the elements $\left( y_{t},\boldsymbol{x}%
_{t}\right) $ such that $y_{t}>w_{n}.$ Given that $S$ has $n_{0}\leq n$
elements, and assuming that these are ordered as $\tau _{1},...,\tau
_{n_{0}}$, a proper way to write the subsample is $\left\{ \left( y_{\tau
	_{i}},\boldsymbol{x}_{\tau _{i}}\right) ,i=1,2,...,n_{0}\right\} .$ However, to
reduce notational complexity, we write $\left\{ \left( y_{\tau },%
\boldsymbol{x}_{\tau }\right) ,\tau =1,2,...,n_{0}\right\} $ with the implicit
understanding that $\tau =1$ is the first element of $S,$ $\tau=2$ the second
element of $S,$ and so on. In the case of the exact Pareto distribution, the size of this subsample is $n_{0}:=\left\lfloor \kappa
n\right\rfloor, $ with $\kappa \in \left( 0,1\right] $ and $\left\lfloor
.\right\rfloor $ being the floor function, and $w_{n}$ is
the $(\left( 1-\kappa \right)100)^{th}$ percentile of $\{y_{t}\}$.

Hence, from (\ref{Pareto_CDF})  $u_\tau=1-(y_\tau/\mathit{w}_n)^{-\alpha(\boldsymbol{x}_\tau, \boldsymbol{\beta})},$ where $u_\tau=F(y_\tau|\boldsymbol{x}_\tau,\mathit{w}_n)$ is uniformly distributed, $u_\tau\sim U\left( 0,1\right)$.  Furthermore, since  $y_\tau=F^{-1}(u_\tau|\boldsymbol{x}_\tau,\mathit{w}_n),$ we establish that $y_\tau$ is also Pareto distributed, i.e.,
\begin{equation}
	y_\tau=(1-u_\tau)^{-\alpha(\boldsymbol{x}_\tau, \boldsymbol{\beta})^{-1}}\mathit{w}_n \label{yt}.
\end{equation}

Applying natural logarithms to (\ref{yt}) and rearranging gives,
\begin{equation}
	\ln \left(\frac{y_{\tau}}{\mathit{w}_n}\right) =\alpha(\boldsymbol{x}_\tau, \boldsymbol{\beta})^{-1}v_{\tau}, \label{log_yt}
\end{equation}
where $ v_{\tau}= -\ln (1-u_\tau)$ and $E(v_\tau)=1$.  Moreover, reapplying the logarithmic transformation to (\ref{log_yt}) it follows that, $\ln \left( \ln \left({y_{\tau}}/{\mathit{w}_n}\right) \right) =\ln \left(\left(\alpha(\boldsymbol{x}_\tau, \boldsymbol{\beta})\right)^{-1}\right) +\ln v_{\tau}$. Recalling that $(\alpha(\boldsymbol{x}_{\tau}, \boldsymbol{\beta}))^{-1}=e^{-\boldsymbol{x}_{\tau}^{\prime }\boldsymbol{\beta} }$, then, $\ln \left( \ln \left( {y_{\tau}}/{\mathit{w}_n}\right) \right) = -\boldsymbol{x}_{\tau}^{\prime}\boldsymbol{\beta} +\ln v_{\tau}$, or simply,
\begin{equation}
	-\ln \left( \ln \left( \frac{y_{\tau}}{\mathit{w}_n}\right) \right) =\boldsymbol{x}_{\tau}^{\prime}\boldsymbol{\beta} +a_{\tau},  \label{at}
\end{equation}
where $ a_{\tau}=-\ln v_{\tau}=-\ln \left( -\ln (1-u_\tau)\right)$. Interestingly, $a_\tau$ in (\ref{at}) follows a standard Gumbel distribution\footnote{Also known as type-I generalized extreme value distribution.}, $P\left( a_{\tau}<a\right) =e^{-e^{-a}}$, with $E\left( a_{\tau}\right) =\gamma$,  where $\gamma \approx 0.5777$ is Euler's constant (also known as Euler-Mascheroni constant); see the Appendix for proof. 

Hence, considering $a_{\tau}=\xi_{\tau}+\gamma$ so that $\xi_{\tau}=a_{\tau}-\gamma$ and $E\left(\xi_{\tau}\right) =0$, we can further write (\ref{at}) as, $
-\ln \left( \ln \left({y_{\tau}}/{\mathit{w}_n}\right) \right) =\boldsymbol{x}_{\tau}^{\prime}\boldsymbol{\beta} +\xi_{\tau}+\gamma$,
establishing in this way the new conditional tail index regression framework,
\begin{equation}
	z_{\tau}=\boldsymbol{x}_{\tau}^{\prime }\boldsymbol{\beta} +\xi_{\tau}, \quad \tau=1,...,n_0, \label{TailIndexRegression1}
\end{equation}%
where $z_{\tau}=-\ln \left( \ln \left( \frac{y_{\tau}}{\mathit{w}_n}\right) \right)
-\gamma$. This is a simple linear regression framework from which the parameters that characterize the tail index can be straightforwardly computed by e.g. OLS as will be shown next. 

\subsubsection{The OLS tail index estimator}

To compute $ \widehat{\boldsymbol{\beta}}_{OLS}=\underset{\boldsymbol{\beta}}{\arg\min} \sum_{\tau=1}^{n_0}(z_\tau - \boldsymbol{x}_{\tau}^{\prime }\boldsymbol{\beta})^2$,  from (\ref{TailIndexRegression1}) we consider that the following Assumptions hold.

\newpage

\begin{assumption} \label{Ass1}
	$\left\{ \left( y_{\tau },\boldsymbol{x}_{\tau }\right)\right\} _{\tau =1}^{n_{0}}$ is an ergodic stationary process.
\end{assumption}

\begin{assumption} \label{Ass2}
	$E\left( \left. \xi _{\tau }\right\vert \boldsymbol{x}_{\tau }\right) =0,$ $\tau =1,...,n_{0}.$
\end{assumption}

\begin{assumption} \label{Ass3}
	The $K\times K$ matrix $\boldsymbol{\Sigma} _{xx}=E\left( \boldsymbol{x}_{\tau }\boldsymbol{x}_{\tau }^{\prime }\right) ,$ $\tau=1,,...,n_{0},$ is finite and nonsingular.
\end{assumption}

\begin{assumption} \label{Ass4} Consider that, 
	\begin{itemize}
		\item[(a)] for $j=0\pm 1,...,$ define the $K\times K$
		autocorrelation function of $\boldsymbol{g}_{\tau }=\boldsymbol{x}_{\tau }\varepsilon _{\tau
		} $ as $\boldsymbol{\Gamma} \left( j\right) =E\left( \boldsymbol{g}_{\tau }\boldsymbol{g}%
		_{\tau -j}^{\prime }\right).$ Then $\sum_{j=-\infty }^{\infty }\left\Vert
		\boldsymbol{\Gamma} \left( j\right) \right\Vert <\infty ,$ where $\left\Vert .\right\Vert 
		$ is the L1 norm, and the long-run variance-covariance matrix $\mathbf{V:}%
		=\sum_{j=-\infty }^{\infty }\boldsymbol{\Gamma} \left( j\right) $ is positive definite.
		\item[(b)]  $E\left( \left. \boldsymbol{g}_{\tau }\right\vert \boldsymbol{g}_{\tau
			-j},\boldsymbol{g}_{\tau -j-1},...\right) \overset{qm}{\longrightarrow }0$ as $%
		j\longrightarrow \infty,$ where $\overset{qm}{\longrightarrow }$ denotes convergence in quadratic mean. 
		\item[(c)] $\sum_{j=0}^{\infty }\left( E\left( 
		\mathbf{r}_{j}^{\prime }\boldsymbol{r}_{j}\right) \right) ^{1/2}< \infty,$ where $\boldsymbol{%
			r}_{j}=E\left( \left. \boldsymbol{g}_{\tau }\right\vert \boldsymbol{g}%
		_{\tau -j},\boldsymbol{g}_{\tau -j-1},...\right) -E\left( \left. 
		\boldsymbol{g}_{\tau }\right\vert \boldsymbol{g}_{\tau -j-1},\boldsymbol{g}_{\tau
			-j-2},...\right)$.
	\end{itemize}
\end{assumption}

\begin{remark}
	\textit{Assumption \ref{Ass2}, $E\left( \left. \xi _{\tau
		}\right\vert \mathbf{x}_{\tau }\right) =0,$ implies correct
		specification of the model, but it does not imply that $\boldsymbol{g}_{\tau }$ is a
		martingale difference. Dependence of $\boldsymbol{g}_{\tau }$ on its past history,
		in terms of mean and variance, is allowed. However, according to Assumptions \ref{Ass4}(b) and \ref{Ass4}(c), this dependence vanishes to zero as the lag order $j \longrightarrow \infty $.}
\end{remark}

Hence, considering Assumptions \ref{Ass1} - \ref{Ass4} the following Theorem can be stated (see \citealp{Hong2020} and \citealp{White01}).
\begin{theorem} \label{Th1} Given (\ref{Pareto_CDF}) as $n \rightarrow \infty$ (and $n_0 \rightarrow \infty$) it follows that: \newline
	(i) Under Assumptions \ref{Ass1} to \ref{Ass4}(a) $\boldsymbol{\hat{\beta} }\overset{p}{\longrightarrow }\boldsymbol{\beta};$ \\  (ii) Additionally, if Assumptions \ref{Ass4}(b) and \ref{Ass4}(c) also hold, then $
	\sqrt{n_{0}}\left( \boldsymbol{\hat{\beta}}_{OLS}-\boldsymbol{\beta} \right) \overset{d}{%
		\longrightarrow }N\left( \boldsymbol{0,\Sigma }_{xx}^{-1}\boldsymbol{V\Sigma }%
	_{xx}^{-1}\right)$,	where $\overset{p}{\longrightarrow }$ and $\overset{d}{\longrightarrow }$ denote convergence in probability and in distribution, respectively.
\end{theorem}

It is important to note that the effective convergence rate is being reduced  - it is given by $n_0$, the number of observations in the tail, which is never greater than $n$ (the full sample size); see \cite{NicolauRodriguesStoykov23}.

\begin{remark}\label{Rem3}\textit{In the special case where $\left\{ \boldsymbol{g}_{\tau }\right\} $ is a  martingale difference sequence (m.d.s.) and the errors are homoskedastic, under Assumption \ref{Ass3} the $\mathbf{V}$ matrix in Assumption \ref{Ass4}(a) simplifies to $\mathbf{V}=\boldsymbol{\Gamma} \left( 0\right) =\sigma _{\xi
		}^{2}E\left( \boldsymbol{x}_{\tau }\boldsymbol{x}_{\tau }^{\prime }\right) 
		$ and 
		\begin{equation}
			\sqrt{n_{0}}\left( \boldsymbol{\hat{\beta}}_{OLS}-\boldsymbol{\beta} \right) \overset{d}{%
				\longrightarrow }N\left( \mathbf{0,}\sigma _{\xi }^{2}\mathbf{\Sigma }%
			_{xx}^{-1}\right). \label{OLS_iid} 
	\end{equation}}
\end{remark}

\begin{remark}\label{Rem4}
	\textit{If $y_\tau$ is Pareto distributed and $\alpha(\boldsymbol{x}_\tau,\boldsymbol{\beta})$ is correctly specified then the special case of Remark \ref{Rem3} holds with $\sigma _{\xi }^{2}=\pi^2/6$ given that $\{\xi_t\}$ is a sequence of iid Gumbel distributed random variables. Note that in this case $n_0=n.$}
\end{remark}

\subsubsection{Pareto-type Distributions \label{PtD}}
The assumption that the response variable follows an exact Pareto distribution may be restrictive in empirical applications. Therefore, in this section, will allow for a broader class of heavy tailed distributions. To that end, consider the Pareto-type distribution,
\begin{equation}
F(y_\tau; \boldsymbol{x}_{\tau}, \boldsymbol{\beta}, w_{n})=1-  \left(\frac{y_{\tau}}{w_{n}}\right)^{-\alpha(\boldsymbol{x}_{\tau}, \boldsymbol{\beta})}\mathcal{L}(y_{\tau};  \boldsymbol{x}_{\tau}, \boldsymbol{\beta})  \label{PtD1}
\end{equation}
where $\mathit{w}_n$ is the tail cut off point, $\alpha(\boldsymbol{x}_{\tau}, \boldsymbol{\beta})$  is the tail index, which controls the heaviness of the tail, and $\mathcal{L}(y_{\tau};  \boldsymbol{x}_{\tau}, \boldsymbol{\beta}) =L\left(y_{\tau};  \boldsymbol{x}_{\tau}, \boldsymbol{\beta}\right) /L\left(
w_{n}; \boldsymbol{x}_{\tau}, \boldsymbol{\beta}\right)$ is a slowly varying function at infinity,  such that for any $\mathit{w}_n>0$,  $\mathcal{L}(\mathit{w}_n y_{\tau};\boldsymbol{x}_{\tau}, \boldsymbol{\beta})/\mathcal{L}(y_{\tau};\boldsymbol{x}_{\tau}, \boldsymbol{\beta})\longrightarrow1$ as $y_{\tau}\longrightarrow\infty$. $\mathcal{L}(y_{\tau};  \boldsymbol{x}_{\tau}, \boldsymbol{\beta})$ changes very slowly and does not affect the tail index but might introduce some secondary effects on the tail behavior (see e.g. \citealp{Maetal2018}).

Under certain conditions, which we will present below, specification (\ref{PtD1}) allows for a number of different distributions (up to a scaling factor) as special cases, such as, e.g. the Burr, the Student-t and the $\alpha$-stable distribution (\citealp{BeirlantGoegebeur03}). Importantly, as $y_{\tau} \rightarrow \infty$, specification (\ref{PtD1}) collapses to (\ref{Pareto_CDF}), the exact
Pareto distribution up to a scaling factor.


Following the approach used in section \ref{2.1.1} it follows that, 
\begin{equation}
	1-\left(\frac{y_\tau}{w_{n}}\right)^{-\alpha(\boldsymbol{x}_{\tau}, \boldsymbol{\beta})}\mathcal{L}(y_{\tau};  \boldsymbol{x}_{\tau}, \boldsymbol{\beta}) =u_{\tau}
\end{equation}
which after applying logarithms twice and rearranging gives, 
\begin{equation}
	-\ln \ln \left( \frac{y_{\tau}}{w_{n}}\right)  =\ln \alpha(\boldsymbol{x}_{\tau}, \boldsymbol{\beta}) -\ln \left( -\ln \left( 1-u_{\tau}\right) \right)
	-\ln \left( 1+\frac{\ln \mathcal{L}(y_{\tau};  \boldsymbol{x}_{\tau}, \boldsymbol{\beta})}{-\ln \left(
		1-u_{\tau}\right) }\right)   \label{approx}
\end{equation}
which in line with (\ref{TailIndexRegression1}) we can write as,
\begin{equation}
	z_t =\boldsymbol{x}'_{\tau} \boldsymbol{\beta}+\xi_{\tau} +\epsilon_{\tau}
\end{equation}
where $\epsilon_{\tau} :=-\ln \left( 1+\frac{\ln  \mathcal{L}(y_{\tau};  \boldsymbol{x}_{\tau}, \boldsymbol{\beta})}{-\ln \left( 1-u_t\right) }\right)$ and $-\ln \left( -\ln \left( 1-u_{\tau}\right) \right)=\xi_{\tau}+\gamma$.

Hence, we see from (\ref{approx}) that, to ensure consistency of the OLS estimator, it is necessary that 	$  \mathcal{L}(y_{\tau};  \boldsymbol{x}_{\tau}, \boldsymbol{\beta}) \rightarrow 1$ as $y_{\tau}\rightarrow \infty ,$
which implies that $\epsilon_{\tau} \rightarrow 0$. Note that $-\ln \left( 1-u_{\tau}\right)$ follows an exponential distribution with parameter $\lambda=1$.

Furthermore, if $y_{\tau}=\varphi w_{n},$ with $\varphi >0$, for $w_{n}\rightarrow \infty$, it follows that,
\[
\mathcal{L}(y_{\tau};  \boldsymbol{x}_{\tau}, \boldsymbol{\beta}) =\frac{L\left(y_{\tau};  \boldsymbol{x}_{\tau}, \boldsymbol{\beta}\right) }{L\left(w_n;  \boldsymbol{x}_{\tau}, \boldsymbol{\beta}\right)  }=\frac{L\left(\varphi w_n;  \mathbf{x}_{\tau}, \boldsymbol{\beta}\right)  }{L\left(w_n; \boldsymbol{x}_{\tau}, \boldsymbol{\beta}\right) }\rightarrow 1.
\]

Given that \( \epsilon_{\tau} \) vanishes as \( y_{\tau} \to \infty \), it does not contribute significantly to the error term, and thus $\lim_{y_{\tau} \to \infty} \mathbb{E}[\epsilon_t | \boldsymbol{x}_{\tau}] = 0$ which ensures that, $
\lim_{y_{\tau} \to \infty} \mathbb{E}[\xi_{\tau} + \epsilon_{\tau} | \boldsymbol{x}_{\tau}] = \mathbb{E}[\xi_{\tau} | \boldsymbol{x}_{\tau}] = 0.$ This convergence to zero is crucial for ensuring that the error term becomes negligible, allowing OLS to produce consistent parameter estimates.  Since  $L$ is a slowly varying at infinity, $\ln \left( \mathcal{L}\right) $ is also a slowly varying at infinity but with even slower variation at infinity. As a result, $\epsilon _{\tau}$ which includes a double logarithm of $L$ should remain nearly constant for moderate values of $y_{\tau}$. Consequently, the dependence between $x_{\tau}$ and $\epsilon _{\tau}$ is expected to be minimal. 

These results provide support to the assumption that in these circumstances the underlying CDF of these observations is (approximately) Pareto, validating the extension of the procedure described above to more general settings than just the pure Pareto.  Consequently, we can state that:

\begin{theorem} \label{Th2} Given (\ref{PtD1}), under Assumptions \ref{Ass1} to \ref{Ass4}, as $n \rightarrow \infty$ (and $n_0 \rightarrow \infty \text{ and }  w_n \rightarrow \infty$) it follows that the results in Theorem \ref{Th1} remain unchanged.
\end{theorem}

\newpage

\begin{remark}
	Regarding the tail cut-off point, $w_n$, let us assume that $w_{n}=O\left( n^{\delta }\right) .$ For large $w_{n},$ 
	the number of observations exceeding $w_{n}$ is given by $
	n_{0}=nP\left( y_{t}>w_{n}\right) \sim n^{1-\alpha \delta }$
	assuming that $y_{t}$ follows a Pareto-type distribution and where $\alpha$ is the tail index. To ensure that a
	sufficiently large number of observations remains above the threshold, and 
	that $n_{0}\rightarrow \infty $ (albeit growing at a slower rate than the
	overall sample size $n$), we impose the condition  $0<\alpha \delta <1,$
	which implies $\delta \in \left( 0,1/\alpha \right) .$ 	If the tail index $\alpha $ is unknown (though it typically falls within the
	range of 1 to 4), it is always possible to choose a relatively small positive $\delta ,$ e.g. $\delta =1/6,$ to satisfy the restriction.	
\end{remark}

The Monte Carlo results based on Pareto type distributions provided in Section \ref{MC} corroborate this conclusion.

\subsection{The MLE tail index estimator \label{MLE}}
To illustrate the usefulness of the OLS approach introduced in the previous section, we compare it with the exponential regression based approach of \citet{WangTsai09}. To briefly introduce the method consider the general case of a survival function of a Pareto-type distribution as in  (\ref{PtD1}). Hence, considering, as $y_\tau\longrightarrow\infty$, a suitable tuning parameter $\mathit{w}_n$,  the conditional probability density function of $y_\tau$ can be approximated, up to a scaling factor (in view of the properties of $\mathcal{L}(y_\tau|\boldsymbol{x}_{\tau})$), as 
\begin{equation}
	f(y_\tau|\boldsymbol{x}_\tau,\mathit{w}_n,\boldsymbol{\beta})=\alpha(\boldsymbol{x}_\tau,\boldsymbol{\beta})(y_\tau/\mathit{w}_n)^{-\alpha(\boldsymbol{x}_\tau, \boldsymbol{\beta})-1}\mathit{w}_n^{-1}. \label{density}
\end{equation}

Following \citet{NicolauRodriguesStoykov23}, the average of the log-likelihood function of $y_\tau$, conditional on $\boldsymbol{x}_\tau$  constructed based on (\ref{density}), ignoring, without loss of generality, the constant term $-\ln(\mathit{w}_n)$, is, 
\begin{equation}
	\ell_{n_0}(\boldsymbol{\beta})
	=\frac{1}{n_0}\sum\limits_{\tau=1}^{n_0}(\boldsymbol{x}'_\tau\boldsymbol{\beta}-(\exp(\boldsymbol{x}'_\tau\boldsymbol{\beta})+1)\ln(y_\tau/\mathit{w}_n)), \label{eq: log-lik}
\end{equation}
with corresponding score vector, 
\begin{equation}
	G_{n_0}(\boldsymbol{\beta})\equiv\frac{\partial \ell_{n_0}(\boldsymbol{\beta})}{\partial\boldsymbol{\beta}}
	=\frac{1}{n_0}\sum\limits_{\tau=1}^{n_0}\boldsymbol{x}_\tau'(1-\exp(\boldsymbol{x}'_\tau\boldsymbol{\beta})\ln(y_\tau/\mathit{w}_n)), \hspace{0.25cm} \label{eq: G}
\end{equation}
and Hessian, 
\begin{equation}
	H_{n_0}(\boldsymbol{\beta})\equiv\frac{\partial^2\ell_{n_0}(\boldsymbol{\beta})}{\partial\boldsymbol{\beta}\partial\boldsymbol{\beta}'}
	=-\frac{1}{n_0}\sum\limits_{\tau=1}^{n_0}\boldsymbol{x}_\tau'\boldsymbol{x}_\tau\exp(\boldsymbol{x}'_\tau\boldsymbol{\beta})\ln(y_\tau/\mathit{w}_n),  \label{eq: H}
\end{equation}
which is negative definite. Note that $\mathit{w}_n$ needs to grow at a suitable rate to guarantee that enough observations are available for estimation. Therefore, the solution to $G_{n_0}(\boldsymbol{\hat{\boldsymbol{\beta}}}_{MLE})=0$, where $\boldsymbol{\hat{\beta}}_{MLE}$ is a vector of parameter estimates (if they exist), corresponds to a global maximum.\footnote{In general, a closed-form solution to the maximization problem does not exist unless $\boldsymbol{x}_\tau$ is one dimensional and constant, which leads to the Hill estimator (\citealp{Hill75}), and numerical methods have to be used.} Moreover, under certain regularity conditions, including i.i.d., it can be shown (see \citealp{NicolauRodriguesStoykov23} and \citealp{WangTsai09}) that, $\sqrt{n_0}G_{n_0}(\boldsymbol{\beta}) \overset{d}{\longrightarrow} N(\boldsymbol{0},\boldsymbol{\Sigma}_{xx})$,  $H_{n_0}(\boldsymbol{\beta})\overset{p}{\longrightarrow}-\boldsymbol{\Sigma}_{xx}$, and
\begin{eqnarray}
	\sqrt{n_0}(\boldsymbol{\hat{\beta}}_{MLE}-\boldsymbol{\beta})&\overset{d}{\longrightarrow}&N(\boldsymbol{0},\boldsymbol{\Sigma}_{xx}^{-1}). \label{MLE1}
\end{eqnarray}

\subsection{Estimation Efficiency and Bias\label{Model Efficiency}}
\subsubsection{Estimation Efficiency}
Under correct model specification, it is well known that OLS is generally less efficient than MLE. For illustration in the current context, consider that $y_\tau$ is Pareto distributed as in (\ref{Pareto_CDF}), and that $\left( y_{\tau},\boldsymbol{x}_{\tau}\right) $ are i.i.d. From (\ref{OLS_iid}) in Remark \ref{Rem3} and (\ref{MLE1}) we establish that the lower efficiency of the linear regression OLS estimator when compared to exponential regression MLE, is a consequence of the presence of $\sigma ^{2}_\xi$ in $var\left( \hat{\boldsymbol{\beta}}_{OLS}\right)$; see (\ref{OLS_iid}). It is however, important to note that in small samples the OLS estimators are centered under general conditions, whereas MLE is biased (simulation results in Section \ref{MC} confirm this statement; see also \citealp{rytgaard_1990}). This is an important feature in the tail index estimation context as samples from Pareto-type distributions can generally be relatively small.  

Moreover, we will show analytically (in this section) and via Monte Carlo simulation (in section \ref{MC}) that, although MLE may be more efficient when the model is correctly specified, this advantage diminishes or reverses in favor of OLS when variables are omitted, especially if the omitted variables have a significant impact on the tail index.

\subsubsection{Model Misspecification}\label{2.3.2}
We also note that there is an interesting contrast in the efficiency/consistency between MLE and OLS under model misspecification. For illustration, consider a Pareto distributed response variable as in (\ref{Pareto_CDF}), but now $\alpha(\boldsymbol{X}_\tau,\boldsymbol{\Theta})=\exp \left( \boldsymbol{X}_{\tau}^{\prime }\boldsymbol{\Theta} \right)=\exp \left( \boldsymbol{x}_{\tau}^{\prime }\boldsymbol{\beta} +\theta x_{\tau}^{\ast }\right)$, where $\boldsymbol{X}_\tau=(\mathbf{x}_\tau, x_{\tau}^{\ast })'$, $\boldsymbol{\Theta}=(\boldsymbol{\beta}, \theta)'$, $\boldsymbol{x}_{\tau}$ is a $K \times 1$ vector of covariates including a constant, $\boldsymbol{\beta}$ is the corresponding vector of parameters, and $x_{\tau}^{\ast }$ is a scalar mean zero random variable independent of $\boldsymbol{x}_{\tau}$ with the corresponding slope parameter $\theta$. In contrast to the previous section we consider, when estimating the tail index, that $x_{\tau}^{\ast }$ is omitted and that the tail index is incorrectly assumed to be $\alpha(\boldsymbol{x}_{\tau},\boldsymbol{\beta}) =\exp \left(\boldsymbol{x}_{\tau}^{\prime }\boldsymbol{\beta} \right)$ instead of the correct $\alpha(\boldsymbol{X}_\tau,\boldsymbol{\Theta})=\exp \left( \boldsymbol{x}_{\tau}^{\prime }\boldsymbol{\beta} +\theta x_{\tau}^{\ast }\right)$.

\bigskip

\noindent \textbf{The MLE approach}

\bigskip

In the MLE approach, the score vector is as in (\ref{eq: G}), however, now it is misspecified as the covariate $x_\tau^*$ is omitted, that is, 
\begin{equation}
	G_{n_0}(\boldsymbol{\beta})
	=\frac{1}{n_0}\sum\limits_{\tau=1}^{n_0}\boldsymbol{x}_\tau'(1-\exp(\boldsymbol{x}_\tau'\boldsymbol{\beta})\ln(y_\tau/\mathit{w}_n)). \label{eq: G1}
\end{equation}

Thus, denoting $\varepsilon _{\tau}^\ast= 1-\exp \left( \boldsymbol{x}_\tau^{\prime }\boldsymbol{\beta} \right) \ln \left(\frac{y_{\tau}}{\mathit{w}_{n}}\right)$, for $\theta \ne 0$ (and $\exp(\theta x_{\tau}^{\ast })\ne1$), it follows that,	$E\left( \left. \varepsilon _{\tau}^\ast\right\vert \boldsymbol{x}_{\tau},x_{\tau}^{\ast
}\right) =E\left( \left. 1-\exp \left( \boldsymbol{x}_\tau^{\prime }\boldsymbol{\beta} \right)
\ln \left( \frac{y_{\tau}}{\mathit{w}_{n}}\right) \right\vert \boldsymbol{x}_{\tau},x_{\tau}^{\ast }\right)
$ $=1-\exp \left( \boldsymbol{x}_\tau^{\prime }\boldsymbol{\beta} \right) (\exp \left( \boldsymbol{x}_\tau^{\prime
}\boldsymbol{\beta} +\theta x_\tau^{\ast }\right))^{-1} =1-\exp \left( -\theta x_\tau^{\ast }\right) \ne 0$,
and consequently, the impact on  the score $G_{n_0}\left( \boldsymbol{\beta} \right) $ of omitting $x_{\tau}^{\ast }$, is, $
E\left( \left. G_{n_0}\left( \boldsymbol{\beta} \right) \right\vert
\boldsymbol{x}_{\tau},x_{\tau}^{\ast }\right) =\frac{1}{n_0}\sum_{\tau=1}^{n_0}E\left(
\left. \boldsymbol{x}_{\tau}\varepsilon _{\tau}^{\ast}\right\vert \boldsymbol{x}_{\tau},x_{\tau}^{\ast }\right)=\frac{1}{n_0}\sum_{\tau=1}^{n_0}\boldsymbol{x}_{\tau}\left( 1-\exp \left( -\theta x_{\tau}^{\ast
}\right) \right)$.
Since, 
\begin{equation}
	E\left( G_{n_0}\left( \boldsymbol{\beta} \right)
	\right) =E\left( E\left( \left. G_{n_0}\left( \boldsymbol{\beta}
	\right) \right\vert \boldsymbol{x}_{\tau},x_{\tau}^{\ast }\right) \right) =E\left(
	\boldsymbol{x}_{\tau}\right) -E\left( \boldsymbol{x}_{\tau}\exp \left( -\theta x_{\tau}^{\ast
	}\right) \right), \label{EG}
\end{equation}%
and considering that $\boldsymbol{x}_{\tau}$ and $x_{\tau}^{\ast }$ are independent, (\ref{EG}) simplifies to,
\begin{equation}
	E\left( G_{n_0}\left( \boldsymbol{\beta} \right) \right) =E\left(
	\boldsymbol{x}_{\tau}\right) \left( 1-E\left( \exp \left( -\theta x_{\tau}^{\ast
	}\right) \right) \right). \label{EG1}
\end{equation}
Hence, for $\theta \ne 0$ and  $E\left( \exp \left( -\theta x_{\tau}^{\ast
}\right) \right) \ne 1$, we observe that (\ref{EG1}) is different from zero, which implies that the
MLE is inconsistent.

Furthermore, considering a parameter estimate  $\bar{\boldsymbol{\beta}} $, such that  $\bar{\boldsymbol{\beta}} $ lies between $\hat{\boldsymbol{\beta}}_{MLE} $ and $\boldsymbol{\beta}$, it follows that $H_{n_0}\left( \bar{\boldsymbol{\beta}}\right) \overset{p}{\longrightarrow }\mathbb{H}\left(
\boldsymbol{\beta}\right)$, but where now,%
\begin{equation}
	\mathbb{H}\left( \boldsymbol{\beta}\right) =-E\left(\boldsymbol{x}_{\tau}\boldsymbol{x}_{\tau}^{\prime }\right) E\left( \exp\left( -\theta x_{\tau}^{\ast }\right) \right); \label{Hessian1}
\end{equation}%
see Appendix for details.

Hence, from a Taylor series expansion of $G_{n_0}\left( \boldsymbol{\hat{\beta}} \right)$ around $\boldsymbol{\beta}$,
$G_{n_0}\left( \boldsymbol{\hat{\beta}} \right)\approx G_{n_0}\left( \boldsymbol{\beta} \right)+ (\hat{\boldsymbol{\beta}}_{MLE}-\boldsymbol{\beta})H_{n_0}\left( \boldsymbol{\beta} \right)$ we establish that, 
$\hat{\boldsymbol{\beta}}_{MLE}-\boldsymbol{\beta}=\left(-H_{n_0}\left( \bar{\boldsymbol{\beta}} \right)\right)^{-1}G_{n_0}\left( \boldsymbol{\beta} \right)$. Consequently, from (\ref{EG1}) and (\ref{Hessian1}),
\begin{equation}
	\hat{\boldsymbol{\beta}}_{MLE}-\boldsymbol{\beta} \overset{p}{\longrightarrow}	\left[E\left( \boldsymbol{x}_{\tau}\boldsymbol{x}_{\tau}^{\prime }\right)\right]^{-1}E\left(
	\boldsymbol{x}_{\tau}\right) \frac{\left( 1-E\left( \exp \left( -\theta
		x_{\tau}^{\ast }\right) \right) \right) }{E\left( \exp \left( -\theta
		x_{\tau}^{\ast }\right) \right) }.
\end{equation}

To better characterize the bias, consider the $K \times 1$ vector $\boldsymbol{e}_1=\left[ 
\begin{array}{c}
	1 \\ 
	\mathbf{0}
\end{array}%
\right]$, where $\mathbf{0}$ is a $(K-1)\times 1$ vector of zeros. Hence, since $\boldsymbol{x}_\tau$ includes an intercept,
\begin{equation}
	E\left( \boldsymbol{x}_{\tau}\boldsymbol{x}_{\tau}^{\prime }\right) \boldsymbol{e}_1=E\left(
	\boldsymbol{x}_{\tau}\right). \label{bias11}
\end{equation}
Pre-multiplying (\ref{bias11}) by $\left[E\left( \boldsymbol{x}_{\tau}\boldsymbol{x}_{\tau}^{\prime }\right)\right]^{-1}$ it follows that,
\begin{equation}
	\boldsymbol{e}_1=\left[E\left( \boldsymbol{x}_{\tau}\boldsymbol{x}_{\tau}^{\prime }\right)\right]^{-1} E\left(
	\boldsymbol{x}_{\tau}\right), \label{bias31}
\end{equation}
since $\left[E\left( \boldsymbol{x}_{\tau}\boldsymbol{x}_{\tau}^{\prime }\right)\right]^{-1} $ exists under Assumption \ref{Ass3}.

Consequently, making use of (\ref{bias31}) we see that,%
\begin{equation}
	\hat{\boldsymbol{\beta}}_{MLE}-\boldsymbol{\beta} \overset{p}{\longrightarrow} \boldsymbol{e}_1 \left(E\left( \exp \left( \theta x_{\tau}^{\ast}\right)\right)-1 \right). \label{bias_MLE11}
\end{equation}

For concreteness, considering $x_{\tau}^{\ast }\sim N\left( \mu ,\sigma ^{2}\right)$, then $E%
\left( \exp \left( \theta x_{\tau}^{\ast }\right) \right) = \exp \left(\theta \mu -\frac{\theta ^{2}\sigma ^{2}}{2}\right)$ and (\ref{bias_MLE11}) is, $
\hat{\boldsymbol{\beta}}_{MLE}-\boldsymbol{\beta} \overset{p}{\longrightarrow} \boldsymbol{e}_1 \left(\exp \left(\theta \mu -\frac{\theta ^{2}\sigma ^{2}}{2}\right)-1\right). $

Moreover, to evaluate the impact on the asymptotic variance of $
\sqrt{n_0}\left( \hat{\boldsymbol{\beta}}_{MLE} - \boldsymbol{\beta} _{0}\right), $ consider, 
\begin{equation}
	Avar\left( \sqrt{n_0}\left( \hat{\boldsymbol{\beta}}_{MLE} - \boldsymbol{\beta} _{0}\right)
	\right) =\mathbb{H}(\boldsymbol{\beta} _{0})^{-1}\boldsymbol{\Sigma}(\boldsymbol{\beta} _{0}) \mathbb{H}(\boldsymbol{\beta} _{0})^{-1}, \label{avar_bias}
\end{equation}
where $\mathbb{H}(\boldsymbol{\beta} _{0})=E\left( \boldsymbol{x}_{\tau}\boldsymbol{x}_{\tau}^{\prime }\right) E\left( \exp
\left( -\theta x_{\tau}^{\ast }\right) \right)$ and
\begin{equation}
	\boldsymbol{\Sigma}(\boldsymbol{\beta}) =E\left( \varepsilon _{\tau}^{\ast 2}\right) E\left(
	\boldsymbol{x}_{\tau}\boldsymbol{x}_{\tau}^{\prime }\right) -E\left( \varepsilon _{\tau}^\ast \boldsymbol{x}_{\tau}\right) 
	E\left( \varepsilon _{\tau}^\ast \boldsymbol{x}_{\tau}^{\prime }\right); \label{AVar}
\end{equation}%
see Appendix for details.

We first need to establish the result for $E\left( \varepsilon _{\tau}^{\ast 2}\right).$ Given that $\ln \left( \frac{y_{\tau}}{\mathit{w}_n}\right) $ is exponentially distributed with
parameter $\alpha(\boldsymbol{X}_\tau,\boldsymbol{\Theta}) $ and $ E\left( \ln \left( \frac{y_{\tau}}{\mathit{w}_n}\right)^{2}\right) =2(\alpha(\boldsymbol{X}_\tau,\boldsymbol{\Theta}))^{-2}$ we have that,
\begin{equation}
	E\left( \left. \varepsilon _{\tau}^{\ast 2}\right\vert \boldsymbol{x}_{\tau},x_{\tau}^{\ast
	}\right) =1-2\exp \left( -\theta x_{\tau}^{\ast }\right) +2\exp \left( -2\theta
	x_{\tau}^{\ast }\right) \label{e2}
\end{equation}%
and 
\begin{equation}
	E\left( \varepsilon _{\tau}^{\ast 2}\right)  =E\left( 1-2\exp
	\left( -\theta x_{\tau}^{\ast }\right) +2\exp \left( -2\theta x_{\tau}^{\ast
	}\right) \right); 
\end{equation}
see Appendix for details.

Therefore, we establish for (\ref{AVar}) that,
\begin{eqnarray}
	\boldsymbol{\Sigma}(\boldsymbol{\beta}) &=&\left( 1-2\exp \left( -\theta x_{\tau}^{\ast }\right) +2\exp \left( -2\theta
	x_{\tau}^{\ast }\right) \right) E\left( \boldsymbol{x}_{\tau}\boldsymbol{x}_{\tau}^{\prime }\right) 
	\notag \\
	&&-\left( 1-E\left( \exp \left( -\theta x_{\tau}^{\ast }\right)
	\right) \right) ^{2}E\left( \boldsymbol{x}_{\tau}\right) E\left(
	\boldsymbol{x}_{\tau}\right)^{\prime }. 
\end{eqnarray}

Consequently, as shown in the Appendix, (\ref{avar_bias}) is,
\begin{equation}
	\mathbb{H}(\boldsymbol{\beta})^{-1}\Sigma(\boldsymbol{\beta}) \mathbb{H}(\boldsymbol{\beta})^{-1} = E\left(\boldsymbol{x}_{\tau}\boldsymbol{x}_{\tau}^{\prime }\right)^{-1}M-B \label{HSigmaH}
\end{equation}
where  $B:={[E\left( \varepsilon_{\tau}^{\ast}\right)]^2}{[E\left(\exp \left( -\theta x_{\tau}^{\ast }\right) \right)]^{-2}}\boldsymbol{e}_1\boldsymbol{e}_1 ' $  (which can
be ignored as it is only affecting the constant term), and
\begin{equation}
	M=\frac{1-2E\left( \exp \left( -\theta x_{\tau}^{\ast }\right)
		\right) +2E\left( \exp \left( -2\theta x_{\tau}^{\ast }\right)
		\right) }{[E\left( \exp \left( -\theta x_{\tau}^{\ast }\right) \right)]
		^{2}}, \label{XXi}
\end{equation}%
which if $x_{t}^{\ast }\sim N\left(0,1\right)$, as assumed in the Monte Carlo section below, simplifies to,\footnote{ In the more general case where $x_{t}^{\ast }\sim N\left( \mu ,\sigma ^{2}\right) $ then $
	M=2\left( e^{\theta ^{2}\sigma ^{2}}-e^{\theta \mu -\frac{\theta ^{2}\sigma
			^{2}}{2}}\right) +e^{\theta \left( 2\mu -\theta \sigma ^{2}\right) }. $},
\begin{equation}
	M=2\left( e^{\theta ^{2}}-e^{-\frac{\theta ^{2}}{2}}\right) +e^{-\theta^2}. \label{M} 
\end{equation}

\noindent \textbf{The OLS approach}

Regarding the OLS estimator, consider the regression framework in (\ref{TailIndexRegression1}) such that,
\begin{equation}
	z_{\tau}=\boldsymbol{x}_{\tau}^{\prime }\boldsymbol{\beta} +v_{\tau}, \label{TailIndexRegression22}
\end{equation}%
where $z_{\tau}=-\ln \left( \ln \left( \frac{y_{\tau}}{\mathit{w}_n}\right) \right)-\gamma,$ and the error term,  because of the omission of $x_{\tau}^{\ast}$, is now $v_\tau=\xi_\tau+\theta x_{\tau}^{\ast}.$ In the simple case, where $x_{t}^{\ast }\sim N\left(0,1\right)$, it follows that,
\begin{eqnarray}
	\widehat{\boldsymbol{\beta}}_{OLS} &= &\boldsymbol{\beta} + \left(\sum_{\tau=1}^{n_0} \boldsymbol{x}_{\tau}\boldsymbol{x}_{\tau}^{\prime }\right)^{-1} \sum_{\tau=1}^{n_0} \boldsymbol{x}_{\tau}v_{\tau} \notag \\ 
	&= &\boldsymbol{\beta} + \left(\sum_{\tau=1}^{n_0} \boldsymbol{x}_{\tau}\boldsymbol{x}_{\tau}^{\prime }\right)^{-1} \sum_{\tau=1}^{n_0} \boldsymbol{x}_{\tau}x_{\tau}^{\ast}\theta + \left(\sum_{\tau=1}^{n_0} \boldsymbol{x}_{\tau}\boldsymbol{x}_{\tau}^{\prime }\right)^{-1} \sum_{\tau=1}^{n_0} \boldsymbol{x}_{\tau}\xi_{\tau} \notag \\
	&\overset{p}{\longrightarrow} & \boldsymbol{\beta},
\end{eqnarray}
since $\frac{1}{n_0}\sum_{\tau=1}^{n_0} \boldsymbol{x}_{\tau}x_{\tau}^{\ast}\overset{p}{\longrightarrow}0$ and $\frac{1}{n_0}\sum_{\tau=1}^{n_0} \boldsymbol{x}_{\tau}\xi_{\tau}\overset{p}{\longrightarrow}0.$

For the variance of 
$\sqrt{n_{0}}\left( \boldsymbol{\hat{\beta}}_{OLS}-\boldsymbol{\beta }\right)$ note first that as $n_{0}\longrightarrow \infty,$
$
\left( {\sum_{\tau=1}^{n_{0}}\boldsymbol{x}_{\tau}\boldsymbol{x}_{\tau}^{\prime }}/{n_{0}}%
\right) ^{-1}\overset{p}{\longrightarrow }\boldsymbol{\Sigma} _{xx}^{-1}.
$
Given that $\boldsymbol{x}_{\tau}v_{\tau}$ is a stationary and ergodic m.d.s. it follows that (c.f. \citealp{Billingsley1961}),
$
\frac{1}{\sqrt{n_0}}\sum_{\tau=1}^{n_{0}}\boldsymbol{x}_{\tau}v_{\tau}\overset{d}{\longrightarrow }N\left( 
\mathbf{0,}E\left( v_{\tau}^{2}\boldsymbol{x}_{\tau}\boldsymbol{x}_{\tau}^{\prime
}\right) \right). 
$
The expression $E\left( v_{\tau}^{2}\boldsymbol{x}_{\tau}\boldsymbol{x}_{\tau}^{\prime }\right) $ can be simplified, since $v_{\tau}$  is independent of $%
\boldsymbol{x}_{\tau},$ $\xi_{\tau}$ is independent of  $x_{\tau}^{\ast },$ and $E(x_{\tau}^{\ast 2})=1$ as, 
\begin{eqnarray}
	E\left( v_{\tau}^{2}\boldsymbol{x}_{\tau}\boldsymbol{x}_{\tau}^{\prime }\right)  &=&%
	E\left( v_{\tau}^{2}\right) \boldsymbol{\Sigma} _{xx} = E\left( \xi_{\tau}^{2}+\theta ^{2}\left( x_{\tau}^{\ast }\right)
	^{2}\right) \boldsymbol{ \Sigma} _{xx} =\left( \sigma _{\xi}^{2}+\theta ^{2}\right)\boldsymbol{ \Sigma} _{xx}. \label{OLSBIAS}
\end{eqnarray}%
Therefore, the asymptotic variance of $\boldsymbol{\hat{\beta}}_{OLS}$ under the misspecification considered is, 
\[
Avar\left( \sqrt{n_{0}}\left( \boldsymbol{\hat{\beta}}_{OLS}-\boldsymbol{\beta} \right) \right) =\left( \sigma _{\xi}^{2}+\theta ^{2}\right) \boldsymbol{ \Sigma} _{xx}^{-1}.
\]

In the presence of omitted variables, the relative efficiency of the OLS
estimator in relation to the exponential regression MLE estimator can be compared using
expressions $M$ (see equation (\ref{M})) and $A=$ $\left( \sigma _{\xi}^{2}+\theta
^{2}\right)$. Since, $M$ grows at an
exponential rate as a function of $\theta $, while $A$ increases
at a quadratic rate, it follows that for relatively large values of $\theta $, $M$ is much
larger than $A$, and so the OLS estimator is more efficient than the MLE
estimator when $\theta $ is relatively large. The efficiency
of the MLE estimator is more sensitive to omitted variables, especially when
the impact of the omitted variable is significant. This result is
corroborated in the Monte Carlo simulation section next.

\section{Monte Carlo Analysis \label{MC}}

In this section we perform a Monte Carlo analysis to evaluate the finite sample properties of the estimators and to validate the theoretical results provided in the previous section.

We generate data as,%
\begin{equation}
	\left. y_{t}\right\vert \boldsymbol{x}_{t}\sim F^{-1}\left(u_t|\boldsymbol{x}_t, \mathit{w}_{n}\right), \label{DGP}
\end{equation}
where $\boldsymbol{x}_{\tau}=(1, x_{2t}, x_{3t})'$, and  $\mathit{w}_n$ the tail cut-off point. The covariates $x_{2t}$ and $x_{3t}$ are independent random variables, $\forall s,t$, and are generated as $x_{2t}\overset{iid}{\sim }U\left( 0,1\right)$ and $x_{3t}\overset{iid}{\sim }N\left( 0,1\right)$. To generate $y_t$, we consider two cases for the cumulative distribution function $F(.)$ in (\ref{DGP}):  i) a Pareto distribution; and ii) a Burr distribution such that $F\left( x\right) =1-\left( 1+x^{-\alpha \rho }\right)^{1/\rho },$ with $\rho =-1$, which is of the Pareto-type class. In both cases the tail index is $\alpha(\boldsymbol{x}_t,\boldsymbol{\beta})=\exp \left( \boldsymbol{x}_{t}^{\prime }\boldsymbol{\beta} \right)$, with $\boldsymbol{x}_t=(1, x_{2t}, x_{3t})'$, and  $\boldsymbol{\beta}=(\beta_1, \beta_2, \beta_3)'$.

The tail index is generated as, 
$
E\left( \alpha \left( \boldsymbol{x}_{t},\boldsymbol{\beta} \right) \right) ={(\left(\exp \left( \beta _{2}\right) -1\right) }/{\beta _{2}})\exp \left(
\beta _{1}+\frac{\beta _{3}^{2}}{2}\right). 
$
$\beta _{1},\beta _{2}$ and $\beta _{3}$ are set to 
ensure that the expected value of $\alpha \left( \boldsymbol{x}_{t},\boldsymbol{\beta} \right) $
falls within a range consistent with values commonly observed for the tail index in the literature (see e.g. \citealp{GabaixIbragimov12} and \citealp{NicolauRodrigues18}). We use two sets of
parameter values: i) $\beta _{1}=0.1,$ $\beta _{2}=1,\beta _{3}=1,$ yielding $E
\left( \alpha \left( \boldsymbol{x}_{t},\boldsymbol{\beta} \right) \right) =3.13,$ and ii) $\beta
_{1}=0.1,$ $\beta _{2}=1,\beta _{3}=0.64,$ such that $E\left(
\alpha \left( \boldsymbol{x}_{t},\boldsymbol{\beta} \right) \right) =2.33.$ We set $\beta
_{3}=0.64$ to ensure that $M $ (see  (\ref{M})) is approximately $\pi
^{2}/6+\beta_{3}^2$, when $x_{3t} \sim N(0,1)$, a point we shall elaborate on later.  

The Monte Carlo experiments involve several steps. First, we generate a jointly independent 
sample $\left( y_{t},\boldsymbol{x}_{t}\right) $ of size $n$, where $n$ is
either $500$ or 5000, from a heavy-tailed distribution, using the
aforementioned parameter values. Secondly, we estimate $\boldsymbol{\beta}$ using the exponential regression approach based on MLE and based on OLS, applied to the framework introduced above, and compute the observed discrepancies between the exact and the estimated values. Lastly, we repeat these steps 5000 times. In Tables \ref{tab:Table1} - \ref{tab:Table3} we report the \textit{mean} and \textit{rmse} which are computed as,%
\begin{eqnarray*}
	mean_{k,j} &=&\sum_{i=1}^{5000}\hat{\beta}_{k,i}^{j}/5000, \text{ and } 	rmse_{kj} =\sqrt{\sum_{i=1}^{5000}\left( \hat{\beta}_{k,i}^{j}-\beta
		_{k}\right) ^{2}/5000},
\end{eqnarray*}%
respectively, where $ k=1,2,3$; $j$ denotes either MLE or OLS, and $\hat{%
	\beta}_{k,i}^{j}$ is the estimate ($j=OLS$ or $j=MLE$) of the parameter $\beta _{k}$ in
the $i$-th iteration of the  Monte Carlo procedure.

\subsection{Correctly Specified Tail Index Regression}

We first address the case where the tail index regression model is correctly specified. 
The results in Table \ref{tab:Table1} for the cases where $\beta _{1}=0.1,$ $\beta _{2}=1,\beta _{3}=1$ and $\beta _{1}=0.1,$ $\beta _{2}=1,\beta _{3}=0.64$ show that both the MLE and the OLS estimators are consistent when the model is correctly specified. However, as expected, the MLE is more efficient than the OLS estimator. Their relative efficiency, quantified by the ratio $rmse(OLS)/rmse(MLE)$, is approximately $\sigma_\xi=\pi/\sqrt{6}$, as predicted in Section \ref{Pareto}. 

\spacingset{1.0}
\begin{table}[h!]  
	\caption{MLE and OLS estimates when the tail index regression is correctly specified \\ - Data is generated from a \textbf{Pareto Distribution}}  
	\begin{center}  
		\scalebox{0.9}{	\begin{tabular}{clcccccc}          
				\toprule
				&  & \multicolumn{3}{c}{\scriptsize $\beta _{1}=0.1,$ $\beta _{2}=1,\beta _{3}=1$} & \multicolumn{3}{c}{\scriptsize$\beta _{1}=0.1,$ $\beta _{2}=1,\beta _{3}=0.64.$} \\          
				&       & $MLE$& $OLS$ & $ratio$ & $MLE$ & $OLS$ & $ratio$ \\    
				\midrule
				\multicolumn{8}{c}{$n=500$ } \\
				$\beta_1$ & $mean$  & 0.104 & 0.101 &       & 0.104 & 0.101 &  \\          
				& $rmse$  & 0.091 & 0.115 & 1.265 & 0.091 & 0.115 & 1.265 \\   
				$\beta_2$ & $mean$  & 1.002 & 0.998 &       & 1.002 & 0.998 &  \\          
				& $rmse$  & 0.157 & 0.199 & 1.269 & 0.115 & 0.199 & 1.270 \\   
				$\beta_3$ & $mean$  & 1.002 & 1.002 &       & 0.641 & 0.642 &  \\          
				& $rmse$  & 0.045 & 0.057 & 1.285 & 0.045 & 0.057 & 1.286 \\ \midrule          
				\multicolumn{8}{c}{$n=1000$ } \\
				$\beta_1$ & $mean$  & 0.100 & 0.098 &       & 0.100 & 0.098 &  \\          
				& $rmse$  & 0.063 & 0.080 & 1.283 & 0.063 & 0.080 & 1.283 \\   
				$\beta_2$& mean  & 1.003 & 1.002 &       & 1.003 & 1.002 &  \\          
				& $rmse$  & 0.109 & 0.140 & 1.280 & 0.109 & 0.140 & 1.280 \\    
				$\beta_3$ & $mean$  & 1.001 & 1.000 &       & 0.640 & 0.640 &  \\          
				& $rmse$  & 0.032 & 0.040 & 1.268 & 0.032 & 0.040 & 1.269 \\      \midrule    
				\multicolumn{8}{c}{$n=5000$ } \\
				$\beta_1$ & $mean$  & 0.100 & 0.099 &       & 0.100 & 0.099 &  \\          
				& $rmse$  & 0.029 & 0.037 & 1.280 & 0.029 & 0.036 & 1.280 \\   
				$\beta_2$ & $mean$  & 1.001 & 1.001 &       & 1.000 & 1.000 &  \\          
				& $rmse$  & 0.049 & 0.063 & 1.288 & 0.049 & 0.063 & 1.287 \\    
				$\beta_3$ & $mean$  & 1.000 & 1.000 &       & 0.641 & 0.640 &  \\          
				& $rmse$  & 0.014 & 0.018 & 1.265 & 0.014 & 0.018 & 1.267 \\    \bottomrule
		\end{tabular}}
		\label{tab:Table1}
	\end{center}
	\small{\textbf{	  Note:} $ratio=rmse(OLS)/rmse(MLE).$}
\end{table}
\FloatBarrier
\spacingset{1.9}
We also examined the properties of the OLS estimator in situations where
the distribution is of the Pareto-type. As mentioned in Remark \ref{Remark1} of Section \ref{Pareto}, this scenario
requires the selection of the tail cut-off point to determine the observations to be used in the estimation. The Monte Carlo simulations in this case followed the steps described above, with two exceptions: 1) the observations for the response variable $y_\tau$ were simulated using a Burr distribution; and 2) an intermediate step was introduced that involved obtaining the tail cut-off using the discrepancy measure
described in Remark \ref{Remark1}  in Section \ref{Pareto}.

Table \ref{tab:Table2}'s primary conclusion is that the relative efficiency between the
estimators remains in favor of MLE, although the difference is not as
pronounced as in the Pareto case. This outcome is due  to the fact that the OLS estimator incorporates more observations in the tails than the MLE estimator.

\spacingset{1.0}
\begin{table}[htbp]  
	\begin{center}
		\caption{rmse of MLE and OLS estimates when the tail index regression is correctly specified ($\beta_1=0.1$, $\beta_2=1$, and $\beta_3=1$)  - Data is generated from a \textbf{Burr Distribution.}}        
		\scalebox{0.9}{		\begin{tabular}{lccccccccc}    
				\toprule
				& \multicolumn{3}{c}{$n=500$}     & \multicolumn{3}{c}{$n=1000$}  & \multicolumn{3}{c}{$n=5000$}  \\          
				& \multicolumn{1}{c}{$MLE$} & \multicolumn{1}{c}{$OLS$} & \multicolumn{1}{c}{$ratio$} & \multicolumn{1}{c}{$MLE$} & \multicolumn{1}{c}{$OLS$} & \multicolumn{1}{c}{$ratio$} & \multicolumn{1}{c}{$MLE$} & \multicolumn{1}{c}{$OLS$} & \multicolumn{1}{c}{$ratio$} \\   \midrule
				$\beta_1$ & 0.264 & 0.322 & 1.221 & 0.229 & 0.277 & 1.211 & 0.159 & 0.177 & 1.115 \\    
				$\beta_2$  & 0.200 & 0.242 & 1.207 & 0.180 & 0.223 & 1.238 & 0.151 & 0.191 & 1.268 \\    
				$\beta_3$  & 0.380 & 0.492 & 1.297 & 0.332 & 0.402 & 1.212 & 0.231 & 0.274 & 1.186 \\    
				$k^{*}$ & 0.242 & 0.256 &   & 0.198 & 0.229 &  & 0.101 & 0.133 &  \\  \bottomrule   \end{tabular} } 
		\label{tab:Table2}
	\end{center}
	
	\small{\textbf{Note:} $ratio=rmse(OLS)/rmse(MLE)$ and $k^{*} \in (0,1)$ corresponds to the value that minimizes de Discrepancy measure (see Section 4.2.1 for an illustration) and determines $\hat{\mathit{w}}_n^*$ as the $(\left( 1-\kappa \right)100)^{th}$ percentile of $\{y_{t}\}$.}
\end{table}

\FloatBarrier

\spacingset{1.9}

\subsection{Misspecified Tail Index Regression}

To evaluate potential efficiency issues, we rerun the analysis, but with $x_{3t}$ omitted from the tail index regression specification. We observe from Table \ref{tab:Table3} that the MLE of $\beta _{1}$
exhibits significant bias, with the bias persisting even as the sample size increases, hence indicating that the MLE is inconsistent (as established in Section \ref{2.3.2}). This occurs with $x_{3t}$ independent of $x_{2t},$ which is surprising at first sight. In standard regression models, omitting an independent variable, with zero mean, from a regression does not lead to endogeneity and inconsistency, potentially only to inefficiency of the estimators. Additionally, while the MLE of $\beta _{2}$ is consistent, it lacks the precision of the OLS estimator. 

\bigskip
\spacingset{1.0}
\begin{table}[t!]  
	\begin{center} 
		\caption{MLE and OLS estimates when the tail index regression is incorrectly specified (omission of $x_{3t}$) - Data is generated from a \textbf{Pareto Distribution}}    
		\scalebox{0.9}{	\begin{tabular}{rlcccccc}    
				\toprule
				&  & \multicolumn{3}{c}{ \small $\beta _{1}=0.1,$ $\beta _{2}=1,\beta _{3}=1$} & \multicolumn{3}{c}{\small $\beta _{1}=0.1,$ $\beta _{2}=1,\beta _{3}=0.64.$} \\          
				&       & $MLE$& $OLS$ & $ratio$  & $MLE$& $OLS$ & $ratio$  \\    
				\midrule
				\multicolumn{8}{c}{$n=500$ } \\
				$\beta_1$ & $mean$  & {-0.390} & {0.101} &       & {-0.100} & {0.101} &  \\          
				& $rmse$  & {0.524} & {0.145} & {0.277} & {0.238} & {0.128} & {0.538} \\  
				$\beta_2$& $mean$  & {1.003} & {1.000} &       & {1.003} & {0.999} &  \\          
				& $rmse$  & {0.320} & {0.252} & {0.786} & {0.221} & {0.221} & {1.000}  \vspace{0.25cm}\\   \midrule 
				\multicolumn{8}{c}{$n=1000$ } \\
				$\beta_1$ & mean  & {-0.399} & {0.098} &       & {-0.105} & {0.098} &  \\          
				& $rmse$  & {0.516} & {0.101} & {0.197} & {0.238} & {0.089} & {0.399} \\    
				$\beta_2$& $mean$  & {1.007} & {1.003} &       & {1.005} & {1.002} &  \\          
				& $rmse$  & {0.227} & {0.177} & {0.778} & {0.155} & {0.155} & {1.006} \\   \midrule
				\multicolumn{8}{c}{$n=5000$ } 
				\\ 
				$\beta_1$ & $mean$  & {-0.399} & {0.099} &       & {-0.104} & {0.099} &  \\          
				& $rmse$  & {0.502} & {0.047} & {0.093} & {0.208} & {0.041} & {0.197} \\ 
				$\beta_2$& $mean$  & {0.999} & {1.001} &       & {1.000} & {1.000} &  \\          
				& $rmse$  & {0.102} & {0.080} & {0.783} & {0.070} & {0.071} & {1.012} \\     \bottomrule
		\end{tabular}}
		\label{tab:Table3}
	\end{center}
	\small{\textbf{	  Note:} $ratio=rmse(OLS)/rmse(MLE).$}
	
\end{table}

\FloatBarrier

\bigskip
\spacingset{1.9}
Hence, the omission of variables, even those uncorrelated with the model's
explanatory variables, lead to a dual impact on the properties of MLE. It
results in a substantial bias in the intercept estimate, and reduces the efficiency of the slope estimators relative to their OLS counterparts. The inconsistency of the intercept estimator is particularly serious if the objective is to obtain an estimate of the tail index $\alpha\left(\boldsymbol{x}_{t},\beta \right) .$ On the other hand, the consistency of the OLS estimator is a result of $x_{2t}$ being independent of the omitted variable. For $\beta _{1}=0.1,$ $\beta _{2}=1,\beta _{3}=0.64$ the MLE of $\beta _{2}$ matches the precision of its OLS counterpart, with their relative efficiency being roughly one. This	occurs because the parameter of the omitted variable is defined such that $M\simeq (\pi ^{2}/6+\beta_{3}^2)$. As discussed in Section \ref{Model Efficiency}, when $M=(\pi ^{2}/6+\beta_3^2)$, both estimators exhibit similar efficiency for the slope coefficients. However,	the inconsistency issue of the MLE for the intercept remains. These findings validate the theoretical results from Section \ref{Model Efficiency}, highlighting the critical dependence of the MLE's efficiency on the magnitude of the omitted variable's parameter.

In our simulations if $\beta_3>0.64$ the OLS estimator is more efficient than the MLE estimator (see left panel of Table \ref{tab:Table3}). This explains why the rmse of the OLS estimator is smaller than that of the MLE in the case of $\beta_3=1$. The greater the impact of the omitted variable, i.e., the greater the value of $\beta_3$, the more efficient the OLS estimator is when compared to MLE.

\section{Empirical Application \label{EmpiricalApp}}

In this section, we analyse the conditional tail index dynamics of commodities' returns. Commodities play an important role in the understanding of supply and demand dynamics, price movements, and shaping market trends. The availability and pricing of commodities can have far-reaching effects on the overall economy, influencing inflation rates, trade balances, and even geopolitical relations (see e.g. \citealp{GarrattPetrella2022}). In addition, understanding and managing commodities' tail risk is relevant for investors, financial institutions and policy makers due to the potential occurrence of  extreme and unexpected price movements. In what follows we provide an empirical analysis of the left- and right-tail dynamics of commodities' returns, as both tails play an equally important role; see e.g.  \citet{Ammannetal2023}.

\subsection{Data}
We consider 23 price series belonging to five categories of commodities: 1) energy; 2) industrial metals; 3) precious metals; 4) agricultural; and 5) livestock. The daily commodity price series are sampled from 02/02/1989 to 01/02/2024, and are collected from Refinitiv. We analyze the returns distributions' left- and right-tail dynamics of crude oil,	brent crude, unleaded gasoline,	heating oil, gas oil, and natural gas from the energy category; aluminium,	copper,	lead,	nickel, and  zinc from the industrial metals category; gold and silver from the precious metals category, wheat,	corn,	soybeans,	cotton,	sugar,	coffee, and cocoa from the agricultural category; and feeder cattle,	live cattle, and lean hogs from the livestock category.

	Analyses of commodity market developments have usually focused on macroeconomic conditions in industrial countries as the principal factors affecting commodity prices (\citealp{Alquist2020}). This evidence has guided our choice of the following covariates to be used in the tail index regression: 
	\textbf{(i)} the S\&P 500 log returns, $\Delta sp500_\tau$, which is used as a barometer of broader economic conditions and corporate profitability;  \textbf{(ii)} the yield curve slope, $slope_\tau$, which signals shifts in economic growth expectations, inflationary pressures, and monetary policy outlook; \textbf{(iii)} the nominal Emerging Market Economies US dollar index, $EME_\tau$, which measures the value of the US dollar against a basket of currencies from emerging market economies; \textbf{(iv)} the CBOE Volatility index, $vix_\tau$, often referred to as the "fear gauge", which provides insights into market sentiment and volatility expectations in equity markets; \textbf{(v)} the CBOE Skew index, $skew_\tau$, which is a measure of the perceived stock market tail risk; \textbf{(vi)} the NBER recession indicator for the US, $NBER_\tau$, which provides insights into broader economic conditions that may influence commodity markets; \textbf{(vii)} a stock market bear state indicator, $bear_\tau$, constructed using the \citet{LundeTimmermann2004} data-based algorithm, which identifies periods of declining asset prices and negative investor sentiment and serves as an early warning system for deteriorating market conditions; \textbf{(viii)} the ICE Bank of Atlanta US High Yield Index Effective Yield, $HY_\tau$, which tracks the performance of US dollar denominated below investment grade rated corporate debt publicly issued in the US domestic market; \textbf{(ix)} the daily news-based Economic Policy Uncertainty Index, $EPU_\tau,$\footnote{For details on the construction of EPU see Baker, Scott R., Bloom, Nick and Davis, Stephen J., Economic Policy Uncertainty Index for the United States [USEPUINDXD], retrieved from FRED, Federal Reserve Bank of St. Louis; https://fred.stlouisfed.org/series/USEPUINDXD, March 5, 2024.} which measures the level of uncertainty surrounding economic policy decisions and their potential impacts on the economy; \textbf{(x)} the total commodities index returns, $commodity_\tau$. 
	
	These variables display common effects on commodities tail risk, which are associated with increased market volatility, risk aversion, economic downturns, credit market conditions, and geopolitical or policy uncertainty. However, they differ in terms of their underlying drivers, time horizons, directness of impact, and systemic versus specific risks they represent. Figure \ref{Regressors} in Appendix B provides plots of all covariates described in (i) - (ix).  
	

	\subsection{Tail Index Estimation} 
	To analyze the tail dynamics, we will focus on both the left- and right-tails of commodities  returns' distribution.  Our interest centers in the estimation of the tail index parameter, $\alpha(\boldsymbol{x}_{i\tau},\boldsymbol{\beta})$, as a function of the vector of covariates, $\boldsymbol{x}_{i\tau}:=(intercept, commodity_\tau$, $\Delta sp500_\tau, slope_\tau, EME_\tau, vix_\tau, \text{ } skew_\tau, NBER_\tau,$ $bear_\tau,  \text{ } HY_\tau, EPU_\tau)',$ and the relevant vector of slope parameters, $\boldsymbol{\beta}$, is estimated from,
	\begin{equation}
		z_{i\tau}=\boldsymbol{x}_{i\tau}^{\prime }\boldsymbol{\beta}_i +\xi_{i\tau}, \quad i=1,...,23, \label{TailIndexRegression1_EA}
	\end{equation}%
	where $z_{i\tau}=-\ln \left( \ln \left( \frac{y_{i\tau}}{\mathit{w}_{in}}\right) \right)-0.5777.$ The left-tail index is estimated in exactly the same way except that instead of $y_{it}$ we consider the survival function of $-y_{it}$ when constructing the data sample. 
	
	\subsubsection{Tail Threshold Determination}
	As indicated in Remark \ref{Remark1}, one important first step in estimating $\alpha(\boldsymbol{x}_{it},\boldsymbol{\beta})$ is to select a suitable tail threshold $\mathit{w}_{in}$, which indicates the beginning of the (left or right) tail of commodity $i$'s returns  distribution. Hence, to determine $\mathit{w}_{in}$ empirically we make use of the discrepancy measure proposed by \citet{WangTsai09}.  This measure looks to ensure that the sample fraction chosen produces the smallest discrepancy between the empirical distribution of $\{\hat{U}_{i\tau}(\boldsymbol{x}_{i\tau})\}$ and a uniform distribution $U[0,1]$, since $\hat{U}_{i\tau}(\boldsymbol{x}_{it}) \equiv \hat{U}_{i\tau} =  \text{exp}\left(-\text{exp}(\boldsymbol{x}_\tau\boldsymbol{\hat{\beta}})
	\text{ln}\left(\frac{y_{i\tau}}{\mathit{w}_{in}}\right)\right),$ with $\mathit{w}_{in}$ well defined, is approximately $U[0,1]$. 
	
	The discrepancy measure is computed as,
	\begin{equation}
		\hat{D}(\omega_{in}, \boldsymbol{x}_{i\tau}) = \frac{1}{n_0}\sum_{\tau=1}^{n_0}\{\hat{U}_{i\tau}- \hat{F}_n(\hat {U}_{i\tau})\}^2, \label{Discr}
	\end{equation}
	where $n_0=\lfloor \kappa n \rfloor$, $\kappa \in \left(0,1\right]$, $\hat{\omega}_n$ is the $(\left( 1-\kappa \right)100)^{th}$ percentile of $y_{t}$, $\hat{F}_n(.)$ is the empirical distribution of $\{\hat{U}_{i\tau}\}$. If $\hat{U}_{i\tau}$ is indeed $U[0,1]$, then $\hat{F}_n(u)\approx u$ for every $u \in [0,1]$. Accordingly, the value of $\hat{D}(\omega_{in}, \boldsymbol{x}_{i\tau}) $ should be small, which suggests that $\omega_{in}$ can be selected as, %
	$\omega_{in}^*=\underset{\omega_{in}}{\arg \text{min}} \hat{D}(\omega_{in}, \boldsymbol{x}_{i\tau}).$ 
	
\begin{figure}[ht!] 
		\caption{Plots of $\hat{D}(\omega_{in}, \boldsymbol{x}_{i\tau})$ for the determination of the left- and right-tail cut off points for Gas oil, Natural gas, Gold and Coffee. \label{Fig:Discrepancy}}
		\begin{center}
			\includegraphics[width=0.4\textwidth]{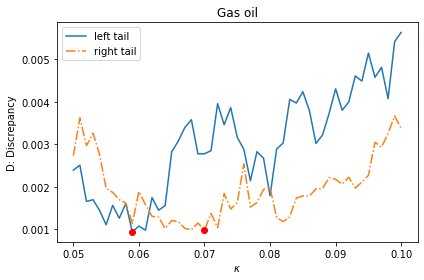}
			\includegraphics[width=0.4\textwidth]{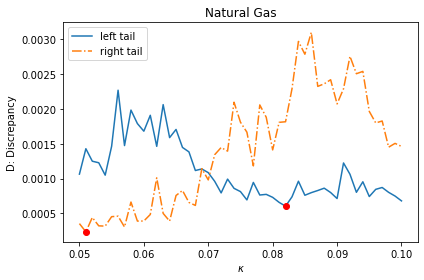}
			\includegraphics[width=0.4\textwidth]{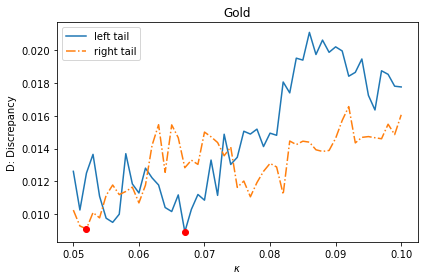}
			\includegraphics[width=0.4\textwidth]{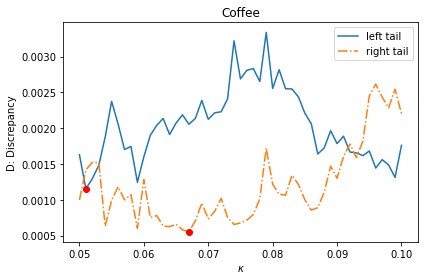}
		\end{center}
	\end{figure}

	Figure \ref{Fig:Discrepancy} provides an illustration of the application of $\hat{D}(\omega_{in}, \boldsymbol{x}_{i\tau})$ in (\ref{Discr}), to the returns series of Gas Oil, Natural Gas, Gold and Coffee, to determine the left- and right-tail cut off points for the corresponding estimation of the parameters in (\ref{TailIndexRegression1_EA}). 
	
	\FloatBarrier
	\vspace{-1cm}
	The red dots in Figure \ref{Fig:Discrepancy}  illustrate the minimum values of $\hat{D}(\omega_{in}, \boldsymbol{x}_{i\tau})$ for the left- and right-tails of the four commodities' return distribution, and the corresponding optimal values of $\kappa$ (and, consequently, $\omega_{in}^*$). Based on ${\omega}^*_{in}$ the sample of returns that compose commodity $i$'s returns distribution's left- and right-tails are determined and (\ref{TailIndexRegression1_EA}) is then used to compute the OLS estimates of $\beta _{ik}$, $i=1,...,N$ and $k=0,1,2,..., K$, with $N=23$ and $K=10$, is the number os commodities  and  the number of regressors (excluding the constant) used in the analysis, respectively. 
	
	\subsubsection{The tail index regression estimation results}
	Once $\omega_{in}^*$ is determined we proceed with the OLS estimation of (\ref{TailIndexRegression1_EA}). Table \ref{Table_Res1}  provides the left- and right-tail index regression estimation results for the 23 commodities under study. The $\beta_{ik}$, $k=0,1,...,K$, estimates in Table \ref{Table_Res1} reveal heterogeneous impacts of the covariates on commodities' tail risk. It is notorious that the impact of the covariates from the energy category on the commodities is, in some cases, different from the impacts on the commodities from the other categories.

	Table \ref{Table_Res1} reveals that Total Commodities index returns, denoted $commodity_\tau$, significantly influence both left- and right-tail indexes of commodities. This finding is expected, considering that $commodity_\tau$ serves as a comprehensive measure of the overall commodity market performance. Notably, the slope parameter of $commodities_\tau$ is  positive in the left-tail and negative in the right-tail. This implies that an increase in $commodities_\tau$ decreases the probability of extreme events in the left-tail while increasing it in the right-tail. 
	
	Results in Table \ref{Table_Res1} show that upward shifts in $\Delta SP500_\tau$ decrease the left-tail index of Gas Oil, Aluminium, and Copper, and the right-tail index of Copper and Live Cattle; and positively impact the left-tail index of Live Cattle and the right-tail index of Gas Oil, Zinc, Soybeans, and Sugar. 
	
	The impact of $slope_{\tau}$ on both tails is strikingly positive, implying a notable decrease in the likelihood of extreme events. We find positive impacts on the left-tails of Heating Oil, Gas Oil, Natural Gas, and Feeder Cattle, as well as on the right-tails of Crude Oil, Heating Oil, Gas Oil, Nickel, Zinc, Live Cattle, and Lean Hogs. We also observe statistically significant negative impacts in the left-tails of Cotton and Coffee, and in the right-tail of Cocoa.

	Increases in $EME_{\tau}$ represent a strengthening of the US dollar against emerging market currencies, often stemming from factors such as economic volatility, geopolitical unrest, or the tightening of monetary policies within emerging markets. From Table \ref{Table_Res1} we observe negative parameter estimates in the left-tails of Copper, Gold, Corn, and Lean Hogs, and in the right-tails of Crude Oil, Brent Oil, Unleaded Gasoline, Heating Oil, Coffee, and Feeder Cattle. There are also positive impacts on the left-tail of Crude Oil and the right-tail of Silver.
	
	Movements in the $vix_{\tau}$ impact commodity returns, particularly during periods characterized by heightened market uncertainty. Table \ref{Table_Res1} reveals that spikes in $vix_{\tau}$ significantly impact the left-tails of Crude Oil, Gas Oil, Natural Gas, Aluminium, Copper, Lead, Nickel, Gold, Cotton, Coffee, and Live Cattle, as well as the right-tails of Brent Crude, Aluminium, Nickel, and Coffee. In both cases, the effect is predominantly negative, exacerbating the probability of extreme events occurring across both tails when volatility escalates.

	Changes in the $skew_{\tau}$ reflect market expectations of extreme negative events. Table \ref{Table_Res1} shows that this variable has negative impacts on the left-tails of Crude oil, Brent oil, and on the right-tails of Unleaded Gasoline and Live Cattle; and positive impacts on the left-tails of Lead, Zinc, and Wheat, and on the right-tails of Zinc, and Gold.

	Interestingly, while energy commodities experience a positive influence from $NBER_{\tau}$ (see left- and right tails of Heating oil, and Gas oil), other commodity categories exhibit a negative impact (see left-tails of Wheat and Cocoa, and right-tails of Copper, Lead, Gold, and Lean Hogs). This divergence in impact can be attributed to the dynamics surrounding economic recessions. During downturns, economic activity typically decelerates, leading to a reduction in energy consumption by both industries and consumers. This decrease in demand exerts downward pressure on energy prices, thereby potentially mitigating tail risk as the likelihood of extreme price fluctuations decreases.

	During periods of heightened market uncertainty and bearish sentiment, investors typically gravitate towards safe-haven assets like government bonds or gold to safeguard their capital. Table \ref{Table_Res1} indicates that increases in $bear_{\tau}$ originate decreases in the left-tail index of Unleaded Gasoline, Zinc, and Cotton, as well as in the right-tail index of Heating Oil, Gas Oil, and Cotton, and leads to increases in the left-tail index of Coffee and the right-tail index of Gold.

	An increase in HY often signals deteriorating economic conditions or heightened investor risk aversion. Table \ref{Table_Res1} highlights the negative impacts of $HY_{\tau}$ on the right-tail index of Crude Oil, Unleaded Gasoline, Heating Oil, and Gas Oil, and on the left-tails of Brent Crude, Heating Oil, and Feeder Cattle. Positive impacts are observed on the left-tail of Lead, Wheat, and Coffee.

	Finally, a rising $EPU_{\tau}$ indicates increasing  economic policy uncertainty, leading investors to opt for hedging strategies or consider reducing their commodity exposure to mitigate potential downside risk. The results in Table \ref{Table_Res1} show that $EPU_{\tau}$ impacts the right-tails of Crude Oil, Unleaded Gasoline, Gold, Silver, Corn, and Feeder Cattle, indicative of its influence on commodity performance during periods of heightened uncertainty, and negative impacts on the left-tails of Crude Oil, Heating Oil, Gas Oil, and Feeder Cattle. Additionally, the results also reveal positive impacts on the left-tails of Aluminum, Copper, and Lead, underscoring the nuanced nature of EPU's influence across different commodity categories.

	\spacingset{1.0}
	\begin{landscape}
		\begin{table}[htbp]
			\centering
			\caption{Left- and right-tail index regression estimates}
			\scalebox{0.7}{\begin{tabular}{L{3cm}C{1.75cm}C{1.75cm}C{1.75cm}C{1.75cm}C{1.75cm}C{1.75cm}C{0.25cm}C{1.75cm}C{1.75cm}C{1.75cm}C{1.75cm}C{1.75cm}C{0.25cm}C{1.75cm}C{1.75cm}}
					\toprule  
					& \multicolumn{6}{c}{\textbf{Energy}} & \multicolumn{1}{c}{} & \multicolumn{5}{c}{\textbf{Industrial Metal}} & \multicolumn{1}{r}{} & \multicolumn{2}{c}{\textbf{Precious Metal}} \\    
					\cmidrule{2-7}\cmidrule{9-13}\cmidrule{15-16}   
					& \multicolumn{1}{c}{\textbf{Crude}} & \multicolumn{1}{c}{\textbf{Brent}} & \multicolumn{1}{c}{\textbf{Unleaded}} & \multicolumn{1}{c}{\textbf{Heating}} & \multicolumn{1}{c}{\textbf{Gas}} & \multicolumn{1}{c}{\textbf{Natural}} & & \multicolumn{1}{c}{\textbf{Alumi.}} & \multicolumn{1}{c}{\textbf{Copper}} & \multicolumn{1}{c}{\textbf{Lead}} & \multicolumn{1}{c}{\textbf{Nickel}} & \multicolumn{1}{c}{\textbf{Zinc}} &  & \multicolumn{1}{c}{\textbf{Gold}} & \multicolumn{1}{c}{\textbf{Silver}} \\ 
					& \multicolumn{1}{c}{\textbf{Oil}} & \multicolumn{1}{c}{\textbf{Crude}} & \multicolumn{1}{c}{\textbf{Gasoline}} & \multicolumn{1}{c}{\textbf{Oil}} & \multicolumn{1}{c}{\textbf{Oil}} & \multicolumn{1}{c}{\textbf{Gas}} &  & & & & & & & & \\ 
					\cmidrule{2-16}
					& \multicolumn{15}{l}{\textbf{Left-Tail}}  \\    
					\cmidrule{1-7}\cmidrule{9-13}\cmidrule{15-16}   
					$constant$ &\textbf{5.834}\tmark[{\makebox[0pt][l]{***}}]  & \textbf{5.454}\tmark[{\makebox[0pt][l]{***}}]  & 1.522 & \textbf{3.273}\tmark[{\makebox[0pt][l]{***}}]  & \textbf{2.068}\tmark[{\makebox[0pt][l]{*}}]  & \textbf{2.283}\tmark[{\makebox[0pt][l]{**}}]  & & 1.414 & 0.104 & -2.077 & \textbf{2.764}\tmark[{\makebox[0pt][l]{*}}]  & -0.286 & & 0.088 & \textbf{2.966}\tmark[{\makebox[0pt][l]{**}}]  \\
					$Commodity_\tau$&\textbf{0.470}\tmark[{\makebox[0pt][l]{***}}]  & \textbf{0.526}\tmark[{\makebox[0pt][l]{***}}]  & \textbf{0.333}\tmark[{\makebox[0pt][l]{***}}]  & \textbf{0.363}\tmark[{\makebox[0pt][l]{***}}]  & \textbf{0.203}\tmark[{\makebox[0pt][l]{***}}]  & \textbf{0.116}\tmark[{\makebox[0pt][l]{***}}]  & & 0.046 & \textbf{0.124}\tmark[{\makebox[0pt][l]{**}}]  & \textbf{0.075}\tmark[{\makebox[0pt][l]{**}}]  & 0.057 & \textbf{0.151}\tmark[{\makebox[0pt][l]{***}}]  & & 0.045 & \textbf{0.124}\tmark[{\makebox[0pt][l]{***}}]  \\
					$\Delta sp500_\tau$&-0.026 & 0.009 & 0.061 & -0.036 & \textbf{-0.073}\tmark[{\makebox[0pt][l]{***}}]  & -0.009 & & \textbf{-0.120}\tmark[{\makebox[0pt][l]{**}}]  & \textbf{-0.030}\tmark[{\makebox[0pt][l]{**}}]  & -0.041 & -0.023 & -0.072 & & -0.020 & -0.023 \\
					$slope_\tau$&-0.002 & 0.022 & 0.086 & \textbf{0.240}\tmark[{\makebox[0pt][l]{***}}]  & \textbf{0.147}\tmark[{\makebox[0pt][l]{**}}]  & \textbf{0.141}\tmark[{\makebox[0pt][l]{***}}]  & & -0.070 & 0.051 & -0.126 & 0.036 & -0.077 & & -0.008 & -0.015 \\
					$EME_\tau$&\textbf{0.221}\tmark[{\makebox[0pt][l]{*}}]  & 0.022 & 0.012 & -0.141 & -0.002 & 0.076 & & -0.136 & \textbf{-0.403}\tmark[{\makebox[0pt][l]{**}}]  & -0.216 & -0.107 & 0.030 & & \textbf{-0.369}\tmark[{\makebox[0pt][l]{**}}]  & -0.213 \\
					$vix_\tau$&\textbf{-0.013}\tmark[{\makebox[0pt][l]{*}}]  & -0.003 & 0.009 & -0.013 & \textbf{-0.026}\tmark[{\makebox[0pt][l]{***}}]  & \textbf{-0.026}\tmark[{\makebox[0pt][l]{***}}]  & & \textbf{-0.048}\tmark[{\makebox[0pt][l]{***}}]  & \textbf{-0.016}\tmark[{\makebox[0pt][l]{***}}]  & \textbf{-0.051}\tmark[{\makebox[0pt][l]{***}}]  & \textbf{-0.026}\tmark[{\makebox[0pt][l]{**}}]  & 0.015 & & \textbf{-0.017}\tmark[{\makebox[0pt][l]{*}}]  & -0.005 \\
					$skew_\tau$&\textbf{-0.020}\tmark[{\makebox[0pt][l]{*}}]  & \textbf{-0.016}\tmark[{\makebox[0pt][l]{**}}]  & 0.005 & -0.002 & 0.002 & -0.009 & & 0.002 & 0.012 & \textbf{0.028}\tmark[{\makebox[0pt][l]{**}}]  & -0.010 & \textbf{0.013}\tmark[{\makebox[0pt][l]{*}}]  & & 0.011 & -0.014 \\
					$NBER_\tau$&0.594 & 0.550 & 0.009 & \textbf{0.827}\tmark[{\makebox[0pt][l]{**}}]  & \textbf{0.807}\tmark[{\makebox[0pt][l]{**}}]  & 0.406 & & 0.076 & -0.021 & -0.224 & -0.222 & 0.292 & & 0.249 & -0.378 \\
					$bear_\tau$&0.074 & 0.024 & \textbf{-0.327}\tmark[{\makebox[0pt][l]{*}}]  & 0.158 & -0.276 & -0.100 & & -0.239 & 0.230 & 0.144 & -0.061 & \textbf{-0.476}\tmark[{\makebox[0pt][l]{**}}]  & & -0.131 & -0.197 \\
					$HY_\tau$&-0.029 & \textbf{-0.056}\tmark[{\makebox[0pt][l]{*}}]  & -0.047 & \textbf{-0.083}\tmark[{\makebox[0pt][l]{***}}]  & -0.007 & 0.018 & & 0.074 & -0.011 & \textbf{0.083}\tmark[{\makebox[0pt][l]{**}}]  & 0.055 & -0.063 & & 0.009 & 0.046 \\
					$EPU_\tau$&\textbf{-0.002}\tmark[{\makebox[0pt][l]{***}}]  & -0.001 & 0.000 & \textbf{-0.002}\tmark[{\makebox[0pt][l]{**}}]  & \textbf{-0.002}\tmark[{\makebox[0pt][l]{**}}]  & 0.001 & & \textbf{0.003}\tmark[{\makebox[0pt][l]{***}}]  & \textbf{0.001}\tmark[{\makebox[0pt][l]{***}}]  & \textbf{0.005}\tmark[{\makebox[0pt][l]{***}}]  & 0.001 & 0.000 & & 0.000 & -0.001 \\
					\midrule            
					& \multicolumn{15}{l}{\textbf{Right-Tail}} \\    
					\cmidrule{1-7}\cmidrule{9-13}\cmidrule{15-16}   
					$constant$ & \textbf{3.269}\tmark[{\makebox[0pt][l]{***}}]  & \textbf{3.938}\tmark[{\makebox[0pt][l]{***}}]  & \textbf{4.730}\tmark[{\makebox[0pt][l]{***}}]  & \textbf{4.559}\tmark[{\makebox[0pt][l]{***}}]  & \textbf{2.351}\tmark[{\makebox[0pt][l]{**}}]  & 0.425 & & -0.474 & 1.604 & -0.364 & -0.319 & \textbf{-1.919}\tmark[{\makebox[0pt][l]{**}}]  & & -2.127 & 0.724 \\
					$Commodity_\tau$&\textbf{-0.708}\tmark[{\makebox[0pt][l]{***}}] & \textbf{-0.701}\tmark[{\makebox[0pt][l]{***}}]  & \textbf{-0.369}\tmark[{\makebox[0pt][l]{***}}]  & \textbf{-0.560}\tmark[{\makebox[0pt][l]{***}}]  & \textbf{-0.264} & 0.039 & & \textbf{-0.128}\tmark[{\makebox[0pt][l]{***}}]  & \textbf{-0.160}\tmark[{\makebox[0pt][l]{***}}]  & \textbf{-0.073}\tmark[{\makebox[0pt][l]{*}}]  & 0.002 & \textbf{-0.200}\tmark[{\makebox[0pt][l]{***}}]  & & 0.002 & -0.024 \\
					$\Delta sp500_\tau$&0.064 & -0.008 & 0.001 & -0.016 & \textbf{0.097}\tmark[{\makebox[0pt][l]{*}}]  & -0.028 & & -0.066 & \textbf{-0.068}\tmark[{\makebox[0pt][l]{*}}]  & -0.046 & -0.018 & \textbf{0.119}\tmark[{\makebox[0pt][l]{***}}]  & & 0.053 & 0.001 \\
					$slope_\tau$&\textbf{0.111}\tmark[{\makebox[0pt][l]{*}}]  & 0.105 & 0.049 & \textbf{0.216}\tmark[{\makebox[0pt][l]{**}}]  & \textbf{0.126}\tmark[{\makebox[0pt][l]{**}}]  & 0.096 & & 0.096 & 0.062 & -0.111 & \textbf{0.100}\tmark[{\makebox[0pt][l]{*}}]  & \textbf{0.138}\tmark[{\makebox[0pt][l]{**}}]  & & 0.004 & 0.031 \\
					$EME_\tau$&\textbf{-0.391}\tmark[{\makebox[0pt][l]{*}}]  & \textbf{-0.313}\tmark[{\makebox[0pt][l]{**}}]  & \textbf{-0.193}\tmark[{\makebox[0pt][l]{*}}]  & \textbf{-0.536}\tmark[{\makebox[0pt][l]{***}}]  & -0.208 & 0.346 & & 0.110 & 0.256 & -0.288 & -0.138 & 0.076 & & 0.192 & \textbf{0.643}\tmark[{\makebox[0pt][l]{***}}]  \\
					$vix_\tau$&0.023 & \textbf{0.025}\tmark[{\makebox[0pt][l]{**}}]  & 0.012 & 0.014 & 0.002 & -0.002 & & \textbf{-0.025}\tmark[{\makebox[0pt][l]{*}}]  & 0.000 & -0.014 & \textbf{-0.024}\tmark[{\makebox[0pt][l]{*}}]  & 0.005 & & -0.006 & -0.012 \\
					$skew_\tau$&-0.004 & -0.010 & \textbf{-0.020}\tmark[{\makebox[0pt][l]{***}}]  & -0.013 & -0.003 & 0.006 & & 0.013 & -0.003 & 0.013 & 0.012 & \textbf{0.024}\tmark[{\makebox[0pt][l]{***}}]  & & \textbf{0.029}\tmark[{\makebox[0pt][l]{***}}]  & 0.006 \\
					$NBER_\tau$&0.460 & -0.103 & -0.023 & \textbf{0.674}\tmark[{\makebox[0pt][l]{**}}]  & \textbf{0.630}\tmark[{\makebox[0pt][l]{**}}]  & 0.136 & & -0.504 & \textbf{-0.836}\tmark[{\makebox[0pt][l]{***}}]  & \textbf{-0.592}\tmark[{\makebox[0pt][l]{*}}]  & 0.210 & -0.243 & & \textbf{-0.546}\tmark[{\makebox[0pt][l]{*}}]  & -0.037 \\
					$bear_\tau$&-0.174 & -0.189 & -0.007 &\textbf{ -0.432}\tmark[{\makebox[0pt][l]{*}}]  & \textbf{-0.440}\tmark[{\makebox[0pt][l]{**}}]  & -0.035 & & -0.023 & 0.007 & 0.020 & -0.249 & -0.024 & & \textbf{0.618}\tmark[{\makebox[0pt][l]{**}}]  & -0.240 \\
					$HY_\tau$&\textbf{-0.113}\tmark[{\makebox[0pt][l]{*}}]  & -0.045 & \textbf{-0.078}\tmark[{\makebox[0pt][l]{**}}]  & \textbf{-0.137}\tmark[{\makebox[0pt][l]{***}}] & \textbf{-0.115}\tmark[{\makebox[0pt][l]{***}}] & -0.017 & & 0.074 & 0.021 & 0.048 & 0.010 & -0.050 & & 0.032 & 0.033 \\
					$EPU_\tau$&\textbf{-0.001}\tmark[{\makebox[0pt][l]{**}}]  & -0.001 & \textbf{-0.001}\tmark[{\makebox[0pt][l]{***}}] & -0.001 & 0.000 & 0.000 & & 0.002 & 0.001 & 0.001 & 0.001 & 0.000 & &\textbf{ -0.002}\tmark[{\makebox[0pt][l]{***}}] & \textbf{-0.002}\tmark[{\makebox[0pt][l]{**}}] \\ \bottomrule
			\end{tabular}}
			\label{Table_Res1}
			\bigskip
			
			\footnotesize{\textbf{Note}: *, **, *** refer to statistical significance at the $10\%$, $5\%$ and $1\%$ significance levels, respectively. For inference purposes HAC standard errors are considered.}
		\end{table}%
	\end{landscape}

	\begin{landscape}
		\setcounter{table}{3}
		\renewcommand{\thetable}{\arabic{table}}%
		\begin{table}[htbp]
			\caption{Left- and right-tail index regression estimates (cont.)}
			\centering
			\scalebox{0.7}{\begin{tabular}{L{3cm}C{1.5cm}C{1.5cm}C{1.5cm}C{1.5cm}C{1.5cm}C{1.5cm}C{1.5cm}C{0.25cm}C{1.5cm}C{1.5cm}C{1.5cm}}    
					\toprule
					& \multicolumn{7}{c}{\textbf{Agriculture }}  &       & \multicolumn{3}{c}{\textbf{Livetock}} \\          
					\cmidrule{2-8}\cmidrule{10-12}   
					& \multicolumn{1}{c}{\textbf{Wheat}} & \multicolumn{1}{c}{\textbf{Corn}} & \multicolumn{1}{c}{\textbf{Soybeans}} & \multicolumn{1}{c}{\textbf{Cotton}} & \multicolumn{1}{c}{\textbf{Sugar}} & \multicolumn{1}{c}{\textbf{Coffee}} & \multicolumn{1}{c}{\textbf{Cocoa}} &       & \multicolumn{1}{c}{\textbf{Feeder}} & \multicolumn{1}{c}{\textbf{Live}} & \multicolumn{1}{c}{\textbf{Lean}} \\    
					&  & & & & & &  &       & \multicolumn{1}{c}{\textbf{Cattle}} & \multicolumn{1}{c}{\textbf{Cattle}} & \multicolumn{1}{c}{\textbf{Hogs}} \\   
					\cmidrule{2-12}    
					& \multicolumn{11}{l}{\textbf{Left-Tail}} \\    
					\cmidrule{1-8}\cmidrule{10-12}   
					$constant$&-1.533 & \textbf{2.651}\tmark[{\makebox[0pt][l]{**}}] & \textbf{1.931}\tmark[{\makebox[0pt][l]{**}}] & 1.500 & 0.826 & 1.015 & 0.665 & & \textbf{1.510}\tmark[{\makebox[0pt][l]{*}}] & 0.982 & \textbf{2.841}\tmark[{\makebox[0pt][l]{**}}] \\
					$Commodity_\tau$&\textbf{0.201}\tmark[{\makebox[0pt][l]{***}}] & -0.030 & \textbf{0.094 }\tmark[{\makebox[0pt][l]{**}}]& \textbf{0.114}\tmark[{\makebox[0pt][l]{***}}] & 0.066 & -0.070 & 0.048 & & 0.027 & -0.028 & -0.063 \\
					$\Delta sp500_\tau$&-0.030 & 0.041 & 0.023 & -0.054 & -0.040 & 0.072 & 0.058 & & 0.054 & \textbf{0.071}\tmark[{\makebox[0pt][l]{**}}] & 0.001 \\
					$slope_\tau$&-0.018 & 0.098 & -0.073 & \textbf{-0.120}\tmark[{\makebox[0pt][l]{*}}] & -0.109 & \textbf{-0.231}\tmark[{\makebox[0pt][l]{***}}] & -0.082 & & \textbf{0.225}\tmark[{\makebox[0pt][l]{***}}] & 0.046 & 0.009 \\
					$EME_\tau$&0.214 & \textbf{-0.474}\tmark[{\makebox[0pt][l]{*}}] & 0.092 & 0.143 & -0.199 & -0.287 & -0.009 & & -0.169 & 0.129 & \textbf{-0.495}\tmark[{\makebox[0pt][l]{**}}] \\
					$vix_\tau$&-0.015 & 0.008 & 0.002 & \textbf{-0.018}\tmark[{\makebox[0pt][l]{*}}] & -0.004 & \textbf{-0.024}\tmark[{\makebox[0pt][l]{*}}] & -0.007 & & 0.004 & \textbf{-0.030}\tmark[{\makebox[0pt][l]{***}}] & -0.010 \\
					$skew_\tau$&\textbf{0.021}\tmark[{\makebox[0pt][l]{**}}] & -0.012 & -0.005 & 0.000 & 0.004 & 0.004 & 0.002 & & -0.003 & 0.001 & -0.011 \\
					$NBER_\tau$&\textbf{-0.424}\tmark[{\makebox[0pt][l]{*}}] & -0.329 & -0.274 & 0.249 & -0.543 & -0.346 & \textbf{-0.778}\tmark[{\makebox[0pt][l]{***}}] & & -0.210 & -0.103 & -0.797 \\
					$bear_\tau$&-0.102 & 0.057 & 0.019 & \textbf{-0.478}\tmark[{\makebox[0pt][l]{***}}]& 0.198 & \textbf{0.407}\tmark[{\makebox[0pt][l]{**}}] & 0.198 & & 0.134 & 0.241 & 0.317 \\
					$HY_\tau$&\textbf{0.089}\tmark[{\makebox[0pt][l]{*}}] & -0.063 & -0.033 & 0.019 & 0.029 & \textbf{0.109}\tmark[{\makebox[0pt][l]{*}}] & 0.062 & & \textbf{-0.063}\tmark[{\makebox[0pt][l]{*}}] & 0.022 & 0.051 \\
					$EPU_\tau$&0.002 & 0.000 & 0.000 & 0.001 & 0.001 & -0.001 & 0.002 & & \textbf{-0.001}\tmark[{\makebox[0pt][l]{*}}] & 0.001 & -0.001 \\ \midrule
					& \multicolumn{11}{l}{\textbf{Right-Tail}} \\    
					\cmidrule{1-8}\cmidrule{10-12}   
					$constant$ &\textbf{2.476}\tmark[{\makebox[0pt][l]{**}}] & 0.534 & \textbf{2.736}\tmark[{\makebox[0pt][l]{**}}] & -0.085 & 2.123 & 1.935 & \textbf{2.076}\tmark[{\makebox[0pt][l]{*}}] & & 2.065 & \textbf{2.978}\tmark[{\makebox[0pt][l]{***}}] & -0.186 \\
					$Commodity_\tau$&\textbf{-0.200}\tmark[{\makebox[0pt][l]{***}}] & -0.052 & -0.018 & -0.035 & -0.080 & \textbf{-0.147}\tmark[{\makebox[0pt][l]{***}}] & -0.053 & & 0.019 & 0.031 & \textbf{-0.083}\tmark[{\makebox[0pt][l]{*}}] \\
					$\Delta sp500_\tau$&0.001 & 0.051 & \textbf{0.089}\tmark[{\makebox[0pt][l]{*}}] & -0.047 & \textbf{0.146}\tmark[{\makebox[0pt][l]{**}}] & 0.089 & -0.001 & & -0.071 & \textbf{-0.091}\tmark[{\makebox[0pt][l]{*}}] & 0.094 \\
					$slope_\tau$&0.087 & -0.013 & -0.078 & 0.037 & 0.103 & -0.085 & \textbf{-0.195}\tmark[{\makebox[0pt][l]{**}}] & & 0.120 & \textbf{0.160}\tmark[{\makebox[0pt][l]{**}}] & \textbf{0.187}\tmark[{\makebox[0pt][l]{***}}] \\
					$EME_\tau$&-0.297 & 0.312 & 0.139 & 0.248 & -0.089 & \textbf{-0.391}\tmark[{\makebox[0pt][l]{*}}] & -0.023 & & \textbf{-0.337}\tmark[{\makebox[0pt][l]{*}}] & 0.014 & -0.160 \\
					$vix_\tau$&0.010 & -0.003 & 0.036 & -0.010 & -0.010 & \textbf{-0.025}\tmark[{\makebox[0pt][l]{*}}] & -0.021 & & 0.010 & -0.005 & 0.017 \\
					$skew_\tau$&-0.007 & 0.006 & -0.012 & 0.014 & -0.003 & -0.006 & -0.001 & & -0.005 & \textbf{-0.015}\tmark[{\makebox[0pt][l]{**}}] & 0.008 \\
					$NBER_\tau$&-0.064 & -0.110 & -0.067 & 0.221 & 0.521 & 0.633 & 0.041 & & 0.325 & -0.166 & \textbf{-0.870}\tmark[{\makebox[0pt][l]{**}}] \\
					$bear_\tau$&-0.211 & \textbf{0.416}\tmark[{\makebox[0pt][l]{*}}] & 0.169 & \textbf{-0.316}\tmark[{\makebox[0pt][l]{**}}] & -0.330 & -0.086 & -0.187 & & 0.232 & 0.144 & 0.368 \\
					$HY_\tau$&-0.055 & 0.024 & -0.142 & -0.012 & -0.073 & 0.076 & 0.018 & & -0.036 & -0.002 & -0.006 \\
					$EPU_\tau$&-0.001 & \textbf{-0.002}\tmark[{\makebox[0pt][l]{*}}] & 0.000 & -0.001 & 0.000 & 0.001 & 0.001 & & \textbf{-0.003}\tmark[{\makebox[0pt][l]{***}}] & -0.001 & -0.001 \\ \bottomrule
			\end{tabular}} 
			\label{tab:addlabel}
			
			\bigskip
			
			\footnotesize{\textbf{Note}: *, **, *** refer to statistical significance at the $10\%$, $5\%$ and $1\%$ significance levels, respectively. For inference purposes HAC standard errors are considered.}
		\end{table}%
	\end{landscape}

	\spacingset{1.9}

	To illustrate the dynamics of the conditional tail index, Figure \ref{Fig3} and Figure \ref{Fig4} in the Supplementary Material plot the estimates of the left- and right-tail conditional tail index $\alpha \left(\boldsymbol{x}_{i\tau},\boldsymbol{\beta}\right)$, respectively, for four commodities: Gas Oil, Natural Gas, Gold and Coffee. 
	
	\spacingset{1.0}
	\begin{figure}[htp!]
		\centering
		\caption{\textbf{Left-Tail Index Estimates }- $\hat{\alpha}(\boldsymbol{x}_{i\tau},\hat{\boldsymbol{\beta}}).$  A lower value of the tail index indicates a higher probability of extreme values occurring, which in turn translates into a greater level of tail risk.} \label{Fig3}
		\begin{subfigure}[t]{0.35\textwidth}
			\centering
			\includegraphics[width=1\textwidth]{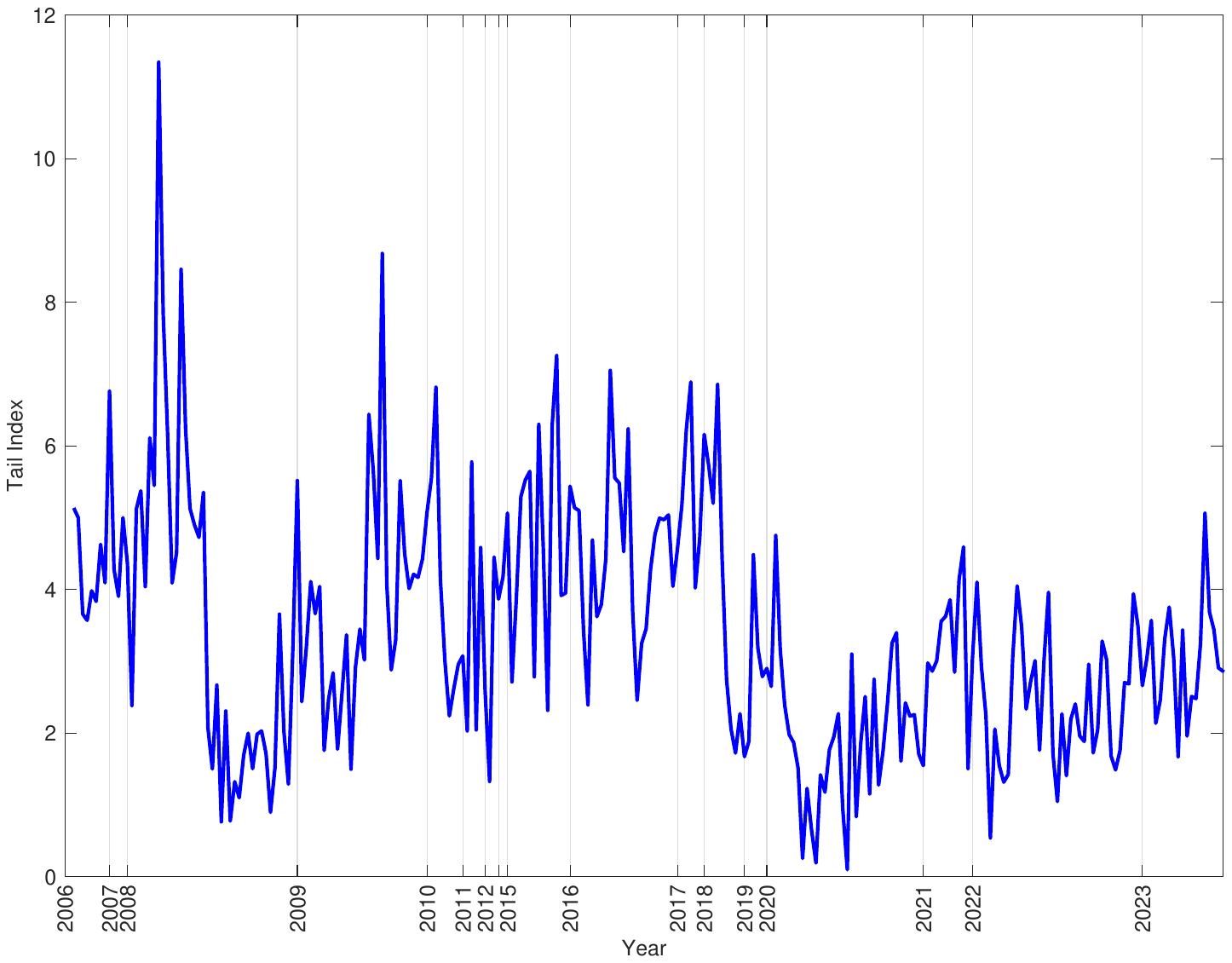}
			\caption{Gas Oil}
		\end{subfigure}
		~
		\begin{subfigure}[t]{0.35\textwidth}
			\centering
			\includegraphics[width=1\textwidth]{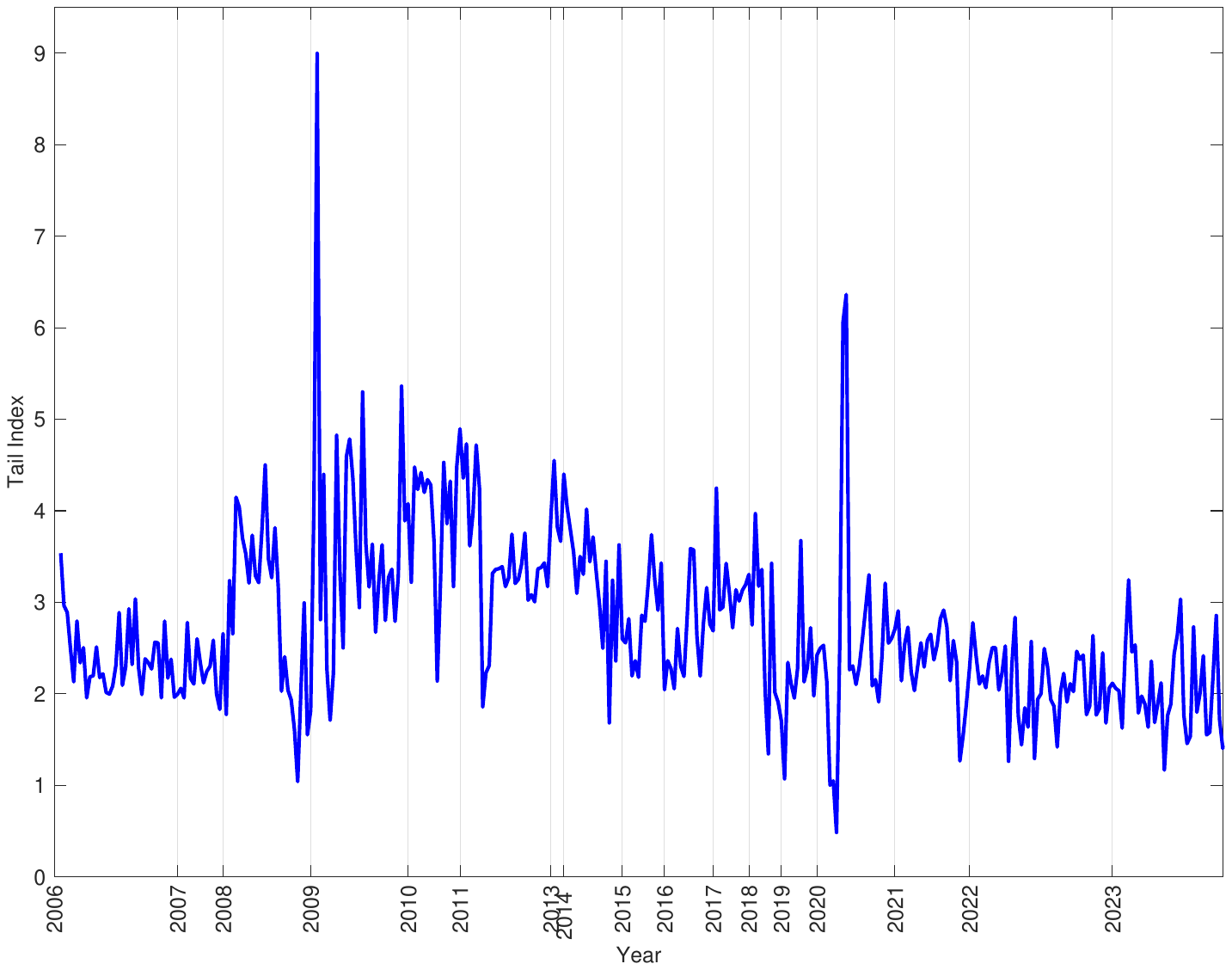}
			\caption{Natural Gas}
		\end{subfigure}
		
		\begin{subfigure}[t]{0.35\textwidth}
			\centering
			\includegraphics[width=1\textwidth]{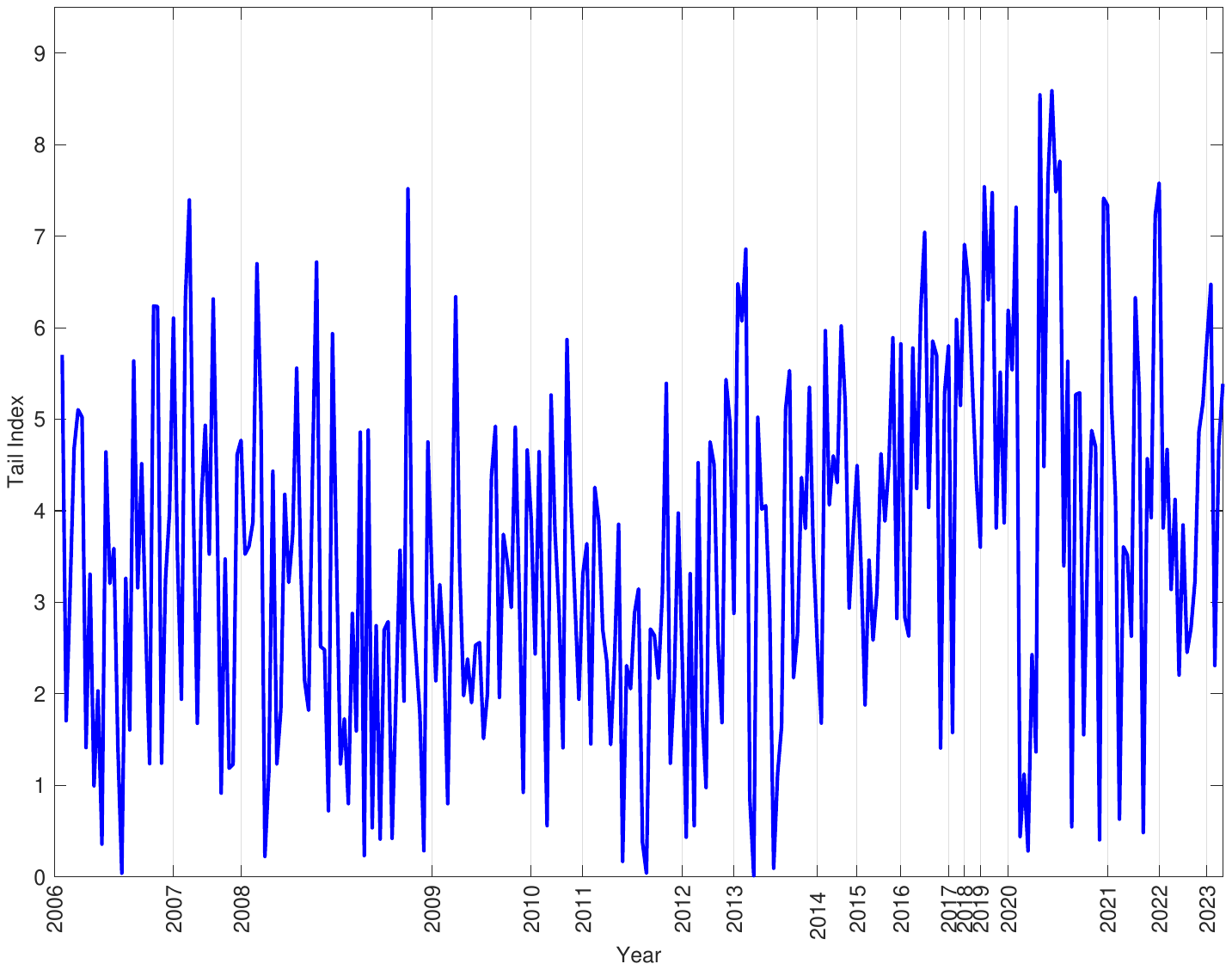}
			\caption{Gold}
		\end{subfigure}
		~
		\begin{subfigure}[t]{0.35\textwidth}
			\centering
			\includegraphics[width=1\textwidth]{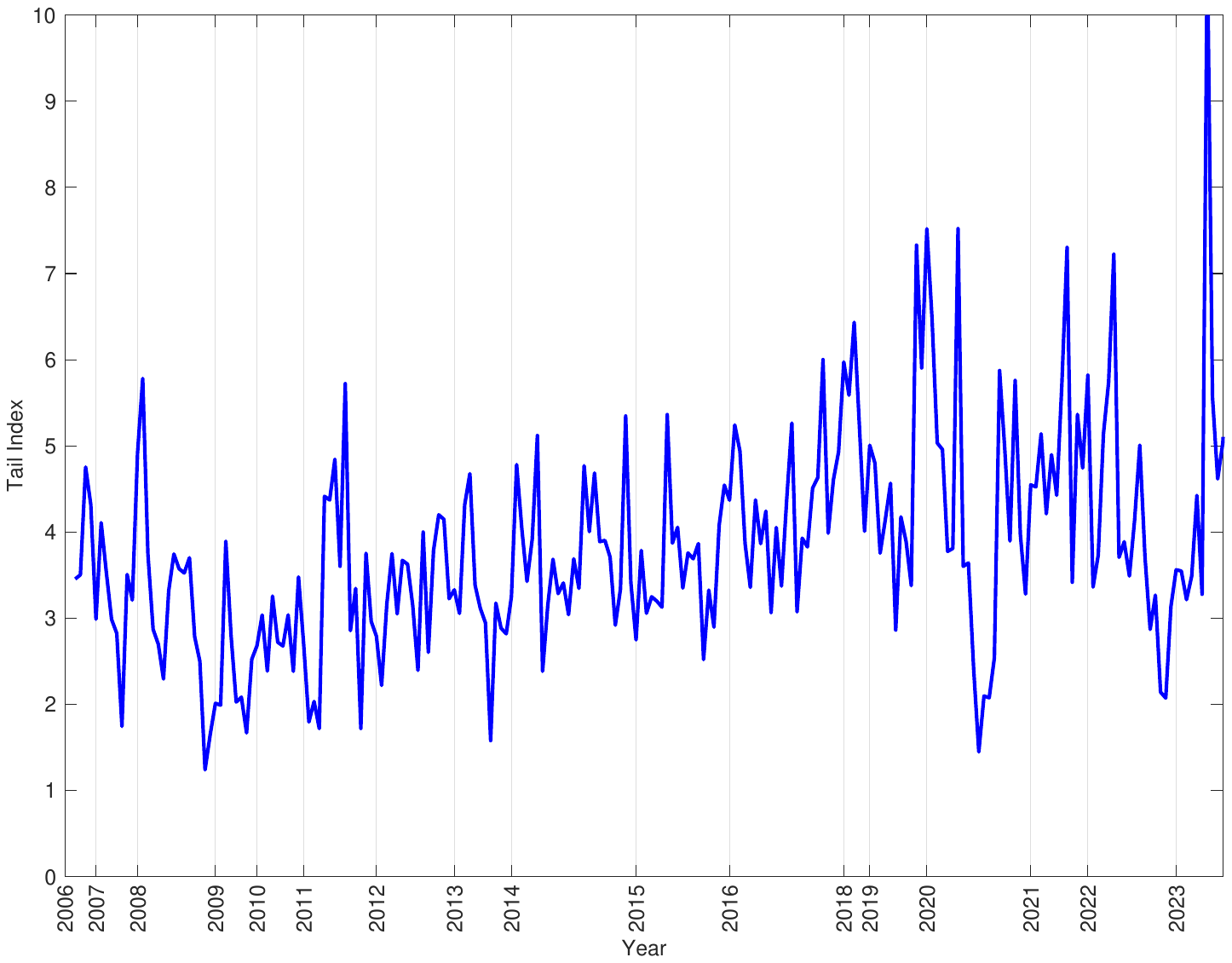}
			\caption{Coffee}
		\end{subfigure}	
	\end{figure}
	
	\FloatBarrier

	\spacingset{1.9}
	Regarding the left-tail index, we observe that for Gas Oil the periods with most volatile tail behavior were 2008, 2009, 2016, 2020, and 2022 (of which 2008 and 2022 were the years with the largest number of extreme events);  for Natural Gas 2006, 2008, 2009, 2012, 2020, 2021, 2022, 2023 (of which 2006, 2022 and 2023 were the years with the largest number of extreme events); for Gold 2006, 2008, 2009, 2011, 2013, 2020 (of which 2006 and 2008 were the years with the largest number of extreme events); and for Coffee 2014, 2015, 2020 and 2022 (of which 2014, and 2023 were the years with the largest number of extreme events). Regarding the right-tail index for Gas Oil the years with more extreme events were 2008, 2009, 2016, 2020, 2022 (of which 2009, 2020 and 2022 were the years with the largest number of extreme events); for Natural Gas 2006, 2009, 2022, 2023 (of which 2022 was the year with the largest number of extreme events); for Gold 2006, 2008, 2009 (of which 2006 and 2008 were the years with the largest number of extreme events); and  finally for Coffee 2010, 2014, 2015, 2019, 2020, 2021, 2022 (of which 2014 was the year with the largest number of extreme events).

	\section{Conclusion \label{conclusion}}
	
	This paper introduces a new framework for estimating the conditional tail index of time series, which is crucial for understanding the risks in extreme market conditions. It employs a simple and flexible method based on OLS regression, offering a practical alternative to more complex methods currently in use. The primary advantage of the OLS approach, as evidenced through both the theoretical discussion and the empirical analysis, lies in its simplicity, which allows both researchers and practitioners to easily assess and manage extreme event risks within the least squares framework. Unlike the MLE estimator, the OLS estimator remains unbiased under general conditions and is robust against model miss-specification. This attribute is particularly valuable when the sample size is small, which is a common scenario in the estimation of the tail index of Pareto-type distributions. Moreover, although MLE may be more efficient when the model is correctly specified, this advantage diminishes or reverses in favor of OLS when variables are omitted, especially if the omitted variables have a significant impact on the tail index. This is pertinent for empirical applications where some relevant variables may not be observable or included in the model.
	
	The empirical application has two aims: to show the applicability of the novel conditional tail index estimation framework in practical market analysis, and to contribute to the existing literature on the determinants of commodities' tail risks. Our analysis highlights the importance of various market and sector-specific variables in influencing the conditional left- and right-tail risks
	of commodities' return distributions. An in-depth analysis of 23 commodity return series across five categories uncovers the diverse impact of these covariates on the conditional tail risk of commodities. We have found that market indicators, volatility and policy uncertainty, economic conditions, and market states have, in general, significant effects on the conditional left and right tails of commodities' returns distributions.
	
	\bibliographystyle{chicago}
	\bibliography{TailIndex, ref_ine}
	
	\newpage
	\begin{center}
		{\large\bf SUPPLEMENTARY MATERIAL}
	\end{center}
	
	\noindent \textbf{\large{Technical Details}}
	
	\noindent \textbf{Proof of result in (\ref{at})}
	
	To show that $a_\tau$ in (\ref{at}) follows a standard Gumbel distribution recall that  $ a_{\tau}=-\ln v_{\tau}=-\ln \left( -\ln (1-u_\tau)\right)$ and $u_\tau\sim U(0,1)$. Hence, it follows that,
	\begin{eqnarray*}
		P\left( a_{\tau}<a\right) &=&P\left( -\ln \left( -\ln (1-u_\tau)\right) <a\right)	\\
		&=&P\left( -\ln (1-u_{\tau}\})>e^{-a}\right) \\
		&=&P\left( 1-u_{\tau}<e^{-e^{-a}}\right) \\
		&=&P\left( u_{\tau}>1-e^{-e^{-a}}\right) \\
		&=&1-\left( 1-e^{-e^{-a}}\right) =e^{-e^{-a}}, \quad a \in \mathbb{R}. \hspace{7cm} \hfil \blacksquare \notag
	\end{eqnarray*}

	\noindent \textbf{Proof of result in (\ref{Hessian1})}
	\begin{eqnarray}
		\mathbb{H}\left( \boldsymbol{\beta}\right) &=&E\left( E\left( \left.
		H_{n}\left( \boldsymbol{\beta} \right) \right\vert \boldsymbol{x}_{\tau},x_{\tau}^{\ast }\right) \right) =-%
		\frac{1}{n_0}\sum_{\tau=1}^{n_0}E\left(\boldsymbol{x}_{\tau}\boldsymbol{x}_{\tau}^{\prime }E\left( \left. \exp \left(\mathbf{ x}_{\tau}^{\prime }\boldsymbol{\beta} \right) \ln \left( \frac{y_{\tau}%
		}{\mathit{w}_{n}}\right) \right\vert \boldsymbol{x}_{\tau},x_{\tau}^{\ast }\right) \right) \notag \\
		&=&-\frac{1}{n_0}\sum_{\tau=1}^{n_0}E\left(\boldsymbol{x}_{\tau}\boldsymbol{x}_{\tau}^{\prime }\frac{\exp \left(\boldsymbol{x}_{\tau}^{\prime }\boldsymbol{\beta} \right) }{\exp \left(\boldsymbol{x}_{\tau}^{\prime }\boldsymbol{\beta}
			+\theta x_{\tau}^{\ast }\right) }\right) \notag \\
		&=&-\frac{1}{n_0}\sum_{\tau=1}^{n_0}E\left( \boldsymbol{x}_{\tau}\boldsymbol{x}_{\tau}^{\prime }\exp
		\left( -\theta x_{\tau}^{\ast }\right) \right) \notag \\
		&=&-E\left( \boldsymbol{x}_{\tau}\boldsymbol{x}_{\tau}^{\prime }\exp \left( -\theta x_{\tau}^{\ast
		}\right) \right) \notag \\
		&=&-E\left(\boldsymbol{x}_{\tau}\boldsymbol{x}_{\tau}^{\prime }\right) E\left( \exp
		\left( -\theta x_{\tau}^{\ast }\right) \right). \hspace{9cm} \hfil \blacksquare \notag
	\end{eqnarray}%
	
	\noindent \textbf{Proof of result in (\ref{AVar})}
	\begin{eqnarray}
		\boldsymbol{\Sigma}(\boldsymbol{\beta})  &=&Avar\left( \sqrt{n_0}G_{n}(\boldsymbol{\beta})
		\right) =\lim var\left( \frac{1}{\sqrt{n_0}}\sum_{\tau=1}^{n}\boldsymbol{x}_{\tau}%
		\varepsilon _{\tau}^\ast\right)  \notag \\
		&=&\lim \frac{1}{n_0}var\left( \sum_{\tau=1}^{n}\boldsymbol{x}_{\tau}\varepsilon
		_{t}^\ast\right) \notag \\
		&=&var\left( \boldsymbol{x}_{\tau}\varepsilon _{\tau}^\ast\right) \text{ i.i.d case} \notag \\
		&=&E\left( \varepsilon _{\tau}^{\ast 2}\boldsymbol{x}_{\tau}\boldsymbol{x}_{\tau}^{\prime }\right) -%
		E\left( \varepsilon _{\tau}^\ast\boldsymbol{x}_{\tau}\right) E\left( \varepsilon
		_{t}^\ast\boldsymbol{x}_{\tau}^{\prime }\right) \notag \\
		&=&E\left( \varepsilon _{\tau}^{\ast 2}\boldsymbol{x}_{\tau}\boldsymbol{x}_{\tau}^{\prime }\right) -%
		E\left( \varepsilon _{\tau}^\ast \boldsymbol{x}_{\tau}\right) E\left( \varepsilon
		_{\tau}^\ast \boldsymbol{x}_{\tau}^{\prime }\right) \notag \\
		&=&E\left( \varepsilon _{\tau}^{\ast 2}\right) E\left(
		\boldsymbol{x}_{\tau}\boldsymbol{x}_{\tau}^{\prime }\right) -E\left( \varepsilon _{\tau}^\ast \boldsymbol{x}_{\tau}\right) 
		E\left( \varepsilon _{\tau}^\ast \boldsymbol{x}_{\tau}^{\prime }\right). \hspace{7cm} \hfil \blacksquare \notag
	\end{eqnarray}%
	
	\noindent \textbf{Proof of result in (\ref{e2})}
	
	Given that $\ln \left( \frac{y_{\tau}}{\mathit{w}_n}\right) $ is exponentially distributed with
	parameter $\alpha $ and $ E\left( \ln \left( \frac{y_{\tau}}{\mathit{w}_n}\right)^{2}\right) =2(\alpha(\mathbf{X}_\tau,\boldsymbol{\Theta}))^{-2}$ we have that,
	\begin{eqnarray*}
		E\left( \left. \varepsilon _{\tau}^{\ast 2}\right\vert \boldsymbol{x}_{\tau},x_{\tau}^{\ast
		}\right) &=&E\left(\left. \left( 1-\exp \left( \boldsymbol{x}_\tau^{\prime }\boldsymbol{\beta} \right)
		\ln \left( \frac{y_{\tau}}{\mathit{w}_{n}}\right) \right) ^{2}\right\vert
		\boldsymbol{x}_{\tau},x_{\tau}^{\ast }\right) \\
		&=&E\left( \left. 1-2\exp \left( \boldsymbol{x}_\tau^{\prime }\boldsymbol{\beta} \right) \ln
		\left( \frac{y_{\tau}}{\mathit{w}_{n}}\right) +\exp \left( \boldsymbol{x}_\tau^{\prime }\boldsymbol{\beta} \right)
		^{2}\ln \left( \frac{y_{\tau}}{\mathit{w}_{n}}\right) ^{2}\right\vert \boldsymbol{x}_{\tau},x_{\tau}^{\ast
		}\right)  \\
		&=&1-2\exp \left( -\theta x_{\tau}^{\ast }\right) +\exp \left( \boldsymbol{x}^{\prime }\boldsymbol{\beta}
		\right) ^{2}\frac{2}{\exp \left( \boldsymbol{x}_{\tau}^{\prime }\boldsymbol{\beta} +\theta x_{\tau}^{\ast
			}\right) ^{2}} \\
		&=&1-2\exp \left( -\theta x_{\tau}^{\ast }\right) +\exp \left( 2\boldsymbol{x}_\tau^{\prime
		}\boldsymbol{\beta} \right) \frac{2}{\exp \left( 2\boldsymbol{x}_{\tau}^{\prime }\boldsymbol{\beta} +2\theta
			x_{\tau}^{\ast }\right) } \\
		&=&1-2\exp \left( -\theta x_{\tau}^{\ast }\right) +2\exp \left( -2\theta
		x_{\tau}^{\ast }\right). \hspace{7cm} \hfil \blacksquare \notag
	\end{eqnarray*}%
	
	\noindent \textbf{Proof of result in (\ref{HSigmaH})}
	\begin{eqnarray*}
		\mathbb{H}(\boldsymbol{\beta})^{-1}\Sigma(\boldsymbol{\beta}) \mathbb{H}(\boldsymbol{\beta})^{-1} &=&\frac{E\left( \boldsymbol{x}_{\tau}\boldsymbol{x}_{\tau}^{\prime }\right)
			^{-1}}{E\left( \exp \left( -\theta \boldsymbol{x}_{\tau}^{\ast }\right) \right) }%
		\left( E\left( \varepsilon _{\tau}^{\ast 2}\right) E\left(
		\boldsymbol{x}_{\tau}\boldsymbol{x}_{\tau}^{\prime }\right) -E\left( \varepsilon _{\tau}^{\ast}\boldsymbol{x}_{\tau}\right) 
		E\left( \varepsilon _{\tau}^{\ast}\boldsymbol{x}_{\tau}^{\prime }\right) \right) \frac{%
			E\left( \boldsymbol{x}_{\tau}\boldsymbol{x}_{\tau}^{\prime }\right) ^{-1}}{E\left( \exp
			\left( -\theta x_{\tau}^{\ast }\right) \right) } \\
		&=&\frac{E\left(\boldsymbol{x}_{\tau}\boldsymbol{x}_{\tau}^{\prime }\right) ^{-1}}{E%
			\left( \exp \left( -\theta x_{\tau}^{\ast }\right) \right) }E\left(
		\varepsilon _{\tau}^{\ast 2}\right) E\left( \boldsymbol{x}_{\tau}\boldsymbol{x}_{\tau}^{\prime }\right) 
		\frac{E\left(\boldsymbol{x}_{\tau}\boldsymbol{x}_{\tau}^{\prime }\right) ^{-1}}{E\left(
			\exp \left( -\theta x_{\tau}^{\ast }\right) \right) } \\
		& & -\frac{E\left(
			\boldsymbol{x}_{\tau}\boldsymbol{x}_{\tau}^{\prime }\right) ^{-1}}{E\left( \exp \left( -\theta
			x_{\tau}^{\ast }\right) \right) }E\left( \varepsilon _{\tau}^{\ast}\boldsymbol{x}_{\tau}\right) 
		E\left( \varepsilon _{\tau}^{\ast}\boldsymbol{x}_{\tau}^{\prime }\right) \frac{E%
			\left( \boldsymbol{x}_{\tau}\boldsymbol{x}_{\tau}^{\prime }\right) ^{-1}}{E\left( \exp \left(
			-\theta x_{\tau}^{\ast }\right) \right) } \\
		&=&\frac{E\left(\boldsymbol{x}_{\tau}\boldsymbol{x}_{\tau}^{\prime }\right) ^{-1}}{[E%
			\left( \exp \left( -\theta x_{\tau}^{\ast }\right) \right)]^2 }E\left(
		\varepsilon _{\tau}^{\ast 2}\right)   \\
		& & -[E\left(
		\varepsilon _{\tau}^{\ast}\right) ]^{2}\frac{E\left( \boldsymbol{x}_{\tau}\boldsymbol{x}_{\tau}^{\prime
			}\right) ^{-1}}{E\left( \exp \left( -\theta x_{\tau}^{\ast }\right)
			\right) }E\left( \boldsymbol{x}_{\tau}\right) E\left( \boldsymbol{x}_{\tau}^{\prime}\right) \frac{E\left( \boldsymbol{x}_{\tau}\boldsymbol{x}_{\tau}^{\prime }\right) ^{-1}}{E\left( \exp \left( -\theta x_{\tau}^{\ast }\right) \right)} \\
		&=& E\left(\boldsymbol{x}_{\tau}\boldsymbol{x}_{\tau}^{\prime }\right)^{-1}\frac{E\left( \varepsilon_{\tau}^{\ast 2}\right)}{[E\left(\exp \left( -\theta x_{\tau}^{\ast }\right) \right)]^{2}}-\frac{[E\left( \varepsilon_{\tau}^{\ast}\right)]^2}{[E\left(\exp \left( -\theta x_{\tau}^{\ast }\right) \right)]^{2}}\boldsymbol{e}_1\boldsymbol{e}_1 ' \\
		&=& E\left( \boldsymbol{x}_{\tau}\boldsymbol{x}_{\tau}^{\prime }\right) ^{-1}\frac{E%
			\left( 1-2\exp \left( -\theta x_{\tau}^{\ast }\right) +2\exp \left( -2\theta
			x_{\tau}^{\ast }\right) \right) }{[E\left( \exp \left( -\theta
			x_{\tau}^{\ast }\right) \right)] ^{2}}-B \\
		&=& E\left(\boldsymbol{x}_{\tau}\boldsymbol{x}_{\tau}^{\prime }\right)^{-1}M-B. \hspace{8cm} \hfil \blacksquare \notag
	\end{eqnarray*}

	\newpage
	
	\noindent \textbf{\large{Additional Figures}}
	
	\setcounter{figure}{0}
	\renewcommand{\thefigure}{B.\arabic{figure}}%
	
\begin{figure}[h]
		\begin{center}
			\caption{Plots of the Total commodities price index, and the price indexes of the five commodity classes considered (Energy, Industrial Metals, Precious Metals, Agriculture, and Livestock).} \label{Fig1}
			\begin{subfigure}[t]{0.3\textwidth}
				\centering
				\includegraphics[width=1\textwidth]{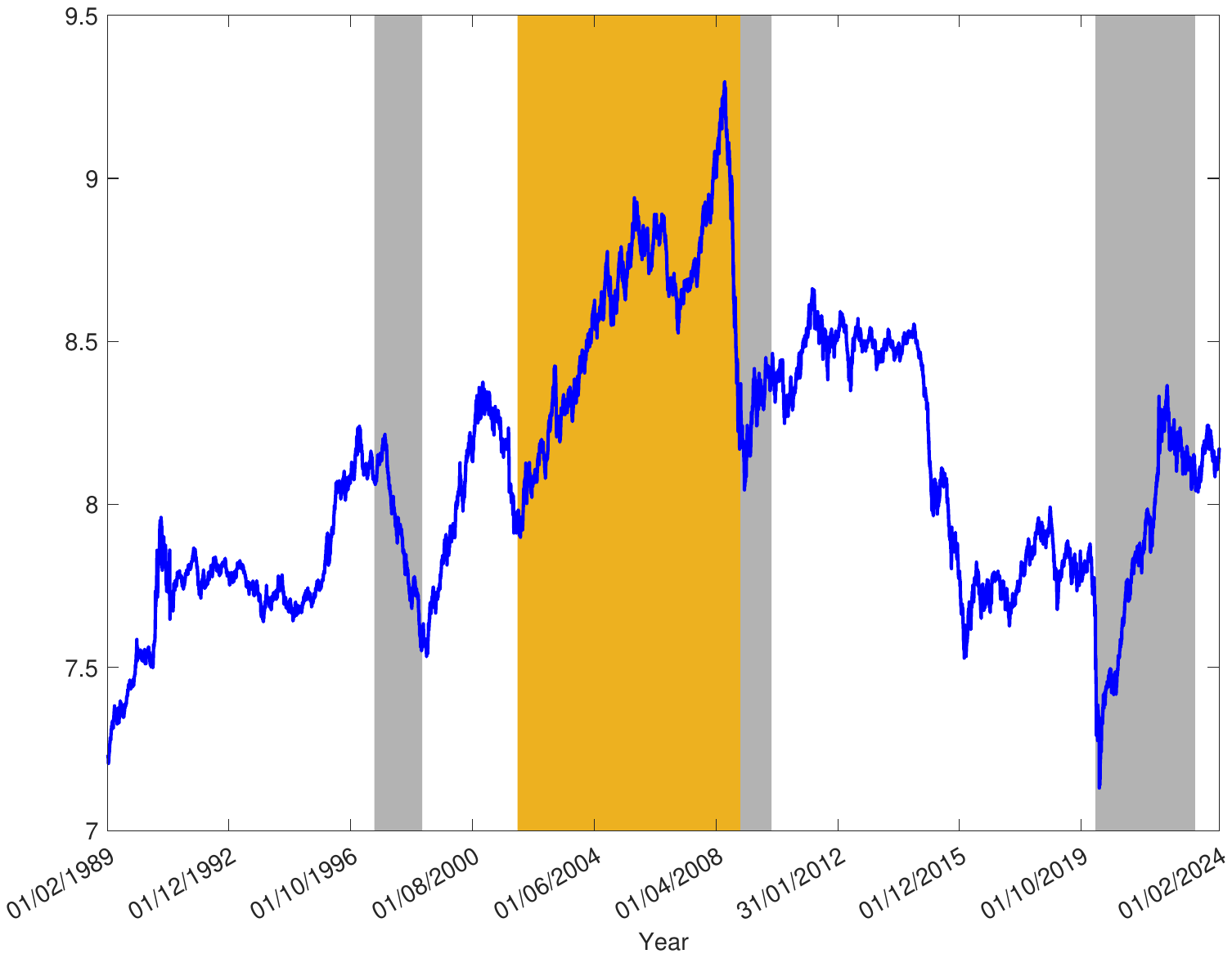}
				\caption{Total Commodities Index}
			\end{subfigure}
			~
			\begin{subfigure}[t]{0.3\textwidth}
				\centering
				\includegraphics[width=1\textwidth]{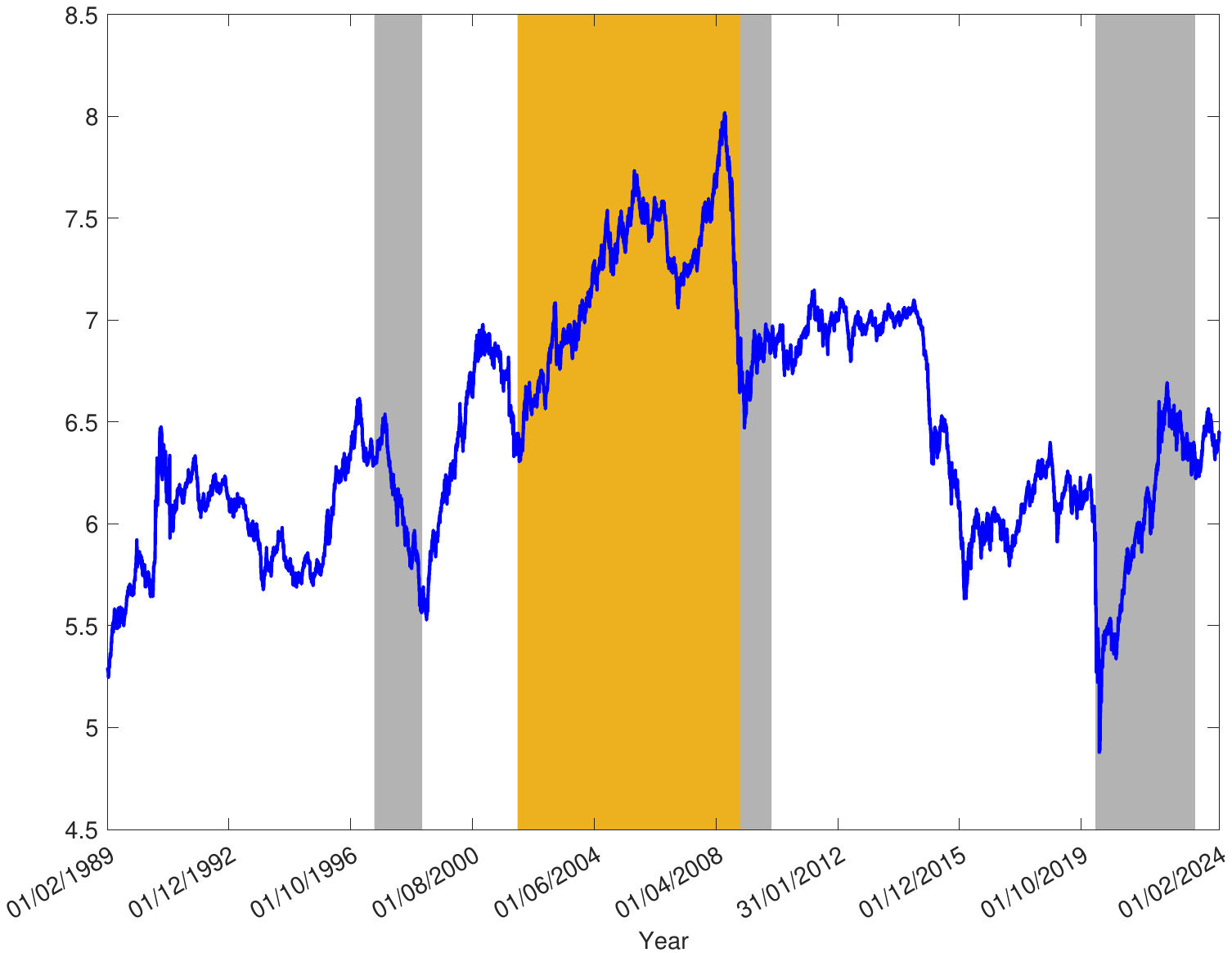}
				\caption{Energy Index}
			\end{subfigure}
			~
			\begin{subfigure}[t]{0.3\textwidth}
				\centering
				\includegraphics[width=1\textwidth]{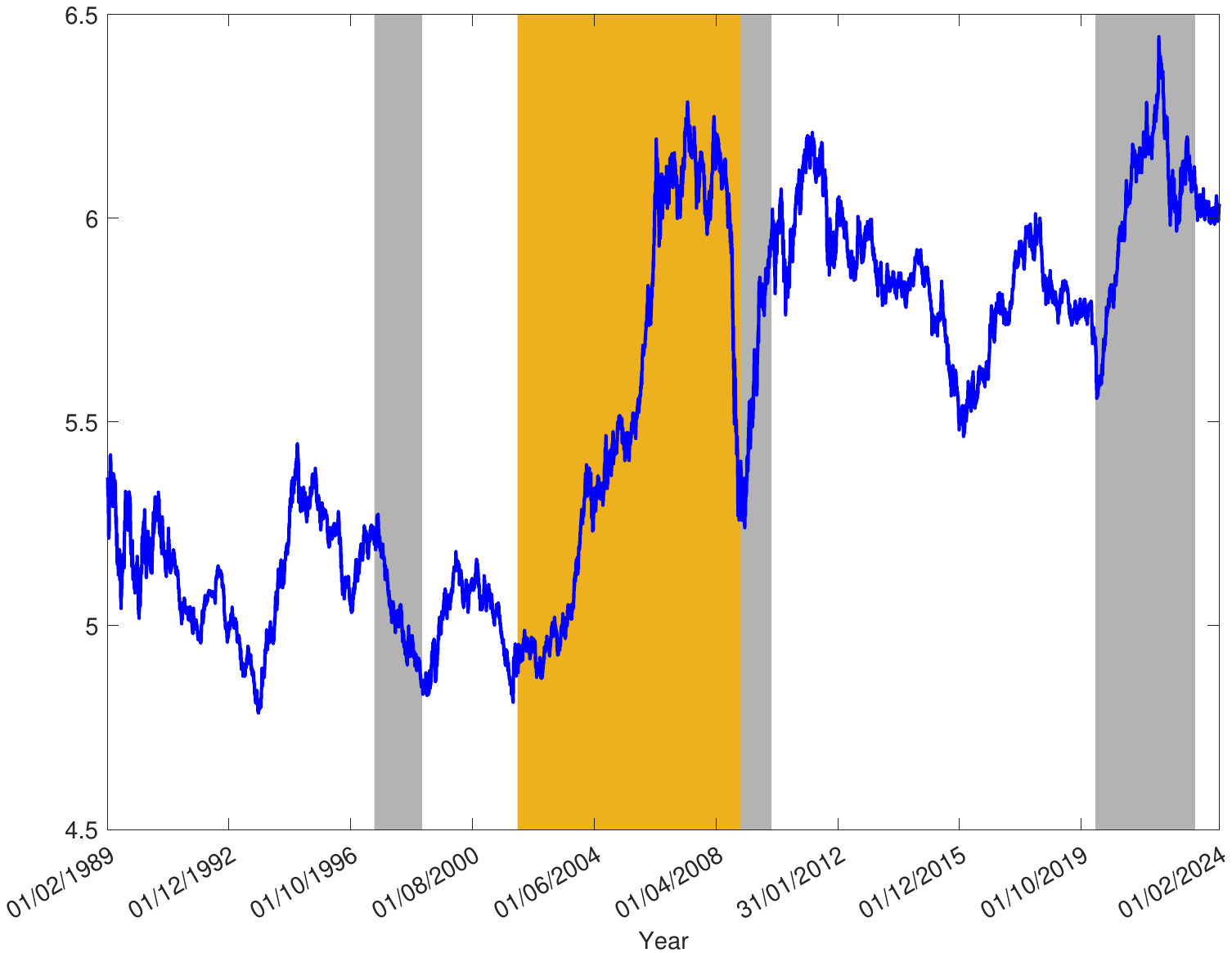}
				\caption{Industrial Metals Index}
			\end{subfigure}
			
			\begin{subfigure}[t]{0.3\textwidth}
				\centering
				\includegraphics[width=1\textwidth]{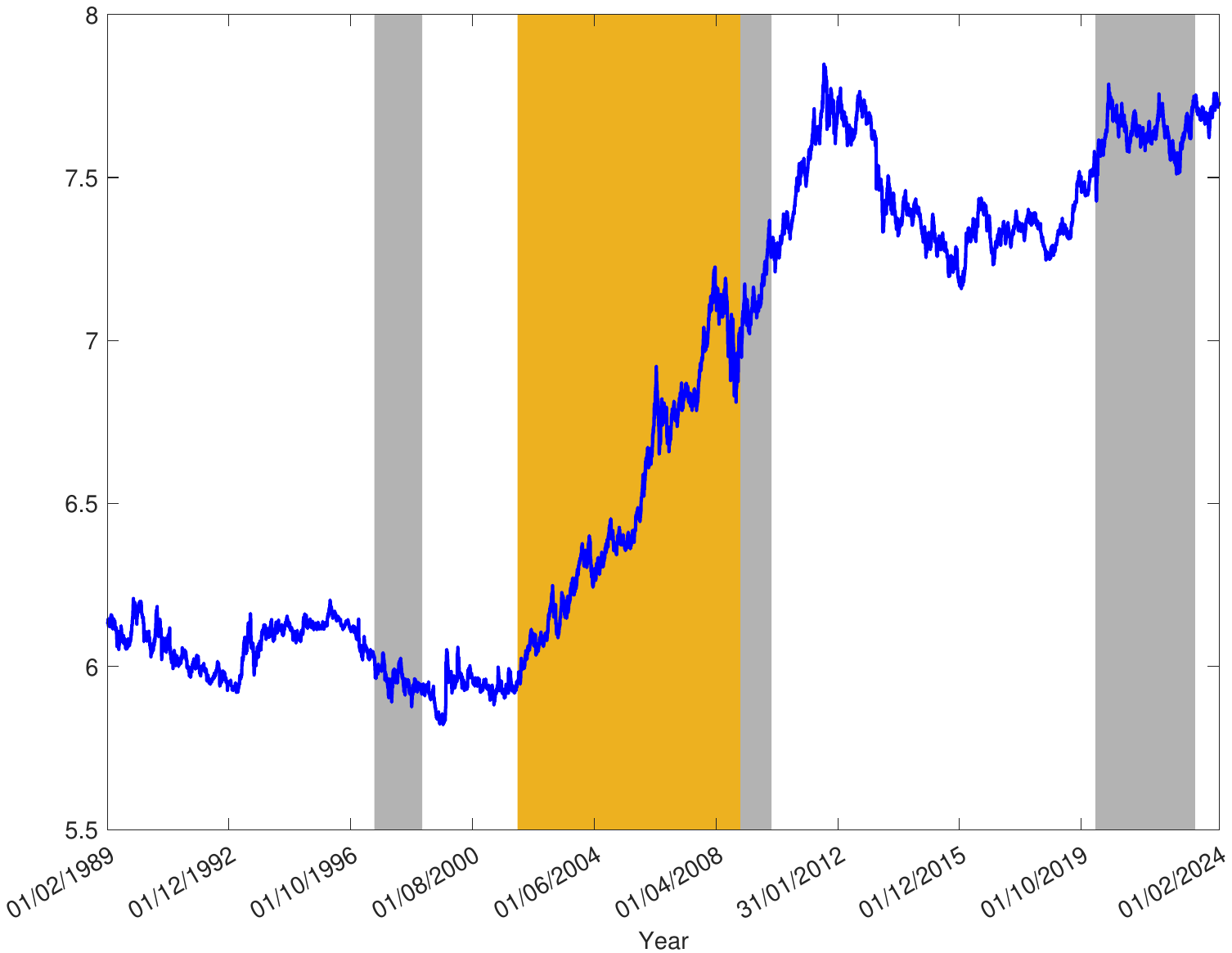}
				\caption{Precious Metals Index}
			\end{subfigure}
			~
			\begin{subfigure}[t]{0.3\textwidth}
				\centering
				\includegraphics[width=1\textwidth]{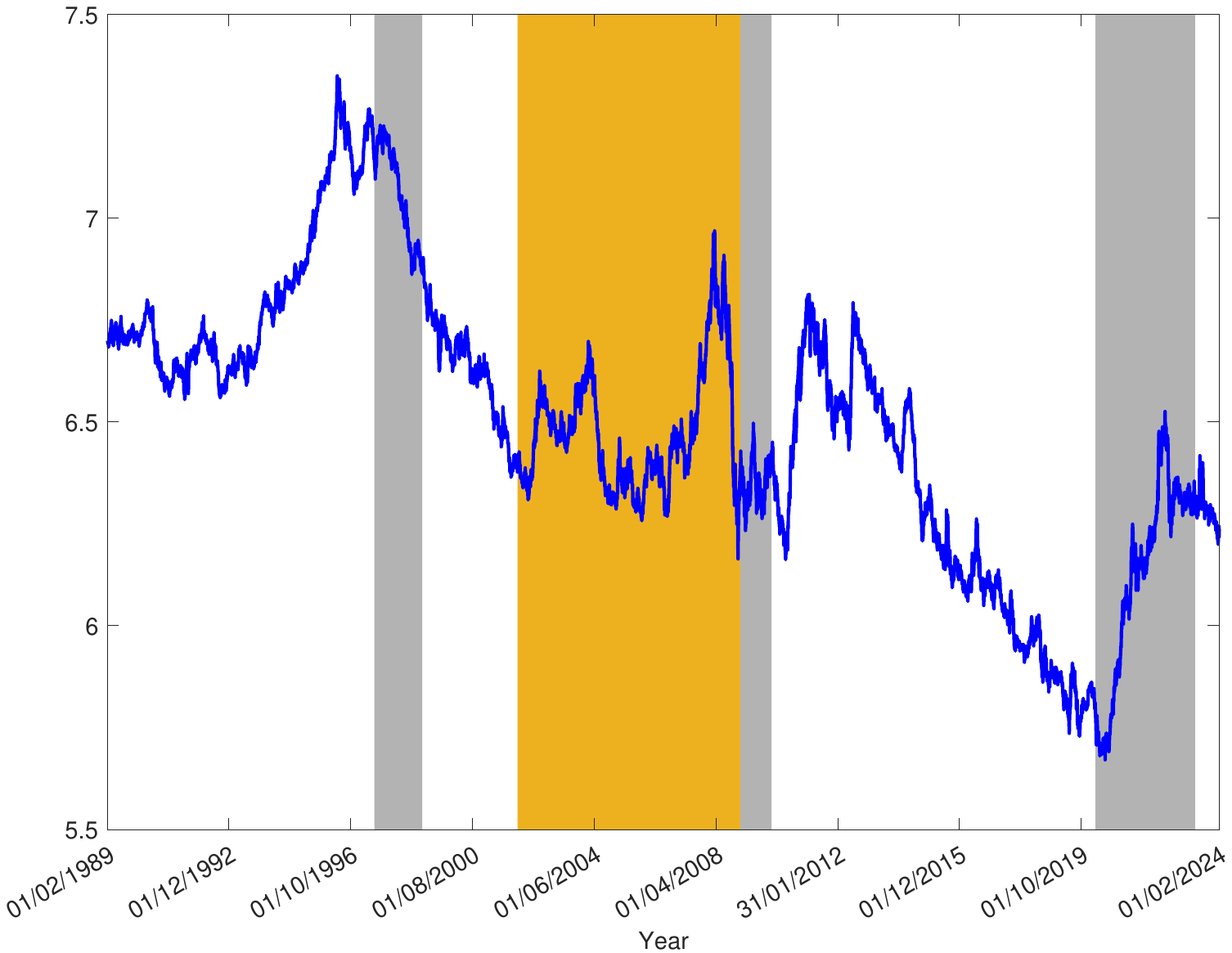}
				\caption{Agriculture Index}
			\end{subfigure}
			~
			\begin{subfigure}[t]{0.3\textwidth}
				\centering
				\includegraphics[width=1\textwidth]{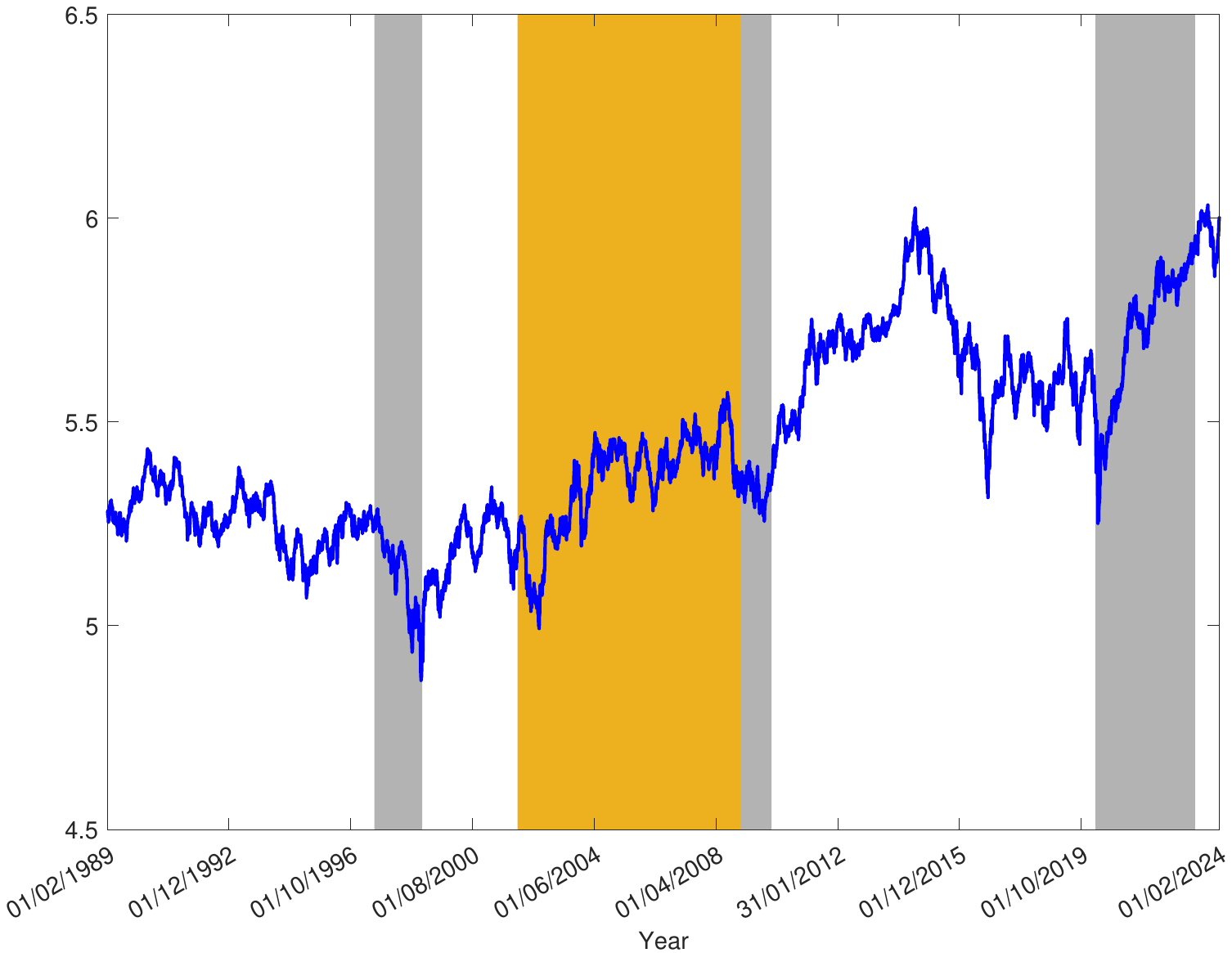}
				\caption{Livestock Index}
			\end{subfigure}
		\end{center}	
		\bigskip
		
		\textbf{Note:} The gray areas correspond to the Asian crisis (1997-1998), the sub-prime crisis (2008-2009), the Covid19 pandemics (2020-2023), and the Russian invasion of Ukrain since Feb 2022. The orange area refers to the commodities supercycle observed between 2002 - 2008.
		
	\end{figure}

	Figure \ref{Fig1} illustrates the dynamics of the commodities indexes considered: the total commodities index and the indexes of the five commodity categories under analysis. 
	
	\spacingset{1.25}
	\FloatBarrier
	\spacingset{1.9}
	
	In general, crises and expansion periods of commodity prices are clearly visible. The graphs in Figure \ref{Fig1} evidence the impact of the Asian crisis (1997-1998), the sub-prime crisis (2008-2009), the COVID19 pandemic (2020) and the Russo-Ukrainian War (Russia's invasion of Ukraine in 2022) on commodities prices (grey areas).  Figure \ref{Fig1} also displays the 2002-2008 commodities supercycle (orange area). This cycle is particularly marked in the precious metals index.\footnote{For instance, Gold prices rose from \$250 per ounce in 2001 to over \$1,800 an ounce in 2011 (an increase of 620\% in ten years), and Silver followed a similar pattern, increasing from \$3.6 per ounce in 2001 to almost \$36 per ounce in 2011; see https://kinesis.money/blog/precious-metals/what-are-commodity-supercycles-affect-precious-metals/} 
	
	\begin{figure}[h!]
		\centering
		\caption{\textbf{Time series plots of covariates}} \label{Regressors}
		
		\begin{subfigure}[t]{0.3\textwidth}
			\centering
			\includegraphics[width=1\textwidth]{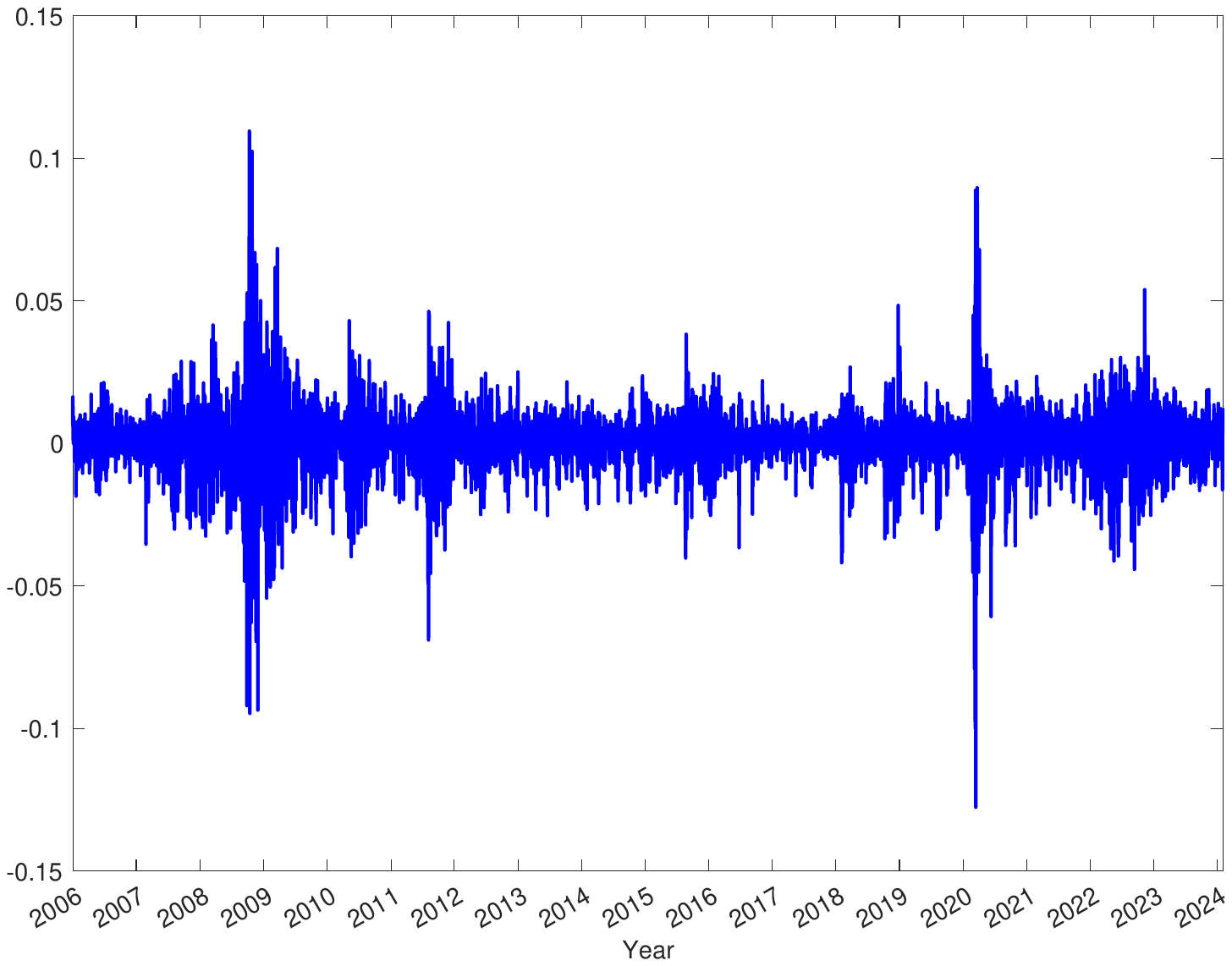}
			\caption{$\Delta S\&P500$}
		\end{subfigure}
		~
		\begin{subfigure}[t]{0.3\textwidth}
			\centering
			\includegraphics[width=1\textwidth]{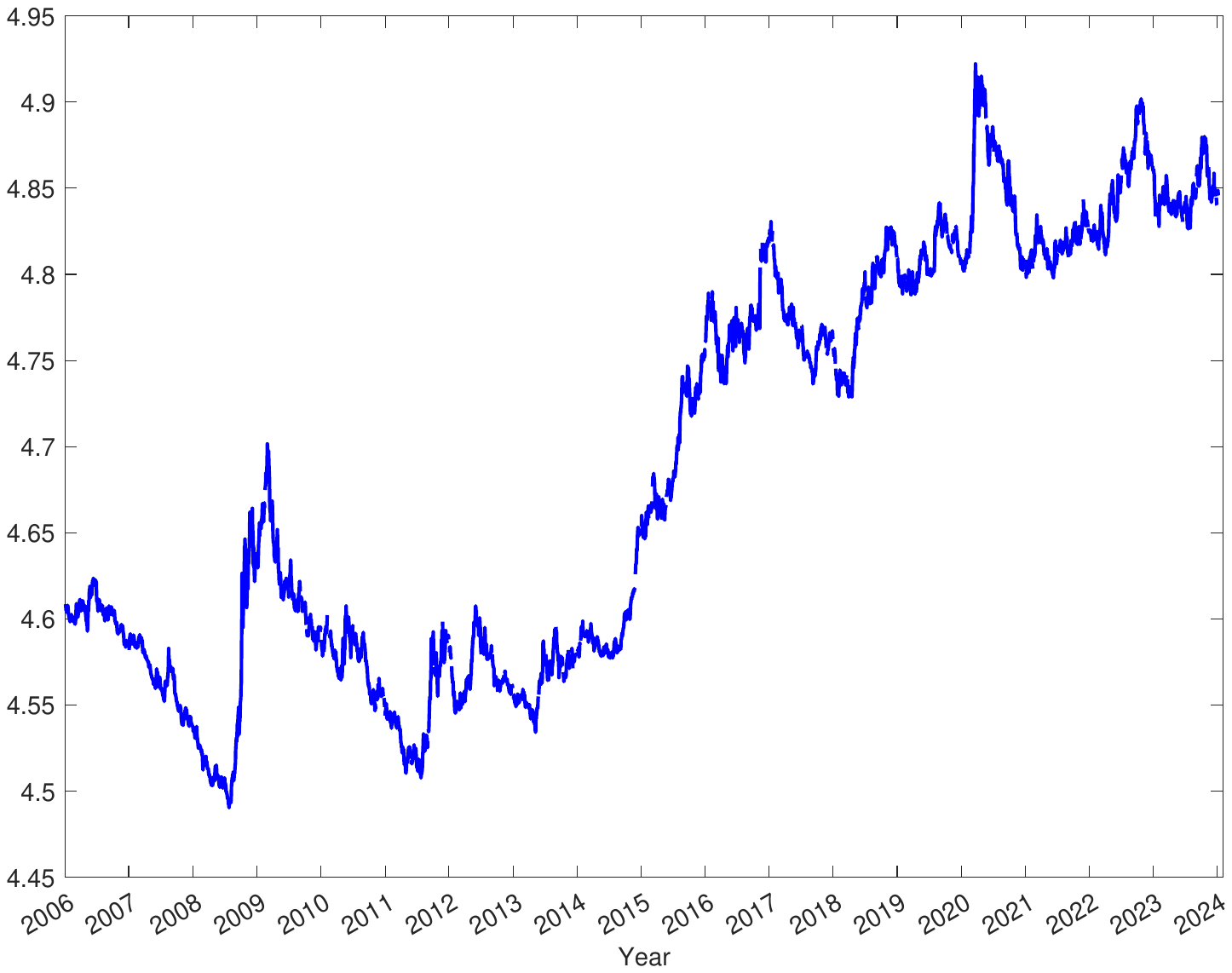}
			\caption{$EME$}
		\end{subfigure}
		~
		\begin{subfigure}[t]{0.3\textwidth}
			\centering
			\includegraphics[width=1\textwidth]{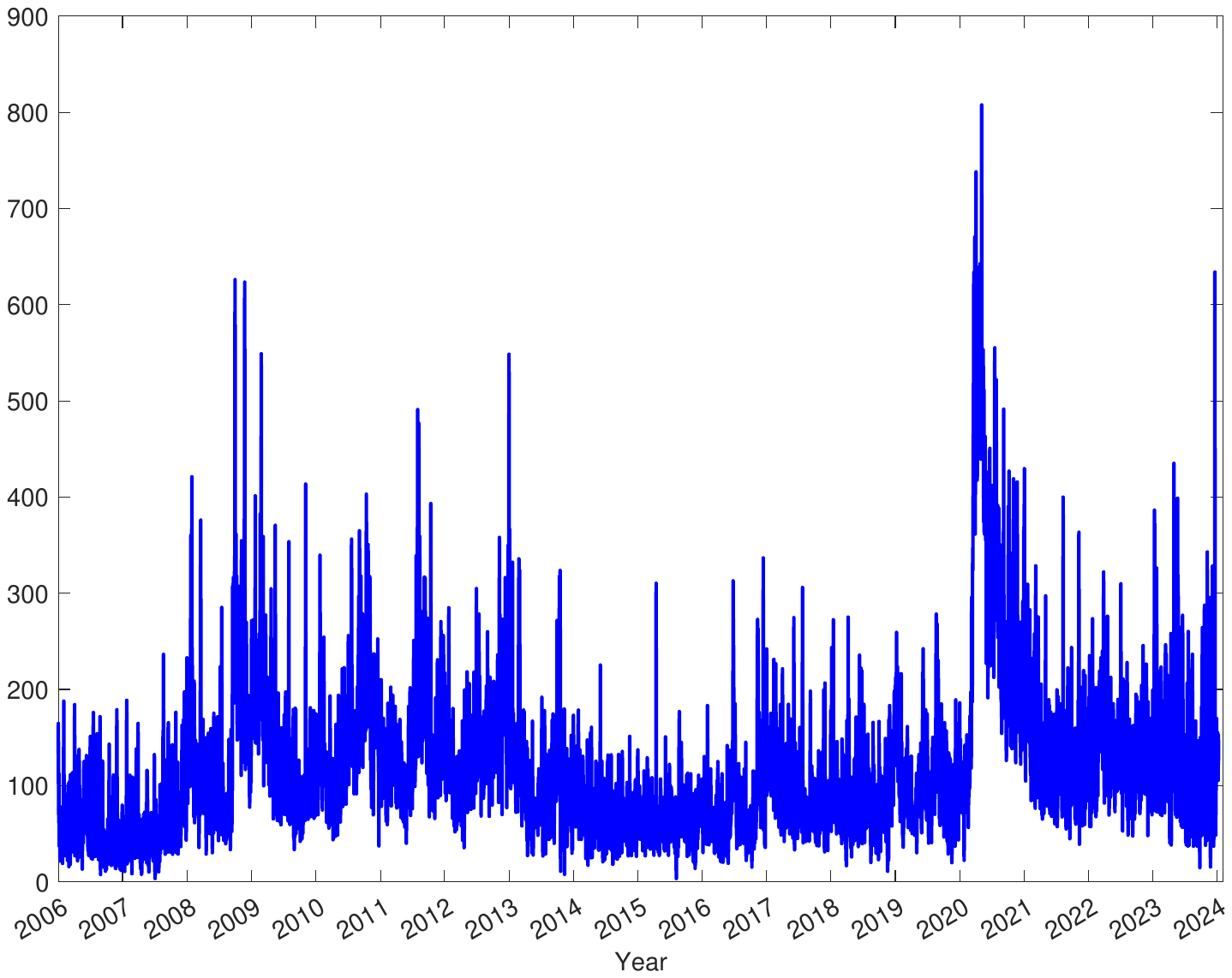}
			\caption{$EPU$}
		\end{subfigure}
		
		\begin{subfigure}[t]{0.3\textwidth}
			\centering
			\includegraphics[width=1\textwidth]{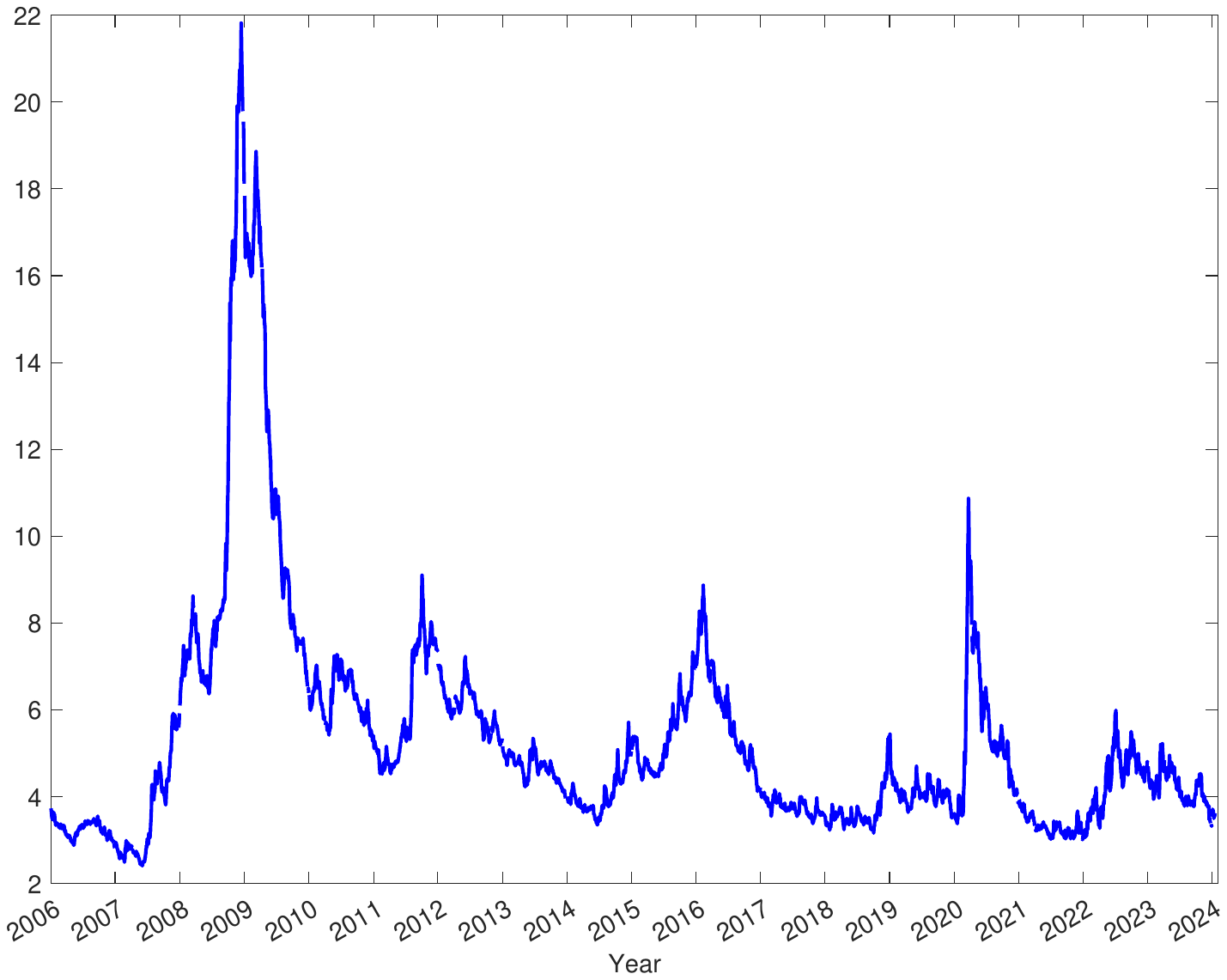}
			\caption{$HY$}
		\end{subfigure}	
		~
		\begin{subfigure}[t]{0.3\textwidth}
			\centering
			\includegraphics[width=1\textwidth]{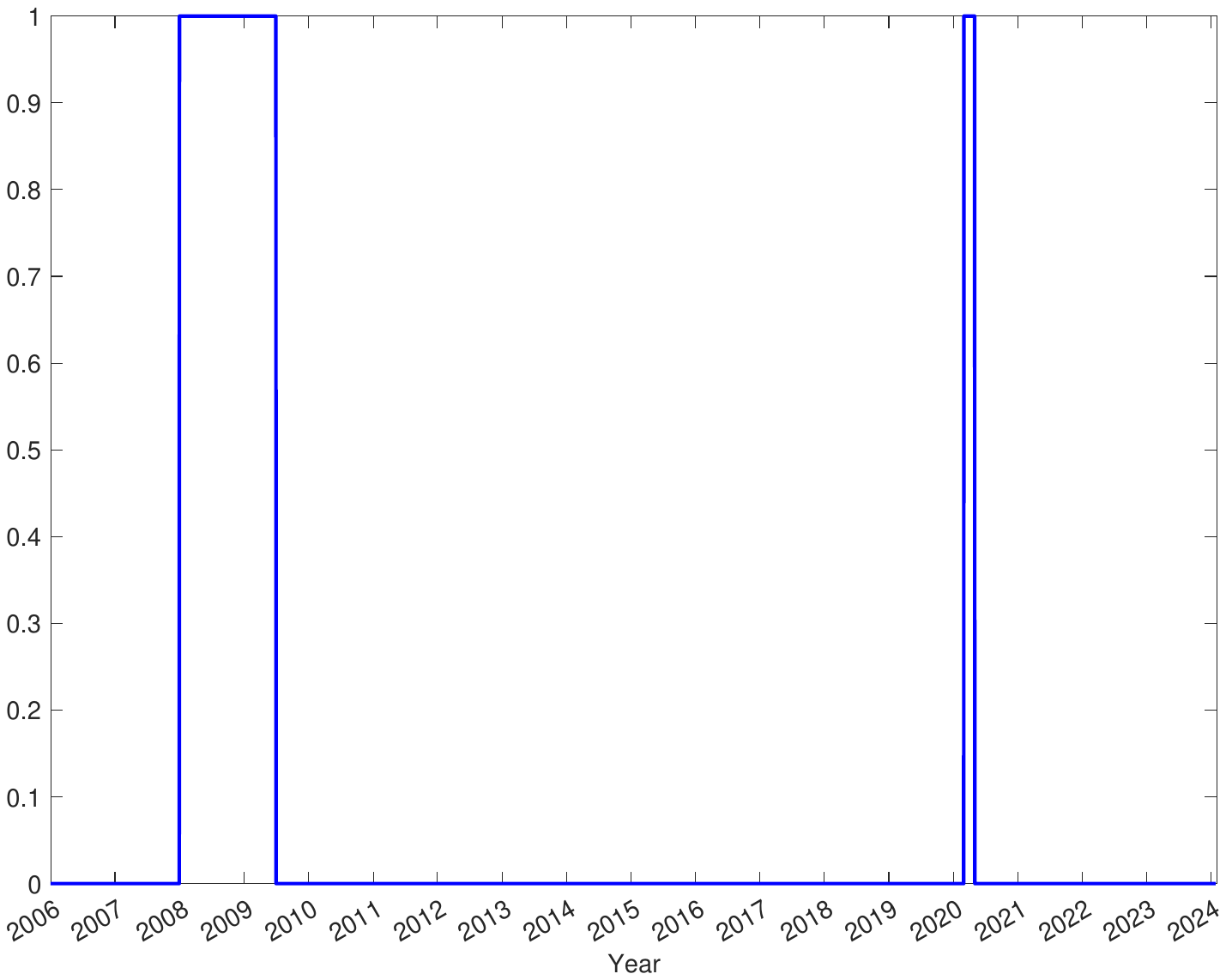}
			\caption{$NBER$}
		\end{subfigure}
		~
		\begin{subfigure}[t]{0.3\textwidth}
			\centering
			\includegraphics[width=1\textwidth]{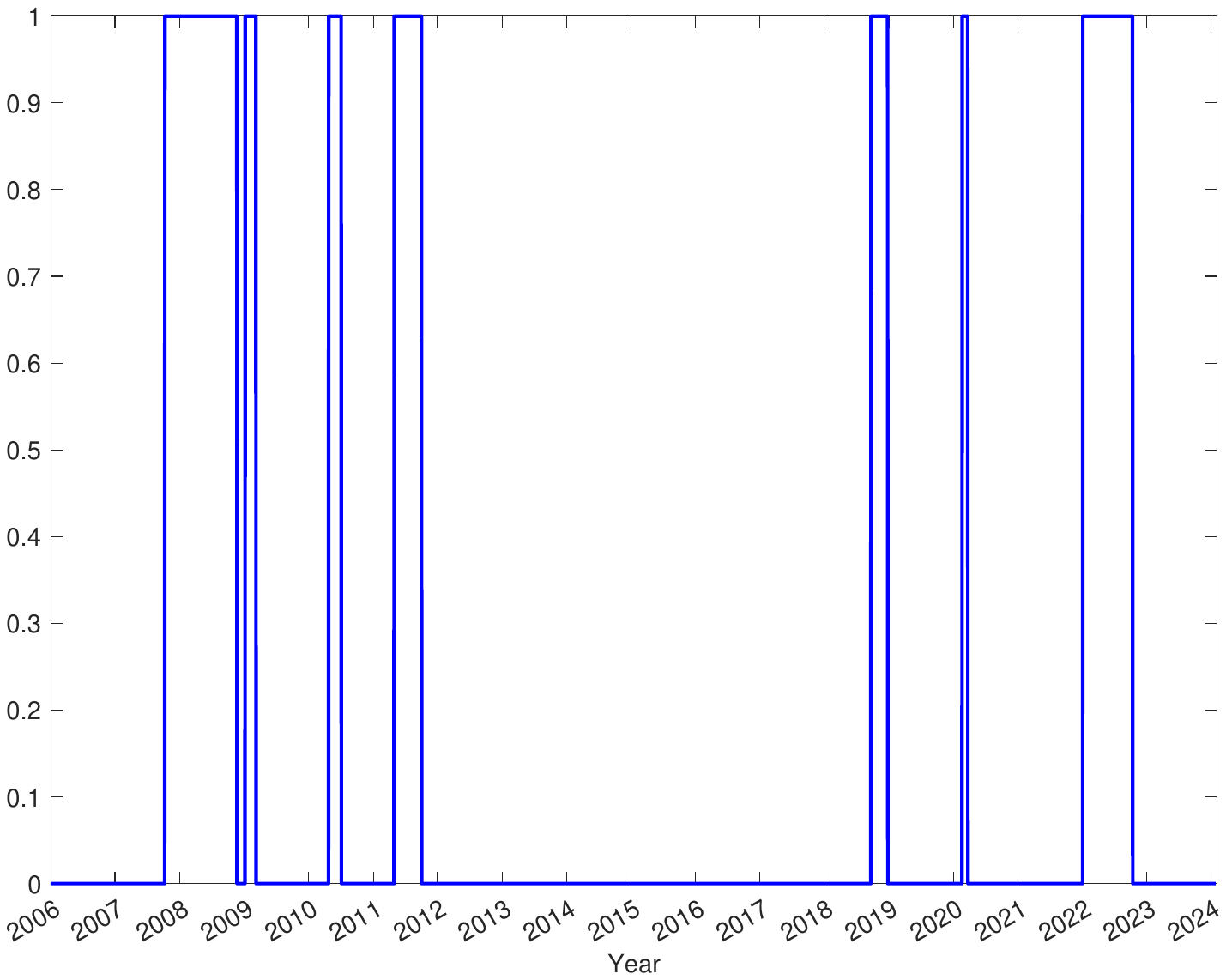}
			\caption{$bear$}
		\end{subfigure}	
		
		\begin{subfigure}[t]{0.3\textwidth}
			\centering
			\includegraphics[width=1\textwidth]{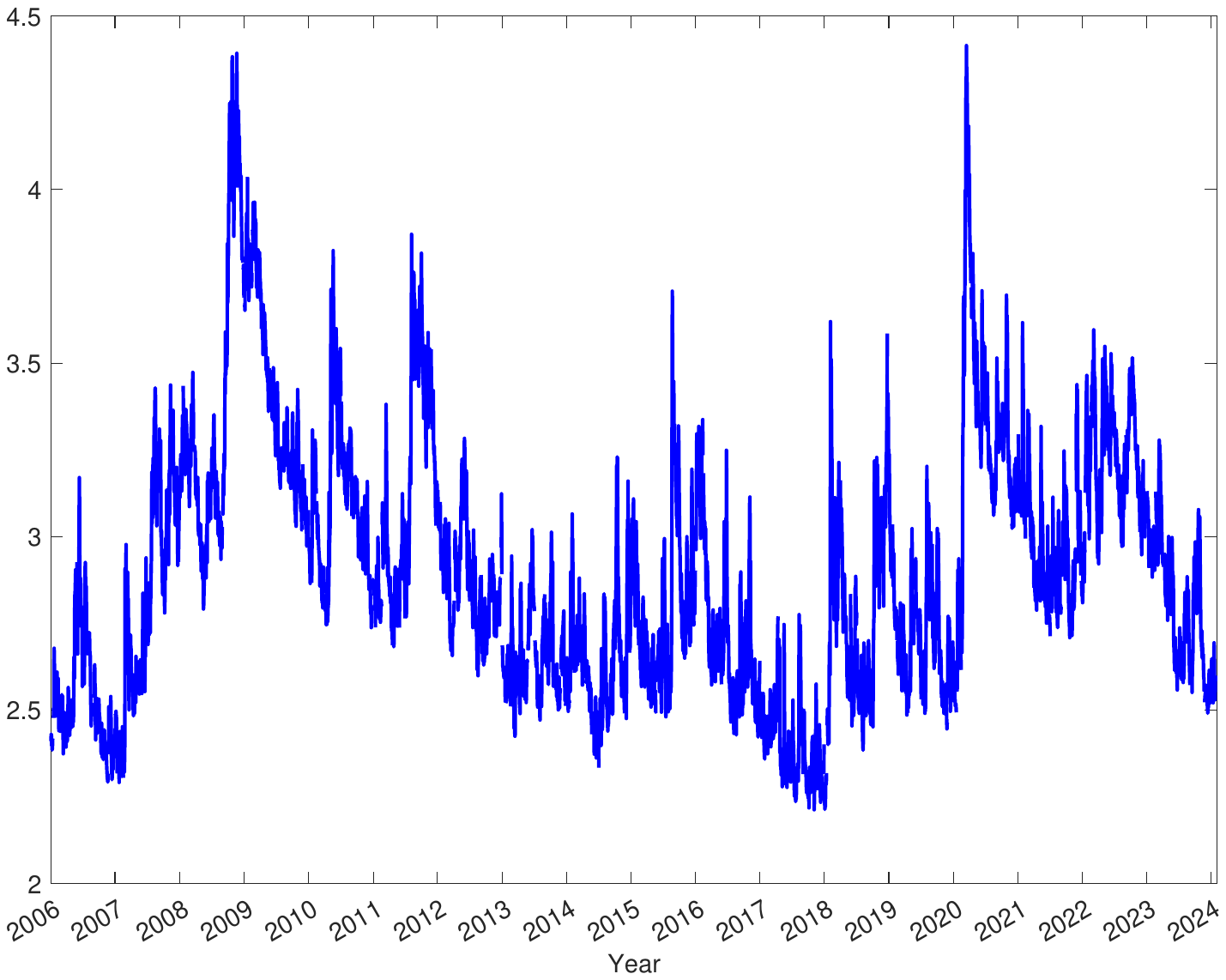}
			\caption{$vix$}
		\end{subfigure}
		~
		\begin{subfigure}[t]{0.3\textwidth}
			\centering
			\includegraphics[width=1\textwidth]{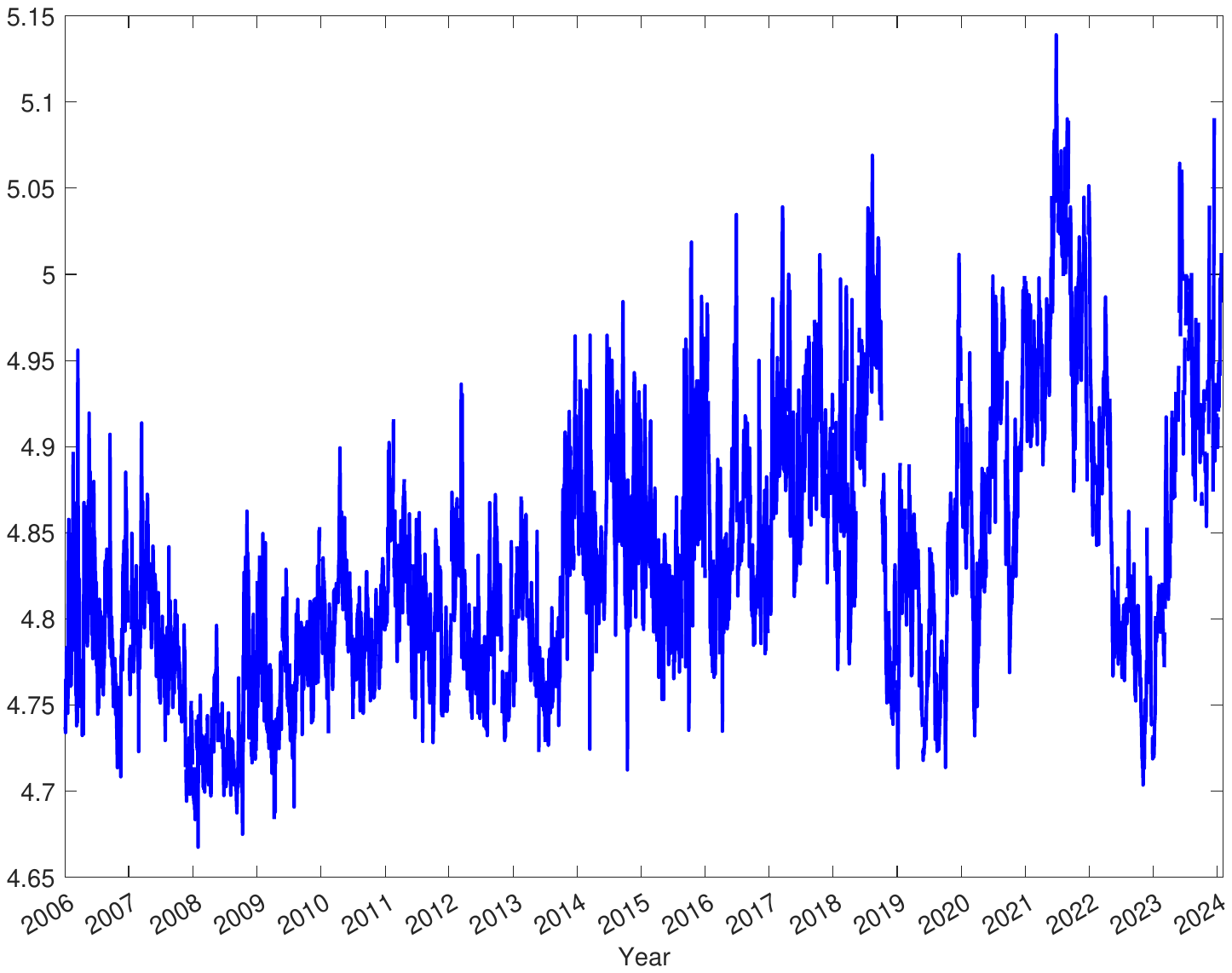}
			\caption{$skew$}
		\end{subfigure}	
		~
		\begin{subfigure}[t]{0.3\textwidth}
			\centering
			\includegraphics[width=1\textwidth]{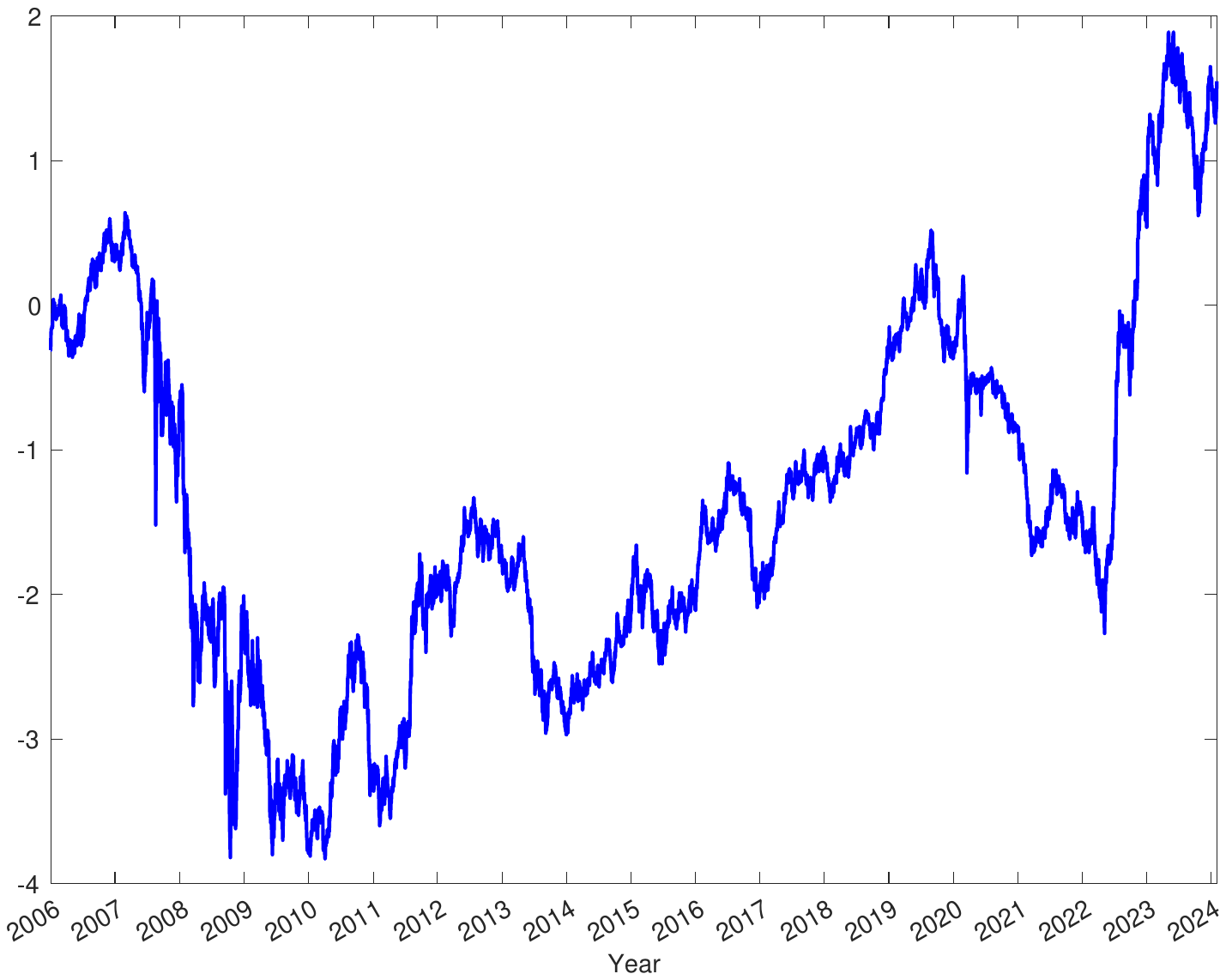}
			\caption{$slope$}
		\end{subfigure}	
	\end{figure}
	\FloatBarrier

	\begin{figure}[t!]
		\centering
		\caption{\textbf{Right-Tail Index Estimates }- $\hat{\alpha}(\boldsymbol{x}_{i\tau},\hat{\boldsymbol{\beta}}).$  A lower value of the tail index indicates a higher probability of extreme values occurring, which in turn translates into a greater level of tail risk.} \label{Fig4}
		\begin{subfigure}[t]{0.4\textwidth}
			\centering
			\includegraphics[width=1\textwidth]{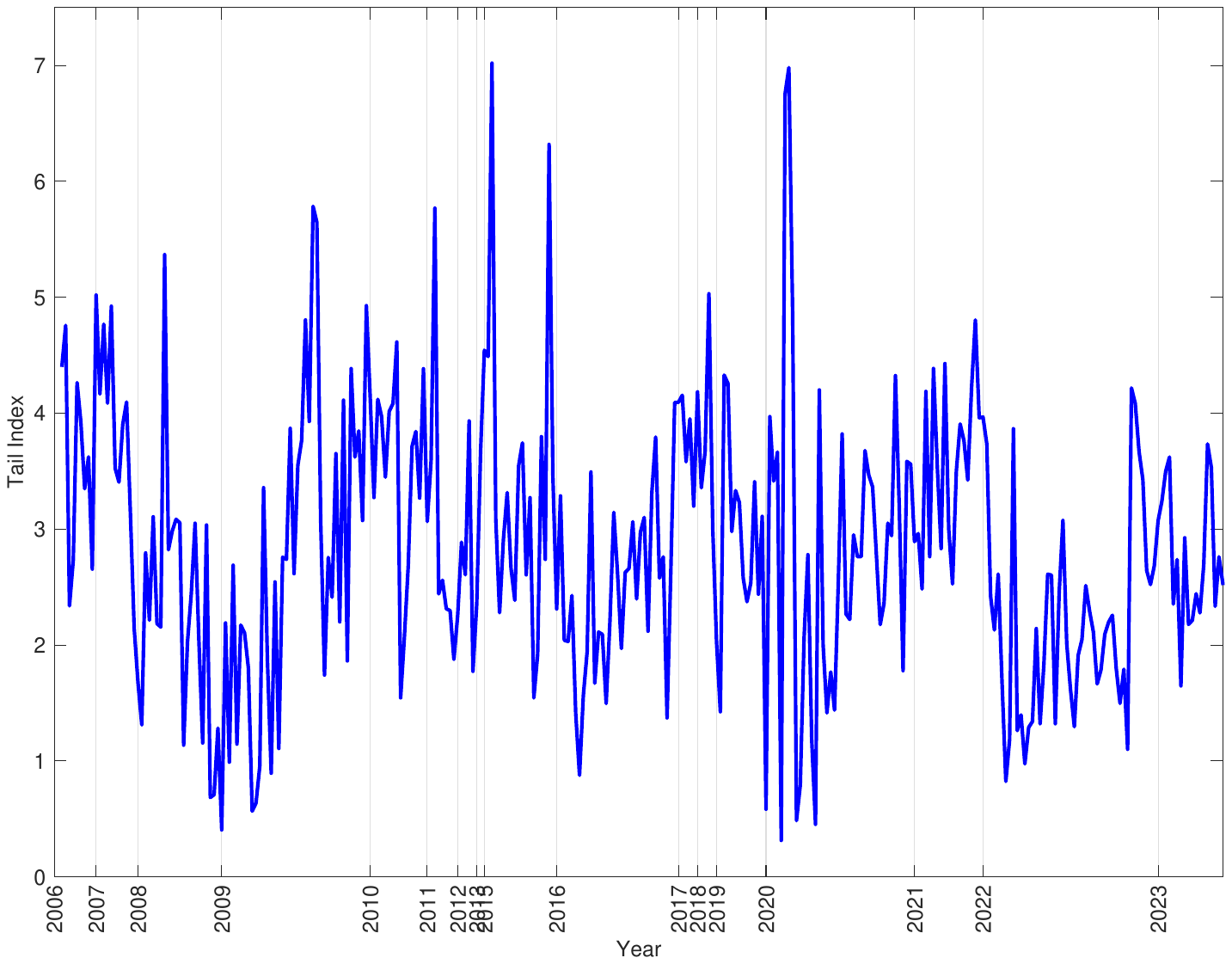}
			\caption{Gas Oil}
		\end{subfigure}
		~
		\begin{subfigure}[t]{0.4\textwidth}
			\centering
			\includegraphics[width=1\textwidth]{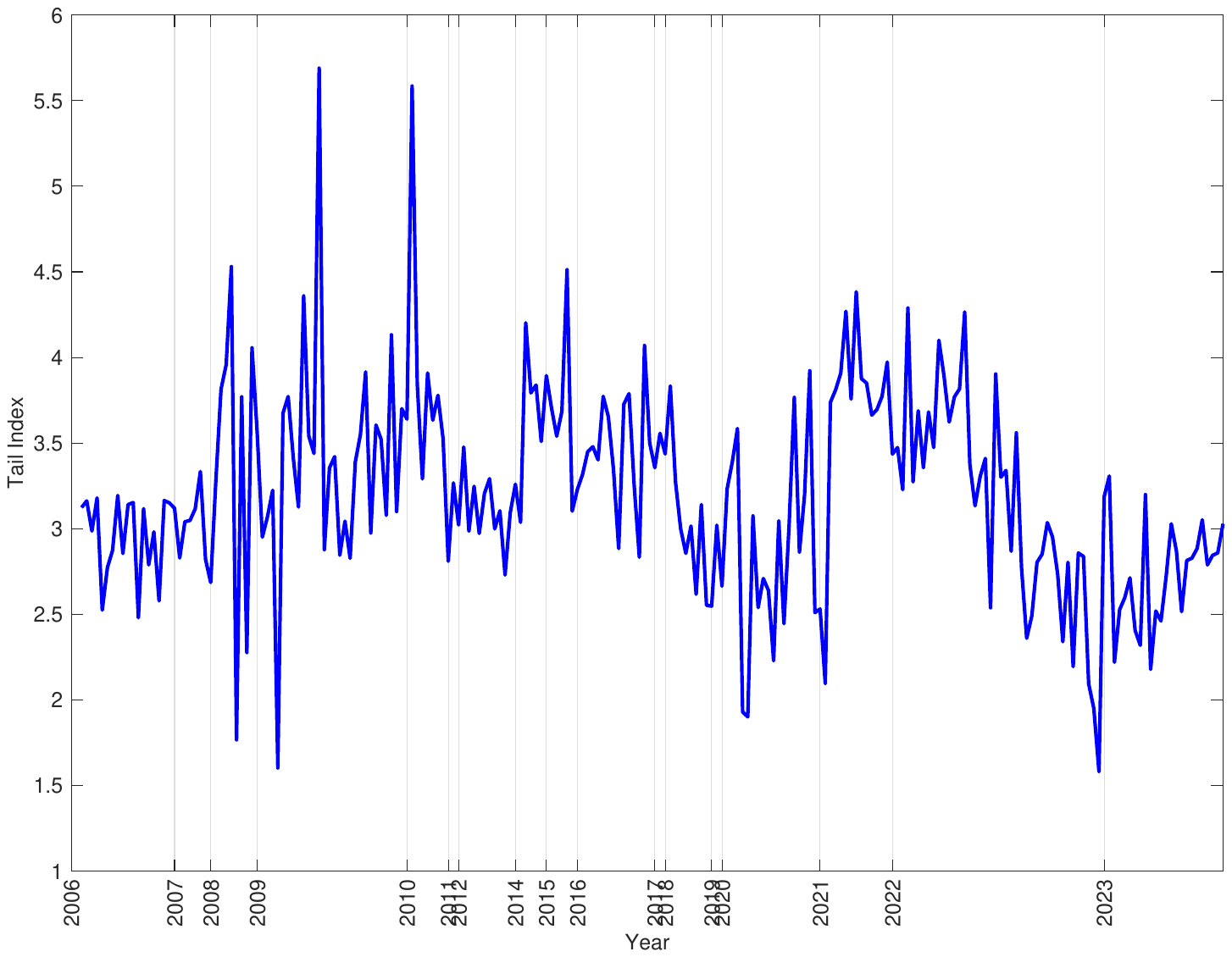}
			\caption{Natural Gas}
		\end{subfigure}
		
		\begin{subfigure}[t]{0.4\textwidth}
			\centering
			\includegraphics[width=1\textwidth]{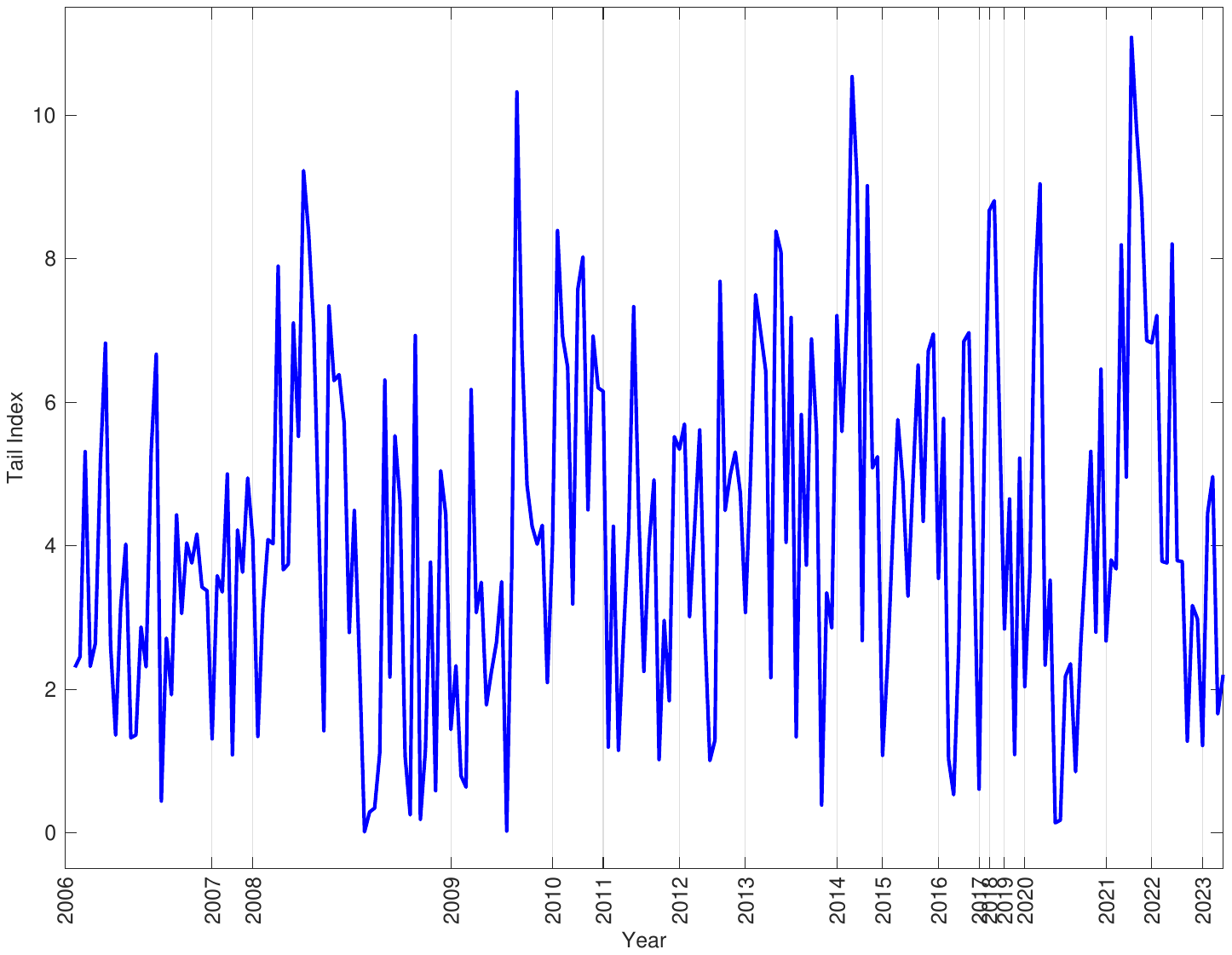}
			\caption{Gold}
		\end{subfigure}
		~
		\begin{subfigure}[t]{0.4\textwidth}
			\centering
			\includegraphics[width=1\textwidth]{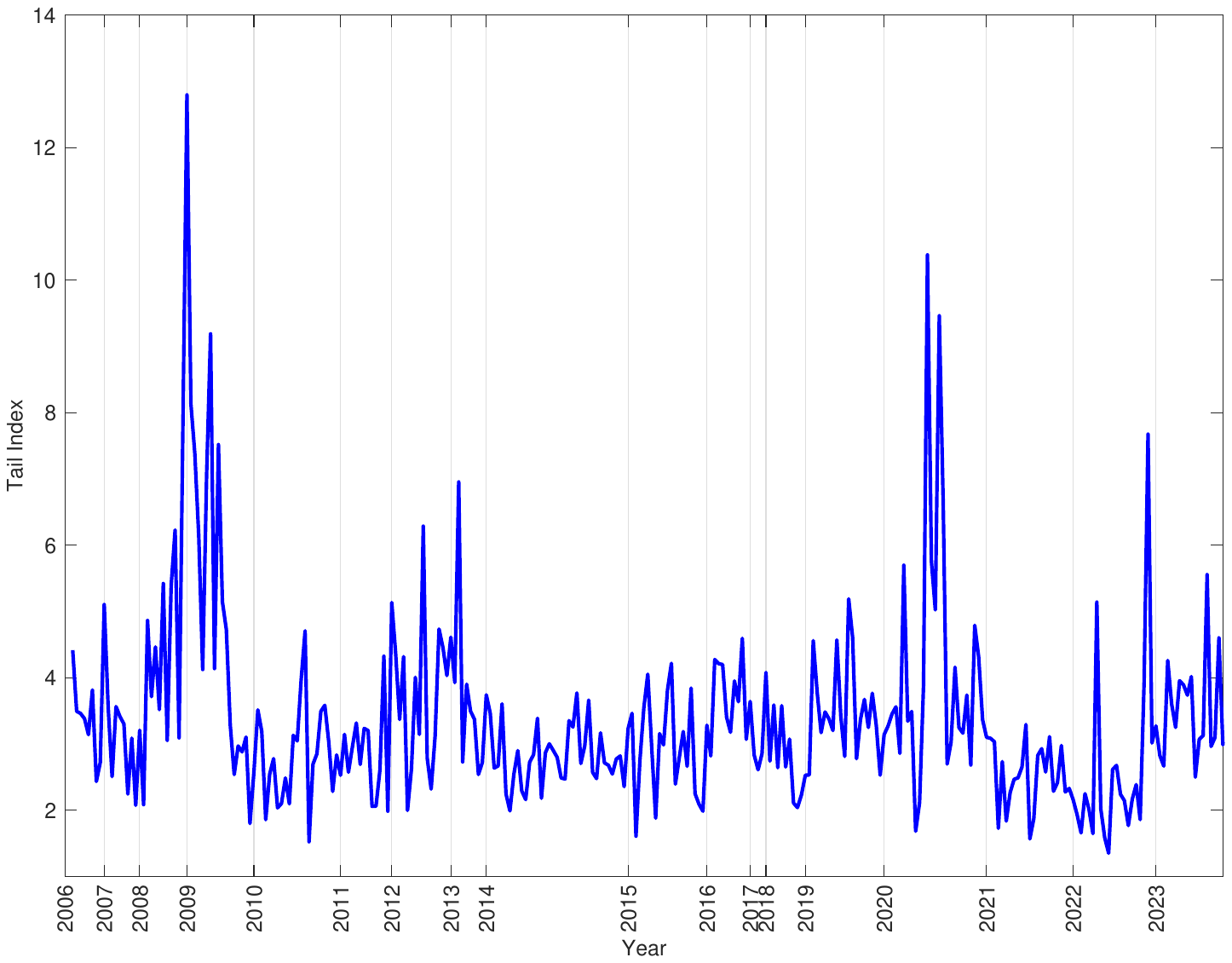}
			\caption{Coffee}
		\end{subfigure}	
	\end{figure}
	
	\FloatBarrier

\end{document}